\documentclass[twocolumn]{aa}
\usepackage{graphicx}
\usepackage[varg]{txfonts}
\usepackage{bm}
\usepackage{soul}
\usepackage[dvipsnames]{xcolor}
\usepackage{verbatim}
\usepackage[version=4]{mhchem}
\usepackage{placeins}
\def\cii{[\ion{C}{II}] }
\def\cplus{\element[+]{C} }
\def\be{\begin{equation}}
\def\ee{\end{equation}}
\def\ba{\begin{eqnarray}}
\def\ea{\end{eqnarray}}
\def\eqi{\begin{equation}}
\def\eqf{\end{equation}}
\def\eqia{\begin{eqnarray}}
\def\eqfa{\end{eqnarray}}

\usepackage{ulem}
\usepackage{multirow}
\usepackage{booktabs}
\usepackage{relsize}
\usepackage{nameref}
\usepackage{array}
\usepackage{float}
\usepackage{stfloats}
\usepackage{amsmath}

\usepackage[colorlinks = true,
            linkcolor = blue,
            urlcolor  = blue,
            citecolor = blue,
            anchorcolor = blue]{hyperref}

\usepackage[nameinlink,noabbrev]{cleveref}
\Crefname{equation}{Eq.}{Eqs.}
\Crefname{eqnarray}{Eq.}{Eqs.}
\Crefname{section}{Sect.}{Sects.}
\Crefname{figure}{Fig.}{Figs.}
\crefname{equation}{Equation}{Equations}
\crefname{section}{Section}{Sections}
\crefname{figure}{Figure}{Figures}

\usepackage{enumitem}

\SetLabelAlign{Center}{\hfil#1\hfil}
\SetLabelAlign{CenterWithParen}{\hfil(\makebox[1.0em]{#1})\hfil}

\newcolumntype{C}[1]{>{\centering\arraybackslash}p{#1}}

\newcommand{\der}{\mathrm{d}}

\newcommand{\Pshot}{P_\mathrm{shot}}
\newcommand{\Pclust}{P_\mathrm{clust}}

\newcommand{\Pwn}{P_\mathrm{WN}}

\usepackage{orcidlink}

\begin{document}

\title{Constraining the \cii luminosity function from the power spectrum of line-intensity maps at redshift 3.6}
\titlerunning{Constraining the \cii luminosity function with the LIM power spectrum}

\author{Elena~Marcuzzo$^{1,}$\thanks{\email{emarcuzzo@astro.uni-bonn.de}}\orcidlink{0009-0005-2491-8507}, Cristiano~Porciani$^{1,2,3,4}$\orcidlink{0000-0002-7797-2508}, Emilio~Romano-D\'iaz$^{1}$\orcidlink{0000-0002-0071-3217}, and Prachi~Khatri$^{1}$\orcidlink{0009-0009-1983-8333}}

\institute{$^{1}$ Argelander-Institut f\"ur Astronomie, Universit\"at Bonn, Auf dem H\"ugel 71, 53121 Bonn, Germany \\ $^{2}$ SISSA, International School for Advanced Studies, Via Bonomea 265, 34136 Trieste, Italy \\ $^{3}$ Dipartimento di Fisica -- Sezione di Astronomia, Universit\`a di Trieste, Via Tiepolo 11, 34131 Trieste, Italy \\ $^{4}$ IFPU, Institute for Fundamental Physics of the Universe, Via Beirut 2, 34151 Trieste, Italy}

\date{Received 1 April 2025 / Accepted 6 July 2025}
\authorrunning{E. Marcuzzo et al.}

\abstract
{Forthcoming measurements of the line-intensity mapping (LIM) power spectrum (PS) are expected to provide valuable constraints on several quantities of astrophysical and cosmological interest.}
{We focus on the \cii luminosity function (LF) at high redshift, which remains poorly constrained, especially at the faint end. As an example of future opportunities, we present forecasts for the Deep Spectroscopic Survey (DSS) that is to be conducted with the Fred Young Submillimeter Telescope (FYST) at $z\simeq3.6$. We also make predictions for hypothetical surveys with a ten times larger sky coverage and/or a sensitivity that is higher by a factor of $\sqrt{10}$. We account for the Lorentzian spectral profile of Fabry-Pérot interferometers and investigate the effect of their increased resolving power $R$ on the constraints.}
{Motivated by the halo-occupation properties of \cii emitters in the \textsc{Marigold} simulations, we used an abundance-matching approach to connect two versions of the ALPINE LF to the halo mass function. The resulting luminosity--mass relation was used in a halo-model framework to predict the PS signal and its uncertainty. Bayesian inference on mock PS data allowed us to forecast constraints on the first two LF moments and Schechter function parameters.}
{Depending on the true LF, the DSS is expected to be able to detect clustering and shot-noise components with signal-to-noise ratios of $\gtrsim2$. At $R=100$, spectral smoothing overwhelms the signal from redshift-space distortions, rendering the associated damping scale $\sigma$ unmeasurable. For $R\gtrsim500$, $\sigma$ can be distinguished from instrumental effects, although the degeneracies with amplitude parameters increase. Joint fits to the PS and LF yield precise constraints on the Schechter normalisation and cutoff luminosity, while the faint-end slope remains uncertain (unless the true value approaches $-2$).}
{An increased survey sensitivity offers greater gains than a wider area. A higher spectral resolution improves the access to physical parameters, but intensifies degeneracies. This highlights key design trade-offs in LIM surveys.}


\maketitle

\section{Introduction}\label{sec:intro}

Line-intensity mapping (LIM) is an emerging observational technique that takes advantage of modern imaging cameras that operate at wavelengths from the far-infrared to the radio regime \citep[see][for recent reviews]{Kovetz+17,Bernal22_review}.
It aims to map fluctuations in the intensity of redshifted radiation that is emitted in specific spectral lines over large areas of the sky without resolving individual sources.
The output consists of a data cube in which the radiation intensity is recorded as a function of the sky position and frequency. 
By assuming a cosmological model, this data cube can be transformed into a three-dimensional map of the line intensity, where the size of each voxel is determined by the angular and spectral resolution of the observations. 
LIM records the cumulative signal from all sources, including the contribution from the faintest galaxies that are missed in traditional flux-limited surveys.

Line-intensity mapping was first proposed for studying the epoch of cosmic reionisation through the 21 cm hyperfine line of atomic hydrogen \citep{Hogan-Rees-79, Scott-Rees-90, Madau+97, Furlanetto+06} in emission or absorption against the cosmic microwave background (CMB).
It was later realised that the 21 cm emission from the post-reionisation Universe might be used as a cosmological probe: Except for a multiplicative normalisation factor and an additive shot-noise term, the power spectrum (PS) of the signal from the neutral hydrogen that is locked up in galaxies and damped Lyman-$\alpha$ systems matches the matter PS on large scales, and it thus encodes cosmological information \citep{wyithe+loeb-07, chang+08}. At the same time, the normalisation and shot-noise terms can be used to constrain the luminosity function (LF) of the emitters.

In addition to the 21 cm transition, LIM has been proposed to be applied to other spectral lines by targeting different regions of the electromagnetic spectrum.  
For instance, it was suggested to employ this technique in the millimetre and centimetre bands in order to detect the cumulative emission from the first galaxies (at redshift $z>10$) through the brightest atomic gas-cooling lines \citep{Suginohara-Suginohara-Spergel-99}.
\citet{Righi+08} estimated the contribution to CMB foregrounds that is generated by redshifted rotational transitions of the CO molecule and the \cii fine-structure line from singly ionised carbon and concluded that LIM experiments would play a key role in reducing theoretical uncertainties.

Later, CO transitions, \cii and, more recently, [\ion{O}{III}] were suggested as possible tracers of the large-scale structure (LSS) of the Universe at high redshift
\citep[e.g.][]{Visbal-Loeb-10, Carilli-11, Lidz+11, Gong+12, Pullen+13, Pullen+18, Breysse+14, Dumitru+19, Padmanabhan19_CII, Padmanabhan22_oct}.
Similarly, the redshifted Ly$\alpha$ line of atomic hydrogen was considered for LIM experiments in the near-infrared \citep{Silva+13, Pullen+14}.

In the past decade, there has been an ever-increasing activity in proposing applications of LIM to miscellaneous topics in astrophysics \citep[e.g.][]{Lidz+09,Gong+12,Visbal+15,Comaschi-Ferrara_16,Breysse-Rahman_17} and cosmology \citep[e.g.][]{Karkare+18,Bernal+19,Azadeh19,Munoz+20,Bauer+21,Bernal+21,Azadeh22}.
This fervid forecasting endeavour provided the basis for developing
about $30$ dedicated instruments\footnote{See \url{https://lambda.gsfc.nasa.gov/product/expt/lim_experiments.html} and references therein.} for LIM from the ground, balloon based, and from space. 

Unlocking the full potential of LIM experiments requires a careful characterisation and mitigation of systematic effects that contaminate the measurements. These include foregrounds and backgrounds with continuous spectra (due to radio-frequency interference, the atmosphere, the Galaxy, the cosmic infrared background, and the CMB, depending on the wavelength) as well as spectral line interlopers (i.e. line emission from different transitions that is redshifted at the same observed frequencies).
Developing efficient foreground-cleaning techniques is a very active research field, and numerous different methods were proposed \citep[e.g.][]{Breysse+15, Silva+15, Sun+18}.
Detections of the LIM signal were originally achieved through cross-correlations with galaxy surveys for the 21 cm line \citep[e.g.][]{Masui+13, Anderson+18, CHIME+22, Wolz+22} and \cii \cite{Pullen+18}.
Recently, a direct detection of the \ion{H}{I} PS at $0.32<z<0.44$ \citep{Paul+23} and tentative detections of the shot-noise PS from rotational CO lines \citep{Keating+16, Keating+20, Ihle+22, Stutzer+24} have been obtained.

In this work, we explore the potential of the LIM PS to constrain the \cii LF at redshift $z>3.5$, when the Universe was younger than 1.8 Gyr.
With the advent of new observational facilities such as the Atacama Large Millimeter/sub-millimeter Array (ALMA) and the Northern Extended Millimeter Array (NOEMA), it is now possible to routinely detect \cii line emission from individual high-redshift galaxies, and thus, to probe the physical conditions of their interstellar medium. It is, however, extremely challenging to conduct wide surveys and collect samples that are statistically representative of the underlying population (see Sect.~\ref{sec:LF} for further details).
Hence, the \cii LF at such early times still remains very poorly constrained, particularly at the faint end.
Knowledge of this quantity, however, would likely allow us to determine the evolution of the cosmic star formation rate density in a way that is unaffected by dust obscuration. In addition, it would provide a stringent test of galaxy-formation models that are able to predict \cii emission \citep[e.g., among others,][]{vallini15, popping+16, Olsen+17, Lagache+18, Lupi+18, Leung+20, Khatri+24_marigold}.

As an example of the forthcoming capabilities that will enable the detection of the \cii LIM signal, we use the specifics of the Deep Spectroscopic Survey (DSS) as a reference setup. This survey will be conducted with the 6-meter Fred Young Submillimeter Telescope (FYST), which is located near the top of Cerro Chajnantor at an elevation of 5600 m in the Atacama desert \citep{CCATprime_collab}.
We also consider the impact of larger sky coverages and/or higher sensitivities.
Moreover, we investigate the effect of changes in the spectral resolving power on the measurement of parameters related to redshift-space distortions.
In all cases, we focus on a narrow redshift interval centred around $z\simeq3.6$.

The paper is organised as follows. 
In Sect.~\ref{sec:halomodel} we outline the halo model for the LIM PS.
The current state of measurements of the \cii LF at high redshift is summarised in Sect.~\ref{sec:CII_emission}, where we also present the analysis of the \textsc{Marigold} simulations and introduce the abundance-matching technique.
In Sect.~\ref{sec:LIM_powerspectrum} we present our predictions for the LIM PS and its uncertainty.
In Sect.~\ref{sec:bayesian_inference_results} we describe our Bayesian-inference pipeline and present the results we obtained from the analysis of mock data.
In Sect.~\ref{sec:conclusions} we finally summarise our findings.

We adopt a flat Friedmann-Lema\^itre-Robertson-Walker cosmological background with dimensionless Hubble constant $h=0.674$ and present-day density parameters $\Omega_\mathrm{m}=0.315$, $\Omega_\mathrm{b}=0.049$, and $\Omega_\Lambda=0.685$ for matter, baryons, and the cosmological constant, respectively.
The PS of primordial density perturbations is characterised by the spectral index $n_\mathrm{s}=0.965$ and the normalisation factor $\sigma_8=0.811$.
We compute the linear PS in the standard $\Lambda$CDM scenario with the Code for Anisotropies in the Microwave Background \citep[\texttt{CAMB}\footnote{\url{https://camb.info/}},][]{Lewis+00}.

\section{Halo model for LIM}
\label{sec:halomodel}
In the absence of absorption and scattering (and neglecting redshift corrections due to peculiar velocities), the specific intensity of radiation detected at frequency $\nu_\mathrm{o}$ along the line of sight $\hat{\mathbf{n}}$ by an observer at redshift zero is
\begin{equation}
    I_\nu(\nu_\mathrm{o},\mathbf{n})=\frac{1}{4\pi} \int_0^\infty \epsilon_\nu[(1+z)\,\nu_\mathrm{o},\hat{\mathbf{n}},z]\,\frac{1}{1+z}\,\frac{\der \chi}{\der z}\,\der z\;,
     \label{eq:radtransf}
\end{equation}
where $\epsilon_\nu(\nu_\mathrm{e},\hat{\mathbf{n}},z)$ is the comoving-volume emissivity at rest-frame frequency $\nu_\mathrm{e}$ due to sources at redshift $z$ and
\begin{equation}
    \frac{\der \chi}{\der z}=\frac{c}{H(z)}
\end{equation}
denotes the comoving radial distance per unit redshift, with $H$ the Hubble parameter.
Considering line emission with a frequency spectrum that can be approximated with a Dirac delta function, we can write
\begin{equation}
    \epsilon_\nu(\nu_\mathrm{e},\hat{\mathbf{n}},z)=
    \rho_L(\hat{\mathbf{n}},z)\,\delta_\mathrm{D}(\nu_\mathrm{e}-\nu_\mathrm{rf})\;,
\end{equation}
where $\rho_L$ denotes the total luminosity emitted per unit comoving volume and $\nu_\mathrm{rf}$ is the rest-frame central frequency of the transition.
Replacing this expression in Eq.~(\ref{eq:radtransf}) gives
\begin{equation}
     I_\nu(\nu_\mathrm{o},\mathbf{n})=\frac{1}{4\pi\nu_\mathrm{rf}}\, \rho_L(\hat{\mathbf{n}},z_*)  \,\frac{\der \chi}{\der z}(z_*)
     =\frac{c}{4\pi H(z_*)\,\nu_\mathrm{rf}}\,
     \rho_L(\hat{\mathbf{n}},z_*) \;,
\label{eq:I-rho}
\end{equation}
which shows that the signal observed at frequency $\nu_\mathrm{o}$ is fully generated at redshift $z_*=\nu_\mathrm{rf}/\nu_\mathrm{o}-1$.
This signal is difficult to isolate from observations because of the presence of much more luminous foregrounds with continuum spectra and various interloper lines.  
Dedicated techniques are being developed to separate the signal from the spectrally smooth foregrounds and mitigate the impact of the interlopers \citep[e.g.][]{Alonso+15,Li+19,Karoumpis+24,Roy-Battaglia_24, Bernal-Baleato_25}. 

\subsection{Mean signal}
The mean specific intensity over the sky is 
\begin{equation}
    \bar{I_\nu}(\nu_\mathrm{o}) =\frac{c}{4\pi H(z_*)\,\nu_\mathrm{rf}}\,\bar{ \rho}_L(z_*)\;,
    \label{mean_specific_intensity}
\end{equation}
where the mean comoving luminosity density $\bar{\rho}_L(z)$ coincides with the first moment of the LF of line emitters at fixed redshift,
\begin{equation}
    \bar{\rho}_L(z)= \int_0^\infty L\,\Phi(L,z)\,\der L\;.
\end{equation}
With a little abuse of notation, in the remainder of this paper, we will write $\bar{I}_\nu(z)$ to indicate $\bar{I}_\nu(\nu_\mathrm{o})$ with $\nu_\mathrm{o}=\nu_\mathrm{rf}/(1+z)$.

\subsection{Power spectrum}
\label{sec:power}
The spatial fluctuations around the mean signal, that is, $\delta I_\nu(\nu_\mathrm{o},\hat{\mathbf{n}})=I_\nu(\nu_\mathrm{o},\hat{\mathbf{n}})-\bar{I}_\nu(\nu_\mathrm{o})$, encode precious astrophysical and cosmological information.
By adopting a fiducial cosmological model, it is possible to convert the pair of observables $(\nu_\mathrm{o},\hat{\mathbf{n}})$ into the position vector $\mathbf{x}=\chi(z_*)\,\hat{\mathbf{n}}$ and thus build a three-dimensional map of $\delta I_\nu$ on the past light cone of the observer.
The information content of the map is then compressed into clustering summary statistics such as the PS.

Assuming that line emission takes place only within dark-matter (DM) haloes provides a particularly convenient framework to model the statistical properties of $\delta I_\nu$.
The key ingredient is the conditional luminosity function (CLF), $\phi(L|M,z)$, which gives the differential distribution of the number of galaxies hosted, on average, within haloes of mass $M$ and redshift $z$, as a function of their line luminosity. By definition,
\begin{equation}
    \Phi(L,z)=\int_0^\infty \phi(L|M,z)\,\frac{\der \bar{n}_\mathrm{h}}{\der M}(M,z)\,\der M\;,
    \label{eq:CLF-def}
\end{equation}
where $\der \bar{n}_\mathrm{h}/\der M$ denotes the halo mass function, i.e. the mean number density of haloes per unit mass. 
For later use, we introduce the moments of the CLF
\begin{equation}
    \eta_n(M,z)
    =\int_0^\infty L^n\,\phi(L|M,z)\,\der L\;,
    \label{eq:CLF-moments}
\end{equation}
with $n\in\mathbb{N}$.
$\eta_0$ gives the mean number of emitters hosted by a DM halo of mass $M$ at redshift $z$, $\eta_1$ gives the mean total luminosity emitted within the halo, and $\eta_2$ gives the mean sum of the squared luminosities of the individual emitters.
Obviously,
\begin{equation}
    \bar{\rho}_L(z)=\int_0^\infty 
    \eta_1(M,z)\,\frac{\der \bar{n}_\mathrm{h}}{\der M}(M,z)\,\der M\;,
\end{equation}
which decomposes the mean comoving emissivity into the contribution from different halo masses.

As commonly done in the literature \citep[e.g.][]{Lidz+11}, 
we computed the large-scale PS of the specific intensity by assuming that:
(i) DM haloes are linearly biased with respect to the underlying matter distribution (i.e. their overdensity $\delta_\mathrm{h}=b_\mathrm{h}\,\delta$ with $b_\mathrm{h}$ a function of $M$ and redshift), (ii) the scales of interest are significantly larger than the virial radii of the relevant haloes, (iii) the surveyed patch of the sky has a small extension compared to the distance to the observer so that we can assume a fixed line-of-sight direction $\hat{\mathbf{n}}$ (distant-observer approximation),
(iv) there is no peculiar-velocity bias, and (v) fluctuations of the CLF and halo counts are Poissonian.
It follows from these assumptions that the redshift-space PS of the specific intensity receives two contributions
\begin{equation}
    P= \Pclust+ \Pshot \; ,
    \label{anisotropic_power_spectrum}
\end{equation}
with $\Pclust$ arising from the clustering of the line-emitting galaxies and $\Pshot$ originating from the fact that they are discrete objects and thus show random fluctuations in their number counts within a finite volume.
Given that the clustering signal only dominates on large scales and that the measurements we consider have relatively large uncertainties, it is sufficient to use linear perturbation theory to model the different components. 

The clustering component can be expressed in terms of the linear matter PS, $P_\mathrm{m}$, as 
\begin{equation}
  \Pclust(k,\mu,z)=\bar{I}_\nu^2(z)\,[b(z)+f(z)\, \mu^2]^2\,\mathcal{D}
  (k,\mu,z)\,P_\mathrm{m}(k,z)\;,
  \label{eq:Pclust}
\end{equation}
where the linear bias coefficient 
\begin{equation}
    b(z)=\frac{1}{\bar{\rho}_L(z)}\,\int_0^\infty 
    \eta_1(M,z)\,b_\mathrm{h}(M,z)\,
    \frac{\der \bar{n}_\mathrm{h}}{\der M}(M,z)\,\der M\;,
    \label{eq:effectivebas}
\end{equation}
$f$ is the growth-rate of structure, and $\mu=\hat{\mathbf{k}}\cdot \hat{\mathbf{n}}$.

The term $\mathcal{D}$ in Eq.~(\ref{eq:Pclust}) is a phenomenological damping factor accounting for the non-perturbative suppression of clustering in redshift space due to velocity dispersion of the line-emitting regions within their host haloes.
This approximation has been first introduced to model galaxy clustering in redshift space \citep[e.g.][]{Peacock-Dodds-1994}. The three most common choices in the literature for the damping function are Gaussian, Lorentzian, and squared Lorentzian shapes:
\begin{equation}
  \mathcal{D}(k,\mu) = \begin{cases}
  \exp(-k^2\mu^2\sigma^2)\;,\\  
  \left[1+(k\mu \sigma)^2\right]^{-1}\;,\\
  \left[1+\displaystyle{\frac{(k\mu \sigma)^2}
  {2}}\right]^{-2}\;,
  \end{cases}
\label{eq:D-FoG-def}
\end{equation}
which all behave as $1-k^2\mu^2\sigma^2$ when $k\to 0$.
Here, the parameter $\sigma$ denotes a typical comoving displacement which should agree, within a factor of order unity, with the pairwise velocity dispersion divided by $aH$. In this work, we used a squared Lorentzian damping function but our conclusions do not change if another of the shapes presented in Eq.~(\ref{eq:D-FoG-def}) is adopted.

The shot-noise component does not depend on $k$ and assumes the redshift-dependent value of 
\begin{equation}
    \Pshot(z)= \frac{\bar{I}^2_\nu(z)}{\bar{n}_\mathrm{eff}(z)}\;,
    \label{shotnoise_component}
\end{equation} 
where the `effective number density'  of emitters satisfies
    \begin{equation}
    \bar{n}^{-1}_\mathrm{eff}(z)=\frac{1}{\bar{\rho}^2_L(z)} \,\int_0^\infty 
    \eta_2(M,z)\,\frac{\der \bar{n}_\mathrm{h}}{\der M}(M,z) \, \der M\;,
    \label{shotnoise_component-2}
\end{equation}
which can also be expressed as \citep{Cheng+19} 
\begin{equation}
\bar{n}_\mathrm{eff}(z)=\frac{\left(\int_0^\infty L\, \Phi(L,z) \,\der L\right)^2}{\int_0^\infty L^2 \,\Phi(L,z) \,\der L} \;.
\label{eq:neff}
\end{equation}

\section{\cii emission}\label{sec:CII_emission}
The \cplus ion is the most abundant form of carbon under many astrophysical conditions. In particular, since the first and second ionisation potentials of carbon (11.26 and 24.38 eV, respectively) bracket the hydrogen ionisation potential (13.6 eV), C$^+$ is present also in regions where hydrogen is neutral.

The ground electronic state of \cplus has two fine structure levels separated by approximately 0.0079 eV (corresponding to a temperature of 91.25 K). 
The associated $^2P_{3/2} - ^2P_{1/2}$ magnetic-dipole transition (hereafter [\ion{C}{II}]) at 157.74 $\mu$m (1900.5369 GHz) is one of the main coolants of the neutral and ionised interstellar medium (ISM). 
Thanks to its long wavelength, \cii radiation can traverse gas and dust with very little attenuation.

Due to telluric water-vapour absorption, \cii emission from the local Universe can only be detected with far-infrared balloon-, aircraft- or space-based observatories.
For cosmological sources with $3.3<z<9.3$, however, the (redshifted) \cii line becomes accessible from the ground (at special high-altitude sites) when it falls in one of the sub-millimetre or millimetre atmospheric windows. 

Recent interferometers such as ALMA and NOEMA allow us to observe \cii at high angular (and spectral) resolution and thus probe the physical conditions of gas in these high-redshift galaxies.

Local and cosmological observations reveal that \cii is one of the brightest emission lines from star-forming galaxies which typically accounts for 0.01\% to 1\% of the total far-infrared (FIR) luminosity \citep[e.g.][]{Stacey+10}.
The precise source of the emission remains unclear as the line can, in principle, arise from a variety of phases of the interstellar medium including molecular, atomic, and ionised gas. 
Depending on the detailed physical conditions of the gas, the line can be easily excited by collisions with electrons, hydrogen atoms, and hydrogen molecules.
At high redshift, the CMB provides a background of continuum radiation (the CMB spectrum peaks at the \cii central wavelength for $z\simeq 5.6$) which leads to an attenuation of \cii emission from low density gas \citep{Goldsmith+12}.

It is widely believed that, at high redshift, \cii should predominantly originate from photon-dominated regions at the boundaries of molecular clouds which are exposed to the ionising flux of nearby young stars \citep{Stacey+10, Pineda+14, Gullberg+15, vallini15, Lagache+18}.

In local, normal, star-forming galaxies, the \cii luminosity correlates with the star formation rate (and metallicity) although with a larger scatter compared with other lines \citep[e.g.][]{delooze14}.
A widespread explanation for this correlation invokes energy balance: namely, in thermal equilibrium, the heating and cooling rates of the gas in the neutral atomic phase of the ISM must match.
The correlation arises from the fact that \cii is the dominant cooling line while the main heating source is collisions with photoelectrons ejected by dust grains and polycyclic aromatic hydrocarbon molecules due to ultraviolet radiation emitted by young massive stars. 
Observations also showed, however, that the [\ion{C}{II}]/FIR luminosity ratio decreases with increasing infrared luminosity \citep{Malhotra+01}, which is expected to be an accurate star-formation tracer as it originates from UV and optical emission from young stars absorbed and re-radiated by dust at longer wavelengths.
This so-called `\cii deficit' is not fully understood yet and casts doubts on the use of \cii as a general star-formation tracer.
Similar correlations (with different normalisations) and trends are seen in high-redshift galaxies \citep[e.g.][]{carniani18, schaerer20}.

\subsection{\cii luminosity function}\label{sec:LF}
\begin{figure}
\centering
\includegraphics[width=0.45\textwidth]{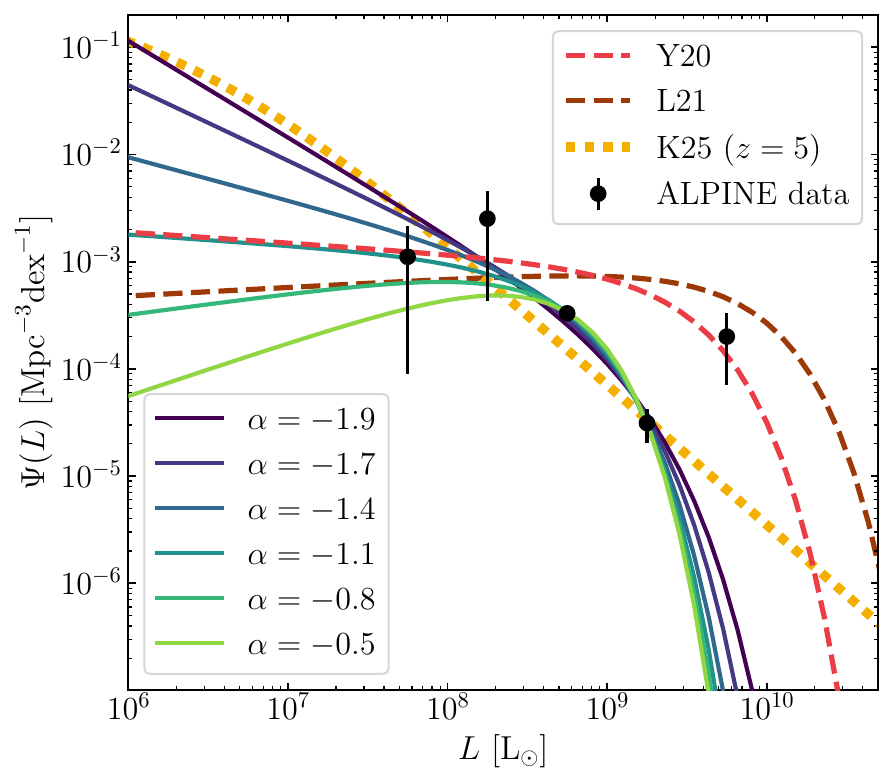}
\caption{\cii LF estimated from 
the targeted ALPINE detections by \citet[][Y20]{yan_20} (black data points and error bars). We show the average between the estimates at redshift $z\sim 4.5$ and 5.5. Our Schechter fits to the data are superposed with different fixed values of the faint-end slope $\alpha$ (purple, blue, and green lines). 
The dashed red line shows the LF fit obtained by Y20 combining multiple datasets at different wavelengths. 
The fit by \citet[][L21]{loiacono_2021} to the serendipitous (and clustered) ALPINE detections is shown with a dashed brown line.
For comparison, the LF from the \textsc{Marigold} numerical simulations by \citet[][K25]{Khatri+24_marigold} at $z=5$ is represented by a dotted gold line.}
\label{fig:fit_LF}
\end{figure}

\begin{table}
    \caption[]{Schechter fits to the observed \cii LF from L21 and Y20.}
    \label{tab:phiparam}
    \centering
    \begin{tabular}{ccccc}
        \hline
        \hline
        \noalign{\smallskip}
        Sample & Ref. & $\log_{10}\frac{{\Psi}_*}{\mathrm{Mpc}^{-3}\mathrm{dex}^{-1}}$  & $\log_{10}\frac{L_*}{L_\sun}$ & $\alpha$ \\
        \noalign{\smallskip}
        \hline
        \noalign{\smallskip}      
        Cluster & L21 & $-3.01^{+0.44}_{-0.61}$ & $9.88^{+0.54}_{-0.55}$  & $-0.92^{+0.56}_{-0.44}$  \\
        Combo & Y20 & $-3.08 \pm 3$ & $9.5\pm 0.6$  & $-1.1 \pm 0.3$  \\
        \noalign{\smallskip}
        \hline
    \end{tabular}
\end{table}

The unprecedented sensitivity of ALMA to \cii emission makes it an ideal tool to conduct follow-up observations of pre-selected galaxies at high redshift.
Because its field of view is small, however, it is very time consuming to carry out untargeted surveys that cover large fractions of the sky. Only a few blind surveys were therefore conducted so far. 
The ALMA Large Program to INvestigate CII at Early Times \citep[ALPINE,][]{lefevre_20,bethermin_20,faisst_20} invested 70 hours of observations in band 7 (275--373 GHz)
to perform targeted observations of 118 main-sequence galaxies (selected by their rest-frame UV luminosity at 1500 \AA$\ $with an absolute-magnitude limit of $M_{1500}<-20.2$) in the redshift range $4.4<z<5.9$ (excluding the range $4.6<z<5.12$ for which the \cii line falls in a low transmission window for ALMA).
It also conducted a blind search within the 118 pointings (covering 24.92 arcmin$^2$ in total) which detected eight secure and four likely \cii emitters.
Eleven of the twelve sources are strongly clustered around the central target in the same pointing. \citet[][hereafter L21]{loiacono_2021} used these emitters to estimate the \cii LF in the `cluster' environment. They parametrised their results in terms of the Schechter function

\begin{equation}
    \Phi(L) = \frac{\der n}{\der L}=\frac{\Phi_*}{L_*} \left(\frac{L}{L_*}\right)^\alpha \exp\left(-\frac{L}{L_*}\right) \; ,
    \label{schechter_function_phi}
\end{equation}
or, equivalently,
\begin{align}
   \Psi(L)= \frac{\der n}{\der \log_{10} L}=\Psi_*\,
    \left(\frac{L}{L_*}\right)^{1+\alpha} \exp\left(-\frac{L}{L_*}\right) \;,
    \label{schechter_function_psi}
\end{align}
where $\Phi_*$ and $\Psi_*= \ln 10 \,\Phi_*$ are normalisation factors, $L_*$ is the characteristic luminosity at which the counts are exponentially suppressed, and $\alpha$ is the slope of the power law describing the low-luminosity regime (without a cutoff at low $L$, the galaxy number density diverges if
$\alpha\leq-1$ but the luminosity density only diverges if $\alpha\leq-2$). 
It turns out that the data poorly constrain $L_*$ and an informative prior ($L_*<10^{10.5} L_\odot$) was used. The best-fit parameters are reported in Table~\ref{tab:phiparam}. We note that $\alpha$ is poorly constrained given the lack of information at faint luminosities.
Based on the ratio between the number of unclustered and clustered sources, L21 estimated that the `field' LF should be a factor of $\sim 11$ lower than the `cluster' one (assuming that the shape is the same).

Another estimate of (and Schechter fit to) the \cii LF in the same redshift range has been presented by \citet[][hereafter Y20]{yan_20}.
This was obtained by combining the serendipitous and targeted \cii ALPINE detections with additional data in the far-IR continuum and for CO line emission \citep{Koprowski_17,decarli_19,riechers_19,gruppioni_20} that were converted into \cii luminosities using empirical scaling relations.
The best-fit parameters are presented in Table~\ref{tab:phiparam} together with their relatively large uncertainties. 
The LF is in agreement with (but slightly lower than) the results by L21 for the cluster sample (see Fig.~\ref{fig:fit_LF}).

The targeted ALPINE detections possibly miss UV-faint but \cii\!\!-bright galaxies. They can therefore only provide a lower limit to the total LF.
On the other hand, the serendipitous detections are scarce and their LF carries large statistical uncertainties. Moreover, they are affected by clustering which leads to a systematic overestimation of the LF.
Given this state of the art, in the remainder of this paper, we follow a twofold strategy.
Namely, we use the fit by Y20 as an upper limit to the LF which we refer to as the optimistic case. Moreover, as a lower limit, we produce our own least-squares fits to the LF of the targeted detections by considering different fixed values of $\alpha$ (see Fig.~\ref{fig:fit_LF}) which we refer to as the pessimistic case.

\begin{figure}
\centering
\includegraphics[width=0.37\textwidth, trim=0 0 0 30, clip]{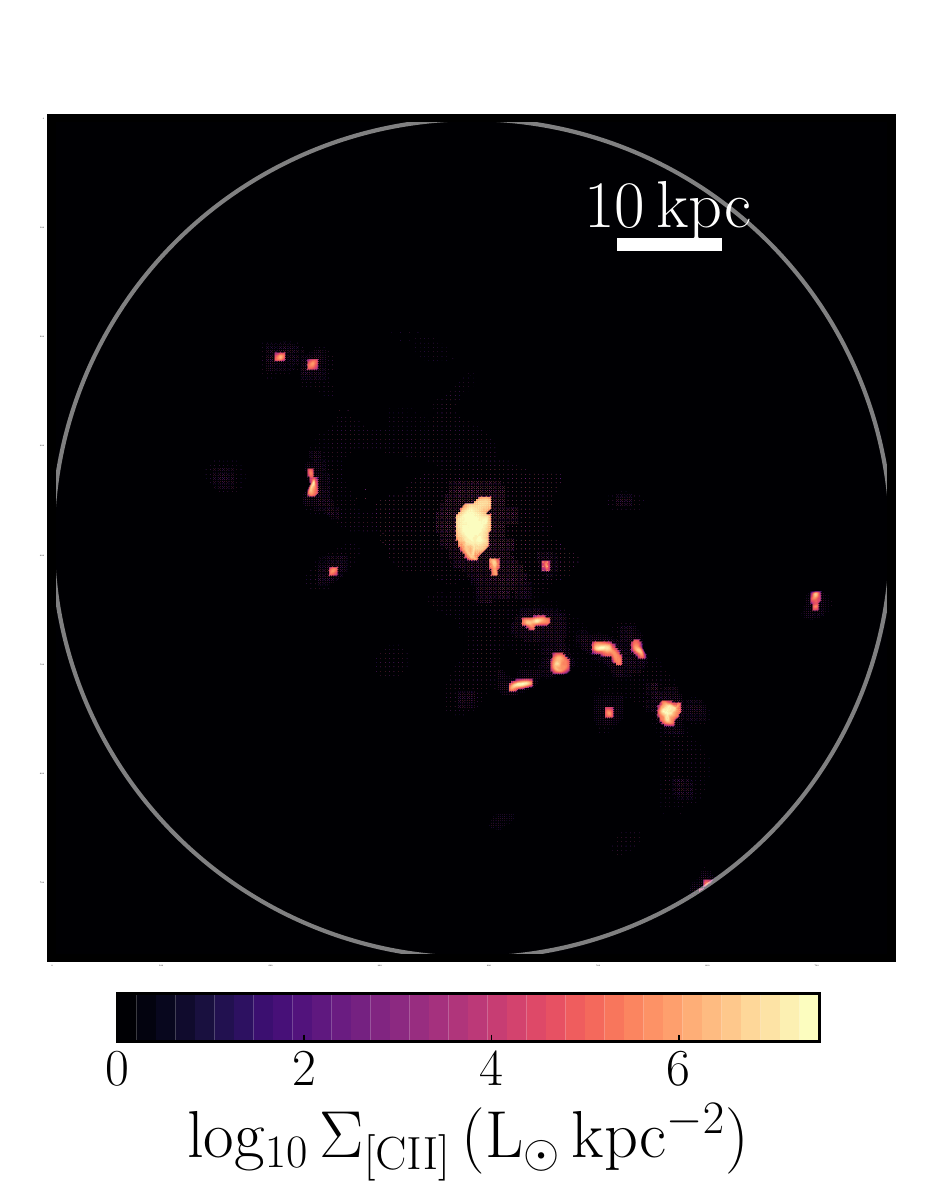} 
\caption{Surface brightness of the \cii emitters hosted by a DM halo of mass $M=5.78\times 10^{11} \ h^{-1}\mathrm{M_\odot}$ in 
the $z=5$ snapshot of the low-resolution \textsc{Marigold} simulation. The circle indicates the virial radius of the halo.}
\label{fig:cii-image}
\end{figure}

\subsection{\cii emitters in the \textsc{Marigold} simulations}

In order to develop insights about the halo-occupation properties of \cii emitters,
we use the \textsc{Marigold} simulations presented in \cite{Khatri+24_marigold}.
\textsc{Marigold} are a suite of cosmological simulations of galaxy formation which account for gravity, adaptive-mesh-refinement fluid dynamics, star formation, stellar feedback, the propagation of the ionising radiation emitted from young stars, and include the HYACINTH module for interstellar chemistry \citep{Khatri+24}. 
Given the computational cost of such an effort, 
the simulations follow the formation of structure until $z=3$ within
periodic cubic boxes of different comoving side lengths $L$ and achieve different spatial resolutions $\Delta x$.
The high-resolution simulation 
has $L=25$ Mpc
and a minimum grid size of $\Delta x=32$ pc. 
The low-resolution simulation, instead, has $L=50$ Mpc and $\Delta x=64$ pc.
\cii emission is computed in post processing
by solving the radiative transfer equation (i.e. without assuming the line is optically thin) as detailed in \cite{Khatri+24_marigold}. 
This calculation can be robustly performed for haloes and subhaloes with $M\geq 10^9 h^{-1}$ M$_\odot$.

\begin{figure}[t]
\centering
\includegraphics[width=0.48\textwidth, trim=0 0 0 60, clip]{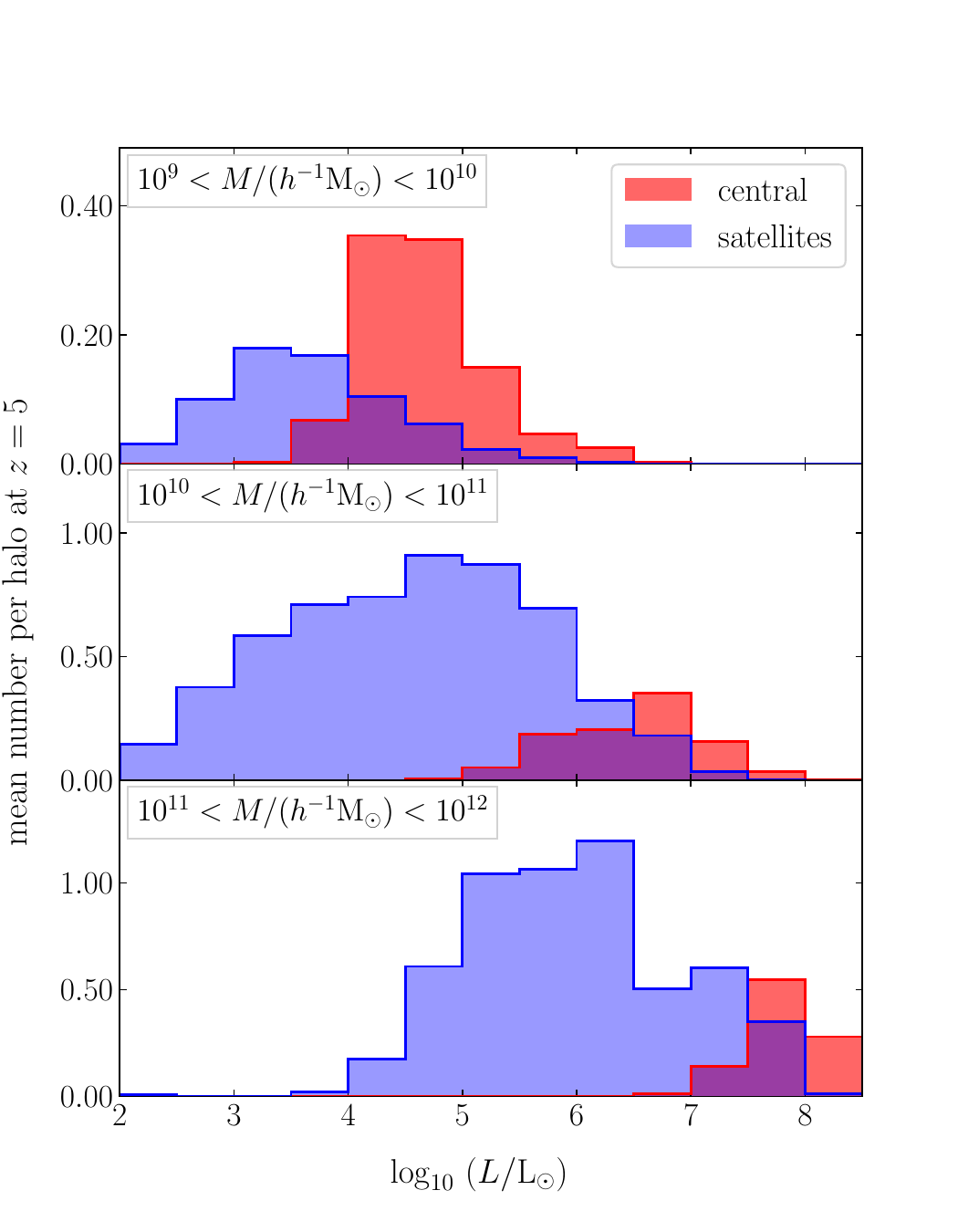} 
\caption{CLF extracted from the \textsc{Marigold} simulations at $z=5$ in three halo mass bins. The contributions from central and satellite \cii emitters are indicated in different colours. The scale of the $y$-axis changes in the different panels. Quantitative information about the CLF is provided in Table~\ref{tab:clfparam}. The top two panels refer to the high-resolution simulation, and the bottom panel is obtained from the low-resolution simulation, which contains more massive haloes.}
\label{fig:CLF-sims}
\end{figure}
\begin{table}
    \caption[]{Properties of simulated central (C) and satellite (S) \cii emitters in different mass bins of their host DM haloes at $z=5$ (see also Fig.\ref{fig:CLF-sims}).}
    \label{tab:clfparam}
    \centering
    \setlength{\tabcolsep}{4pt}

    \begin{tabular}{cccccc}
        \hline
        \hline
        \noalign{\smallskip}
          Halo mass [M$_\odot$] & C/S & $\bar{N}_i$&$\frac{L_i}{L_\mathrm{tot}}$ & $\log_{10} L_{50} \ [\mathrm{L}_\odot]$ & $\log_{10} \frac{L_{80}}{L_{20}}$  \\
        \noalign{\smallskip}
        \hline
        \noalign{\smallskip}
 $10^{9}$--$10^{10}$& C & 1 & 0.87 &4.58 & 0.83\\
                    & S & 0.68 & 0.13&  3.57 & 1.27\\
         \noalign{\smallskip}
        \hline
 $10^{10}$--$10^{11}$& C & 1 & 0.70 &6.56 & 1.09\\
                    & S & 5.58 & 0.30 &4.64 & 2.08\\
        \noalign{\smallskip}
        \hline
  $10^{11}$--$10^{12}$& C & 1 & 0.72 &7.84 & 0.51\\
                    & S & 5.58 & 0.28 &5.93 &1.61\\       
        \noalign{\smallskip}
        \hline
    \end{tabular}
    \tablefoot{From left to right, we list the mean number of objects per halo, the collective fractional contribution to the total halo luminosity, the median luminosity of one emitter, and the ratio between the 80$^\mathrm{th}$ and the 20$^\mathrm{th}$ percentile of the individual emitters. Consistently with Fig.~\ref{fig:CLF-sims}, the top two and the bottom entries refer to the high- and low-resolution \textsc{Marigold} simulations, respectively.}
\end{table}
Fig.~\ref{fig:cii-image} shows a synthetic image of the \cii emitters hosted within a massive DM halo at $z=5$. The central galaxy is the dominant source and is surrounded by more than a dozen substantially fainter emitters. This is a typical situation as evidenced in Fig.~\ref{fig:CLF-sims} where we plot the conditional luminosity function extracted from the simulations in three different mass bins. 
We distinguish between central and satellite \cii emitters. The central ones encompass the region within 0.1 virial radii from the stellar centre of mass of the main galaxy. Satellites extend up to the tidal radius of the subhaloes.
In the most massive bin we consider, $10^{11}\leq M/(h^{-1}\mathrm{M}_\odot)<10^{12}$ (bottom panel\footnote{It is worth mentioning that, at $z=5$ there are only two haloes more massive than this in the whole low-res simulation box.}), 
the central galaxies present a narrow distribution of luminosities (with a median value of $\log_{10} L/\mathrm{L}_\odot =7.84$ and a logarithmic width of 0.51, see Table~\ref{tab:clfparam}) which overlaps
with the range covered by the ALPINE detections.
Each halo contains, on average, 5.58 satellites which follow a very broad distribution of luminosities with a median value of $\log_{10} L/\mathrm{L}_\odot =5.93$.
Their aggregated luminosity only accounts for 28\% of the total \cii emission (see Table~\ref{tab:clfparam}). The integrated contribution from satellites becomes even less important for lower mass bins (top two panels). The results related to these halo masses are influenced by the finite mass resolution of the simulation. For this reason, we examined the contribution of centrals and satellites in the mass bins spanning from $10^9$ to $10^{11} \ h^{-1} \ \mathrm{M_\odot}$ using the high-resolution \textsc{Marigold} simulation.

We note that the median luminosity of the central galaxies scales approximately as $M^\gamma$ with $1.2<\gamma<1.5$ while  satellites show a much shallower slope of $0.2<\gamma<0.7$.
It turns out that, for every \cii luminosity we can probe, at least 80\% of the emitters are central galaxies and this fraction reaches 100\% for the brightest ones. 

\subsection{Abundance matching}
\label{sec:AM}
\begin{figure}
\centering
\includegraphics[width=0.45\textwidth]{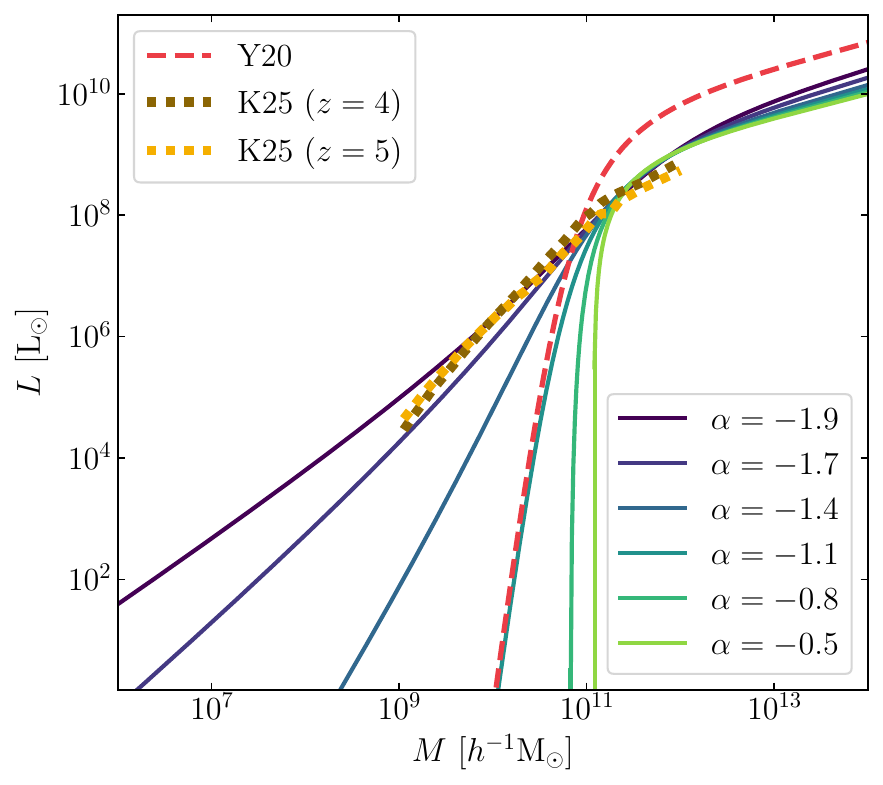}
\caption{\cii luminosity as a function of halo mass obtained via abundance matching. The colours for the solid-line fits to the ALPINE data at $z\simeq 5$ are the same as in Fig.~\ref{fig:fit_LF}. The dashed red line refers to the LF fit obtained by Y20.
The dotted gold and dark gold lines show the actual $\eta_1$ function (i.e. the mean total luminosity per halo) extracted by K25 at $z=5$ and $z=4$, respectively.}
\label{fig:HAM_CDM}
\end{figure}

We now return to discussing about the actual \cii emitters. Based on the simulation results presented above, in the remainder of this paper, we assume that each halo contains only one source and that there is no scatter in the \cii luminosity at fixed mass, i.e. $\phi(L|M)=\delta_\mathrm{D}[L-\mathcal{L}(M)]$, which, once inserted in Eq.~(\ref{eq:CLF-moments}), gives $\eta_n(M,z)=[\mathcal{L}(M)]^n$. 
Further assuming that $\mathcal{L}(M)$ is a monotonic function (always growing with $M$) allows us to determine its inverse function by a simple abundance-matching procedure. In fact, by integrating Eq.~(\ref{eq:CLF-def}) in $L$, we obtain
\begin{equation}
     \int_{L}^{\infty} \Phi (L') \, \der L'=\int_{\mathcal{L}^{-1}(L)}^{\infty} \frac{\der \bar{n}_\mathrm{h}}{\der M'} \, \der M' \;.
    \label{eq:halomodel}
\end{equation}
For instance, this approach has been used in \citet{Padmanabhan18_CO} to model LIM of the CO line \citep[see also][]{Padmanabhan19_CII, Padmanabhan22_dic, Padmanabhan22_oct}.

We evaluated the halo mass function $\der \bar{n}_\mathrm{h}/\der M$ using the fit to numerical simulations by \cite{ST_01} but setting their parameter $q=1$ as in \cite{Schneider+13}.
This requires calculating the variance of the smoothed linear density perturbations, for which we adopted the so-called `smooth-$k$' window function $1/[1+(kR)^\beta]$ \citep{Leo+18} with $\beta=4.8$.
For the smoothing radius, we used $R=R_\mathrm{TH}/3.3$, where $R_\mathrm{TH}$ denotes the comoving Lagrangian radius of a spherical perturbation of mass $M$.
These choices provide an excellent fit to N-body simulations in different cosmological scenarios
\citep{Sameie+19,Bohr+21,parimbelli_21} and allow us to extend our calculations beyond CDM in our future work (Marcuzzo et al., in prep.).

Results based on the halo mass function at redshift $z=5$ (chosen as representative of the ALPINE redshift range) are displayed in Fig.~\ref{fig:HAM_CDM}. At the low-mass end, the halo mass function follows a power-law behaviour with a slope close to $-2$.
If the faint-end slope of the luminosity function satisfies $\alpha > -1$, then the total number density of \cii\!\!-emitting galaxies converges to a finite value. In this case, the abundance-matching procedure imposes a sharp lower cutoff in the luminosity--mass relation $\mathcal{L}(M)$ at the halo mass corresponding to that number density ($\simeq 10^{11}h^{-1}$ M$_\odot$ for the case shown in the figure).
In contrast, when $-2 < \alpha \leq -1$, the cumulative number density of emitters increases more slowly with decreasing luminosity than the halo number density increases with decreasing mass. As a result, the corresponding relation between luminosity and halo mass, $\eta_1(M) = \langle L | M \rangle$, exhibits a smoother transition at low masses rather than an abrupt cutoff. This transition occurs around $M \lesssim 10^{11}\,h^{-1}\,\mathrm{M}_\odot$, where the halo mass function approximates a power law. The steepness of the transition depends on $\alpha$ and approximately scales as $M^{-1/(1+\alpha)}$ in the low-mass regime.
Finally, we point out that our result for $\alpha = -1.7$ shows excellent agreement with the $\eta_1(M,z)$ relations extracted from the \textsc{Marigold} simulations at $z = 5$ and $z = 4$ (shown by the dotted gold and dark gold lines in Fig.~\ref{fig:HAM_CDM}).

\begin{figure}
\centering
\includegraphics[width=0.43\textwidth]{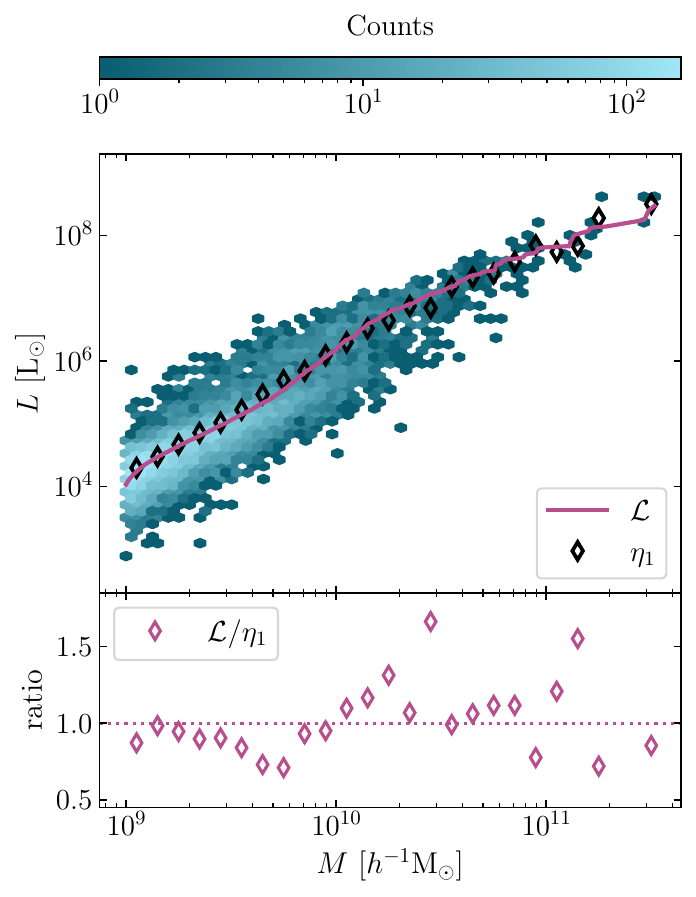} 
\caption{Hexbin scatter plot of (total) \cii luminosity and halo mass for the emitters in the high-resolution \textsc{Marigold} simulation at $z=5$. Solid black symbols show the mean total luminosity  computed in narrow mass bins (which coincides with the $\eta_1$ function also shown in Fig.~\ref{fig:HAM_CDM} as a dotted gold line).
The solid dark pink line shows the function $\mathcal{L}(M)$ obtained applying abundance matching to the simulation output following the steps and assumptions described in Section~\ref{sec:AM}.
Finally, the ratio of the latter two is shown in the bottom panel.
} 
\label{fig:test-am}
\end{figure}

\begin{figure}
\centering
\includegraphics[width=0.43\textwidth]{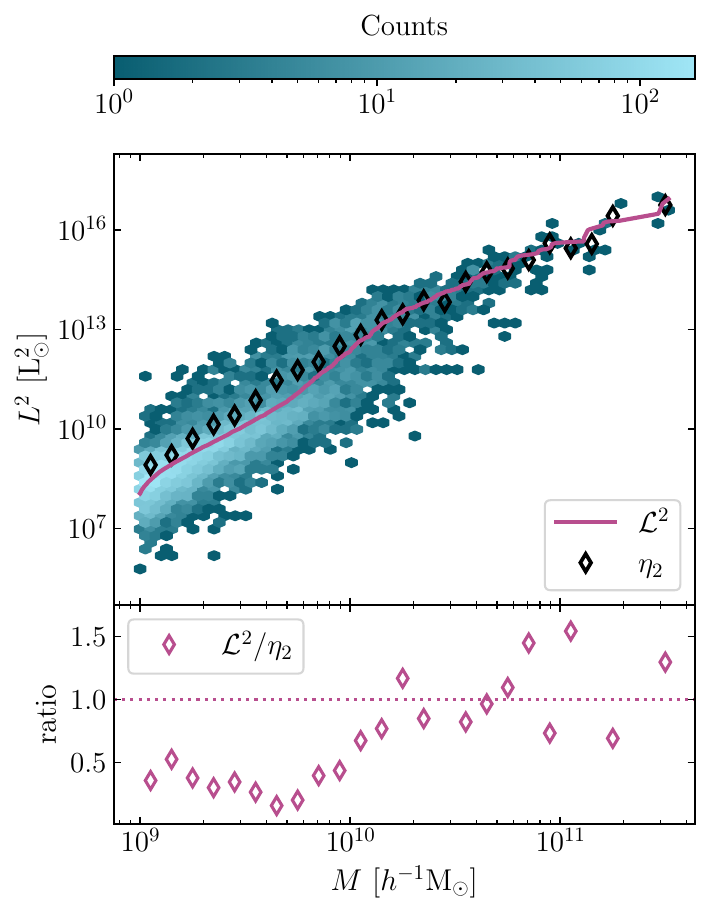} 
\caption{As in Fig.~\ref{fig:test-am}, but for the second moment of the CLF. In this case, the black diamonds and the dark pink line show the functions $\eta_2$ and $\mathcal{L}^2$, respectively.} 
\label{fig:test-am2}
\end{figure}

In Fig.~\ref{fig:test-am}, we use the simulations to directly test how accurate is the function $\mathcal{L}(M)$ determined via abundance matching. 
The blue hexagons in the scatter plot show the total \cii luminosity vs. halo mass. The total luminosity is obtained by summing up the contributions of all the resolved emitters hosted within a single halo. The black symbols indicate the mean (total) luminosity within narrow logarithmic mass bins and thus provide an estimate of the function $\eta_1(M,z=5)$. The result is monotonically increasing with $M$ as we assumed in Sect.~\ref{sec:AM} in order to perform abundance matching.
Finally, the dark pink line shows the $\mathcal{L}(M)$ function obtained by matching individual emitters to main haloes, as in Sect.~\ref{sec:AM}.
The ratio $\mathcal{L}/\eta_1$ is plotted in the bottom panel and shows that abundance matching gives approximately the correct answer.

Fig.~\ref{fig:test-am2} repeats the same analysis but after replacing the total \cii luminosity with the sum of the squares of the luminosities of the individual emitters.
The black symbols here give an estimate of $\eta_2(M,z=5)$ and the dark pink line is $[\mathcal{L}(M)]^2$ (with $\mathcal{L}$ taken from Fig.~\ref{fig:test-am}). 
The bottom panel shows that these two functions agree very well at large masses, while $\mathcal{L}^2$ underestimates the second moment of the CLF by a factor of $\sim2$ for $M<10^{10}\,h^{-1}$M$_\sun$.
This result suggests that our approach might slightly underpredict the amplitude of the shot-noise term in the power spectrum when all halo masses are considered.

In summary, we find that the functions $\mathcal{L}$ and $\mathcal{L}^2$ obtained with abundance matching provide a sound approximation to $\eta_1$ and $\eta_2$. The main reasons for this success are: (i) the total \cii luminosity is dominated by the central galaxy at all halo masses, and (ii) the scatter in luminosity at fixed halo mass remains moderate, although it increases toward lower halo masses, where the effects of bursty star formation may become more pronounced \citep[see also][]{Liu+24}.

\section{LIM power spectrum} \label{sec:LIM_powerspectrum}

\subsection{EoR-Spec on FYST}

As an example of current technology for LIM experiments, we use the specifications of the Epoch of Reionization Spectrometer \citep[EoR-Spec,][]{Nikola+23, Freundt+24} that will be deployed on FYST.
Prime-Cam---one of the two first-generation instruments that will be installed on FYST by the CCAT-prime collaboration---will have an unprecedented mapping speed at the target wavelengths \citep{CCATprime_collab}.
In its cryostat, it will hold up to seven instrument modules (five cameras working at different frequencies and two EoR-spec modules), each with a field of view of 1.3 square degrees.

EoR-Spec consists of an optical system made of four silicon lenses and several filters, a scanning Fabry-Perot interferometer (FPI), and three hexagonal arrays of Microwave Kinetic Inductance Detectors (MKIDs) sensitive to both polarisations.
Over 5 years, this imaging spectrometer will perform the DSS, namely a LIM survey of \cii over two patches\footnote{Further multiwavelength coverage of these fields, including grism spectroscopy from the Euclid mission \citep{mellier+24}, is planned with many telescopes \citep{CCATprime_collab}.} of the sky (4 square degrees each) covering the Extended-COSMOS \citep{Aihara+18} and the Extended Chandra Deep Field South fields \citep{Lehmer+05}, whose first light is expected in 2026.
Observations will be conducted in two frequency bands, 210-315 GHz ($5.033<z<8.050$ for [\ion{C}{II}]) and 315-420 GHz ($3.525<z<5.033$), with a resolving power of $R\sim 100$ over the whole spectral range.
The two frequency intervals are observed simultaneously by picking the second- and third-order fringes of the FPI for the low- and high-frequency bands, respectively.
Two of the MKID arrays will cover the low-frequency band and the third one will cover the high-frequency band.
At any given time, the observed frequency will change as a function of the distance from the centre of the array due to the light incidence angle. Basically, there will be rings of detectors that see the same frequency across the arrays, with increasing frequency outwards \citep[see Fig.~12 in][]{Nikola+22}.
The sequence of telescope sky scans and the FPI frequency scans will be optimised to obtain uniform coverage of the survey volume with a total observing time of $t_\mathrm{surv}\simeq 4000$ hours \citep{Cothard+2020, CCATprime_collab}.
Around 15 FPI steps are needed to fill in all frequencies.

\subsection{Survey characteristics}
\label{sec:survey}

\subsubsection{Survey volume}
Since a statistically significant detection of the PS with EoR-Spec with modest contamination from interlopers is expected only at the highest frequencies \citep[e.g.][]{karoumpis_22, Clarke+24}, we follow previous studies and consider a 40 GHz interval centred around 410 GHz (corresponding to $z\simeq 3.6355$) and thus covering the redshift range $3.42<z<3.87$.
In our reference cosmology, this corresponds to a comoving radial distance in redshift space of $\Delta  r_\parallel\simeq 240 \,h^{-1}$ Mpc.

Setting competitive constraints on the LF of the \cii emitters requires sampling large areas at high sensitivity. For this reason, we consider an abstract future survey that covers a larger area than the currently planned DSS and further discuss how results vary as a function of the survey size and sensitivity.
The only assumption we make is that progress with manufacturing on-chip spectrometers and developing novel readout technologies will allow us to achieve the same sensitivity of DSS. 
The smallest survey area we take in consideration is $\Omega_\mathrm{surv}=16$ sq. deg., a configuration which has been already studied in the literature as it was the planned area of an earlier version of the DSS \citep{karoumpis_22}. 
In this case, the survey extends for $\Delta r_\perp \simeq 332\,h^{-1}$ comoving Mpc in the direction perpendicular to the line of sight (assuming a compact geometry on the sky with angular extension $\Delta \theta\simeq \sqrt{\Omega_\mathrm{surv}}$).

\subsubsection{Spatial resolution of the intensity maps}
\label{sec:resolution}

The finite resolution of the observational setup smooths the measured intensity field over a characteristic angular and spectral scale. 
In Fourier space, this corresponds to a damping of the measured power spectrum, which can be written as
\begin{equation}
    P_\mathrm{obs}(\mathbf{k}, z) = P(\mathbf{k},z)\, W_\perp(k_\perp)\, W_\parallel(k_\parallel)\, W_\mathrm{vox}(\mathbf{k})\;,
    \label{eq:newsmoothing}
\end{equation}
where $W_\perp$ and $W_\parallel$ are window functions that account for resolution effects transverse and parallel to the line of sight, respectively,
and $W_\mathrm{vox}$ accounts for the fact that observations taken at different times are averaged into a single value within each individual spatial and spectral pixels during the map-making process.

The instrument beam, which we assume to be Gaussian, has a full width at half maximum (FWHM) of $\Delta \theta_\mathrm{FWHM} = 33$ arcsec. At the redshift of interest, this angular resolution corresponds to a transverse comoving size of $\Delta_\perp^\mathrm{FWHM} = d_\mathrm{A}(z)\,\Delta \theta_\mathrm{FWHM} \simeq 0.76\,h^{-1}\,\mathrm{Mpc}$, where $d_\mathrm{A}(z)$ is the comoving angular diameter distance (equal to the comoving radial distance in a flat universe).
For a Gaussian beam, the smoothing effect on the power spectrum is described by the transverse window function
\begin{equation}
W_\perp(k,\mu) = 
e^{-k_\perp^2(k,\mu)\,\sigma_\perp^2}=
e^{-(1 - \mu^2)\,k^2\,\sigma_\perp^2}\;,
\end{equation}
where $\sigma_\perp = \Delta_\perp^\mathrm{FWHM}/(2 \sqrt{2 \ln{2}}) \simeq \Delta_\perp^\mathrm{FWHM}/2.355$. In our setup, this yields $\sigma_\perp \simeq 0.32\,h^{-1}\,\mathrm{Mpc}$ at the central frequency, implying that attenuation becomes significant for transverse wavenumbers $k_\perp \gtrsim \sigma_\perp^{-1} \simeq 3.1\,h\,\mathrm{Mpc}^{-1}$. For instance, at $k_\perp=\pi/ \Delta_\perp^\mathrm{FWHM}\simeq 4.13\,h$ Mpc$^{-1}$, the window function evaluates to $W_\perp\simeq 0.17$.

To produce a well-sampled map that faithfully captures all relevant spatial features, individual measurements---each covering overlapping regions of the sky---are combined and interpolated onto a regular grid with pixel size $\delta_\perp$ that is at most half the angular resolution, that is, $\delta_\perp \leq \Delta^\mathrm{FWHM}_\perp/2$, in accordance with the Nyquist-Shannon sampling criterion.
\footnote{The theorem strictly applies to perfectly band-limited signals. However, the LIM signal convolved with a Gaussian beam retains non-zero power at all spatial frequencies, even if that power decays rapidly.}
The angular sampling factor $\eta_\perp = \Delta_\perp^\mathrm{FWHM}/\delta_\perp$ 
quantifies how finely the sky is sampled and corresponds to the number of pixels per resolution element.
Nyquist sampling corresponds to $\eta_\perp = 2$, while values greater than two indicate oversampling, which enhances the fidelity of the reconstructed map without improving its intrinsic resolution. A common choice is $\eta_\perp \simeq 3$ \citep[e.g.][]{Sullivan+24}.
As a result, for each angular dimension on the sky, the accessible wavenumbers in Fourier space are integer multiples of the fundamental mode 
$k_{\mathrm{f}}^\perp= 2\pi/\Delta r_\perp\simeq 0.019\,h$ Mpc$^{-1}$ and
information on the intensity field is available up to the Nyquist wavenumber,
$k_{\mathrm{N}}^\perp=\pi/\delta_\perp= \eta_\perp\,\pi/\Delta_\perp^\mathrm{FWHM}=
4.13\,\eta_\perp\,h$ Mpc$^{-1}$. 
Beyond this limit, the discrete sampling introduces aliasing, rendering signal reconstruction unreliable.

Similar considerations apply to the sampling of fluctuations along the line of sight.
A FPI transmits radiation with a frequency-dependent response that is well approximated by a Lorentzian profile\footnote{We thank A. Dev, C. Karoumpis, and D. Riechers for pointing this out.} which is characterised by the central frequency $\nu_\mathrm{o}$ and a FWHM of $\Delta \nu_\mathrm{o}=\nu_\mathrm{o}/R$.
This frequency width corresponds to a comoving radial resolution of
\begin{equation}
    \Delta_\parallel^\mathrm{FWHM} = \frac{c}{H(z)} \ \frac{\Delta \nu_\mathrm{o}}{\nu_\mathrm{o}}\,(1+z)= \frac{c}{H(z)} \,\frac{1}{R}\,(1+z)\; ,
\end{equation}
which evaluates to $\Delta_\parallel^\mathrm{FWHM}\simeq 24.54\,(100/R) \,h^{-1}$ Mpc at the central frequency of the survey.
The frequency response of the spectrometer suppresses the PS along the line of sight by the factor
\begin{equation}
    W_\parallel(k,\mu)=e^{-k\,|\mu|\,\Delta_\parallel^\mathrm{FWHM}}\;,
    \label{eq:Wparallel}
\end{equation}
which reduces to $W_\parallel\simeq 1-|k_\parallel|\,\Delta_\parallel^\mathrm{FWHM}+
k_\parallel^2\,(\Delta_\parallel^\mathrm{FWHM})^2/2$ when $k_\parallel\to 0$.
In the LIM literature, following the approach of \citet{li_16}, several authors approximate the line-of-sight window function $W_\parallel$ with a Gaussian damping factor of the form $W_\mathrm{G} = e^{- \mu^2 k^2 \sigma_\parallel^2}$, where $\sigma_\parallel \simeq \Delta_\parallel^\mathrm{FWHM}/2.355$. 
While this expression provides a convenient analytical form, it does not accurately capture the behaviour of the true window function for FPI spectrometers. 
In particular, since $W_\mathrm{G} \simeq 1 - k_\parallel^2 \sigma_\parallel^2$ in the limit $k_\parallel \to 0$, it underestimates the damping of the signal on large scales for surveys with moderate spectral resolution, such as those carried out with EoR-Spec.
For instance, at $k_\parallel = 1/\Delta_\parallel^\mathrm{FWHM}$, the exponential function in Eq.~(\ref{eq:Wparallel}) yields a value of $0.37$, while the Gaussian approximation gives $0.65$, substantially overestimating the transmitted power.

To ensure a well-sampled map in the spectral dimension,  the frequency channel spacing $\delta \nu$ should be at most half of the spectral FWHM, i.e. $\delta \nu<\Delta\nu_{\mathrm{o}}/2$ (corresponding to the comoving radial length $\delta_\parallel$).
The spectral sampling factor is defined as
$\eta_\parallel=\Delta_\parallel^\mathrm{FWHM}/\delta_\parallel$.
With this setup, the fundamental wavenumber along the line of sight is
$k_{\mathrm{f}}^\parallel= 2\pi/\Delta r_\parallel\simeq 0.026\,h$ Mpc$^{-1}$ while the Nyquist wavenumber is $k_{\mathrm{N}}^\parallel=\pi/\delta_\parallel>
\eta_\parallel\,\pi/\Delta_\parallel^\mathrm{FWHM}
\simeq 0.13\,(R/100)\,\eta_\parallel \, h$ Mpc$^{-1}$.

Finally, we briefly discuss the voxel window function, which introduces an additional damping factor to the power spectrum. In general, this function can be complex, as it depends on various factors such as the telescope’s scanning strategy, pixel geometry, beam shape, and the spectral response of the instrument (see Appendix~\ref{sec:SCANS} for a detailed discussion).
As a first approximation, and assuming the flat-sky limit with regularly spaced voxels, we model the voxel window function as a product of squared sinc functions along each Cartesian direction:
\begin{equation}
W_\mathrm{vox}(\mathbf{k})= \prod_{i=1}^2
\left[\frac{\sin( k_{\perp,i}\,\delta_\perp/2)}{k_{\perp,i}\,\delta_\perp/2}\right]^2\,
    \left[\frac{\sin( k_\parallel\,\delta_\parallel/2)}{k_\parallel\,\delta_\parallel/2}\right]^2 \;,
    \label{eq:Wvox}
\end{equation}
where $k_{\perp,i}$ are the Cartesian components of the transverse wavevector.
This function remains close to unity for all wavenumbers significantly below the corresponding Nyquist limits. By choosing a typical sampling factor of
$\eta_\perp=\eta_\parallel\simeq 3$, the first zero of the sinc function is pushed well beyond the relevant Fourier modes. We therefore safely neglected this effect and set $W_\mathrm{vox}=1$.

\subsubsection{Direction-averaged power spectrum}

In studies of the large-scale structure of the Universe, the anisotropic power spectrum in redshift space $P_\mathrm{obs}(k,\mu)$ is often expanded in multipole moments with respect to the angle between the wavevector and the line of sight
\begin{align}
P(k,\mu)&=\sum_\ell P_\ell(k)\,\mathcal{L}_\ell(\mu)\;, \\
    P_\ell(k)&=\frac{2\ell+1}{2}\int_{-1}^{1} P(k,\mu)\,\mathcal{L}_\ell(\mu)\,\mathrm{d}\mu\;,
\end{align}
where $\mathcal{L}_\ell$ denotes the Legendre polynomials.
In this study, we focus on the monopole moment of the intensity PS, 
\begin{align}
    P_0(k)=\frac{1}{2}\int_{-1}^{1} P_\mathrm{obs}(k,\mu)\,\mathrm{d}\mu=\frac{\int_{-1}^{1} P_\mathrm{obs}(k,\mu)\,\mathrm{d}\mu}{\int_{-1}^1 \mathrm{d}\mu}\;,
    \label{eq:P0}
\end{align}
which is obtained by averaging $P_\mathrm{obs}(k,\mu,z)$ over all possible values of $\mu$
and, for this reason, is also known as the spherically-averaged power spectrum. Obviously, this quantity can be measured with a higher signal-to-noise ratio (S/N) than $P_\mathrm{obs}$ itself.

\begin{figure}
\centering
\includegraphics[width=0.475\textwidth]{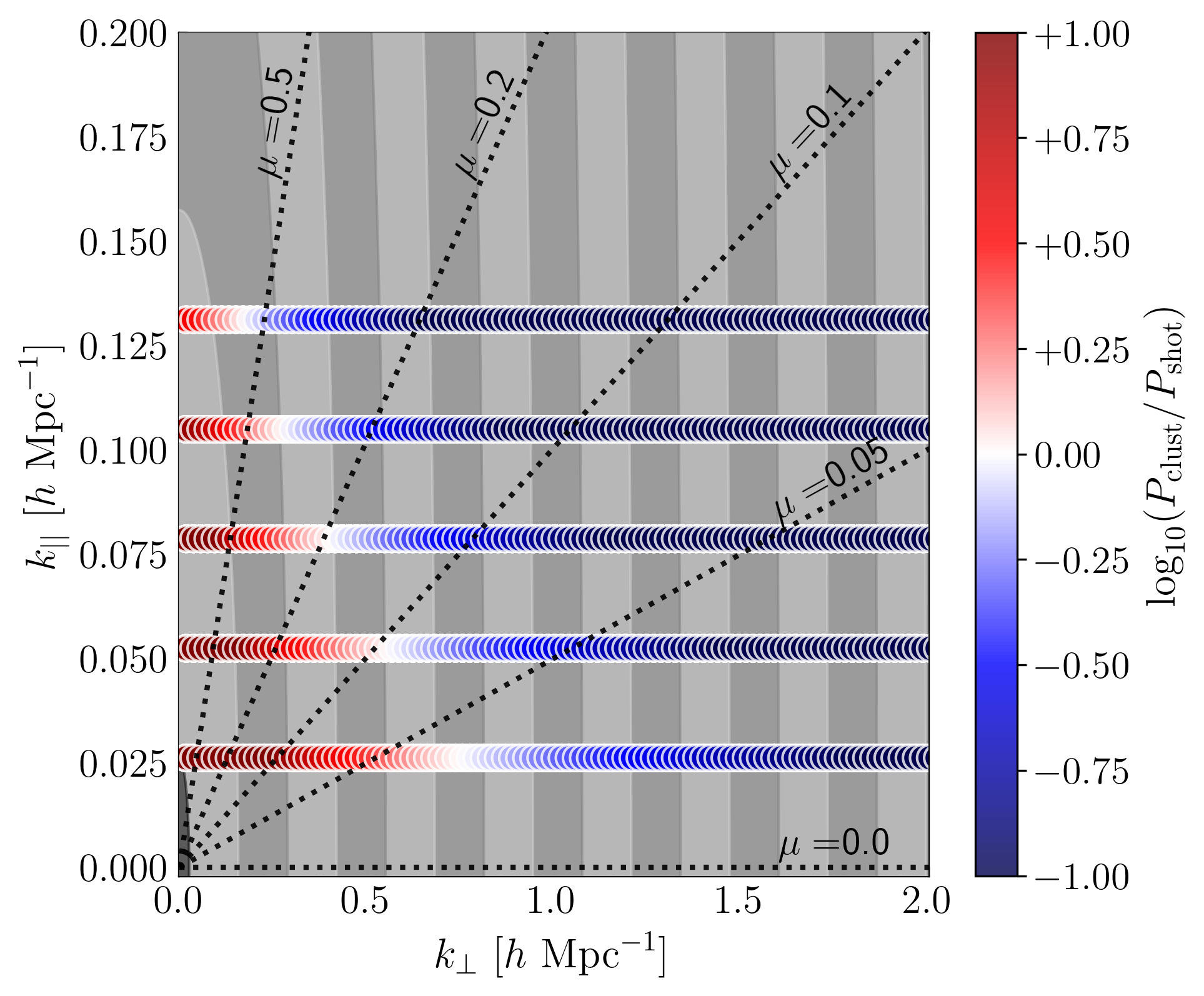}
\caption{Location in the $(k_\perp, k_\parallel)$ plane of the Fourier modes that are available in a 16 sq. deg. survey conducted  with EoR-Spec ($R=100$) at $z\simeq 3.6$ and with $\eta_\perp=\eta_\parallel=1$ (partially overlapping circles). 
The colour indicates the ratio of the corresponding clustering and shot-noise contributions to the PS (for our pessimistic LF with $\alpha=-1.1$).
The light and dark grey bands highlight the bins adopted in our analysis ($\Delta k=5\,k_\mathrm{f}^\parallel\simeq 0.13\,h$ Mpc$^{-1}$). These are annuli but appear as vertical bands due to the strong asymmetry in the scales along the axes. The dotted lines denote fixed values of $\mu=k_\parallel/k$.} 
\label{fig:CCAT_k_asymmetry}
\end{figure}

For surveys that cover a small solid angle on the sky, the line of sight can be considered to be fixed and $P(k,\mu)=P(k,-\mu)$ so that the integrations in Eq.~(\ref{eq:P0}) can be performed between 0 and 1.
Modifications are needed when the instrumental setup or selection effects only allow measurements of $P_\mathrm{obs}$ for a limited range of $\mu$, however.
This applies to the surveys planned with EoR-Spec, where a strong asymmetry between the accessible Fourier modes along and across the line of sight restricts the range over which meaningful averaging can be performed (this is illustrated in Fig.~\ref{fig:CCAT_k_asymmetry} for $R=100$ and sub-Nyquist sampling, i.e. $\eta_\perp=\eta_\parallel=1$).
High values of $\mu$ are only possible for $k\lesssim k_\mathrm{N}^\parallel$, while at much higher wavenumbers all modes have $\mu\simeq 0$ (i.e. $k\simeq k_\perp \gg k_\parallel$). 
We therefore adapted the definition of the monopole moment by introducing the direction-averaged power spectrum,
\begin{equation}
    P_0(k)=\frac{\int_{\mu_\mathrm{min}}^{\mu_\mathrm{max}} P_\mathrm{obs}(k,\mu)\,\mathrm{d}\mu}{\int^{\mu_\mathrm{max}}_{\mu_\mathrm{min}}\mathrm{d}\mu}
    \equiv \langle P_\mathrm{obs}(k,\mu)\rangle_\mu
    \;,
    \label{observedPk_CCAT}
\end{equation}
where the integrals should be replaced by discrete sums when too few modes are available at fixed $k$.
To implement Eq.~(\ref{observedPk_CCAT}) for EoR-Spec and prospective future instruments, we set $\mu_\mathrm{min} = k_\mathrm{f}^\parallel / k$ to approximately account for foreground contamination (see Sect.~\ref{sec:foregrounds}),
and $\mu_\mathrm{max} = \min(1, k_{\mathrm{max}}^\parallel/k)$ 
with $k_{\mathrm{max}}^\parallel=\pi /\Delta_\parallel^\mathrm{FWHM}$ to capture the effects of finite spectral resolution.
We adopted this refined treatment in our forecasts, whereas earlier studies \citep{karoumpis_22, Clarke+24} relied on a simplified formulation \cite[cf. equation 40 in][]{karoumpis_22}.

\subsection{Binning and error budget}\label{subsec:CCAT_errors}
In practice, $P_0$ is estimated within finite bins of size $\Delta k$.
In what follows, we present results obtained using $\Delta k=5 \,k_\mathrm{f}^\parallel$ but we have tested that our conclusions do not depend on this choice.
The number of independent Fourier modes contributing to each bin can be approximately computed by taking the ratio of the $k$-space volume of a bin and the volume of a fundamental cell, $k_\mathrm{f}^\parallel\, (k_\mathrm{f}^\perp)^2$, which gives (see Appendix \ref{sec:Nmodes})
\begin{equation}
    N_\mathrm{m}(k) = \frac{\mathrm{min}(k,k_\mathrm{max}^\parallel) \, k \, \Delta k \,  V_\mathrm{surv}}{4 \pi ^2} \; ,
    \label{eq:nmodes}
\end{equation}
where $V_\mathrm{surv}$ denotes the comoving volume covered by the survey (assumed to be a rectangular cuboid). Only the region with $k_\parallel >0$ is considered as the line intensity is a real-valued quantity and its Fourier modes at $\mathbf{k}$ and $-\mathbf{k}$ are complex conjugates and thus not independent.

Assuming that both the LIM fluctuations and the detector noise can be approximated as Gaussian random fields, the statistical uncertainty associated with the direction-averaged PS is
\begin{equation}
    \sigma_{P_0}(k) = \frac{P_0(k) + \Pwn}{\sqrt{N_\mathrm{m}(k)}}\;,\label{powerspectrum_error}
\end{equation}
where $\Pwn$ denotes the white-noise power spectrum set by the sensitivity of the instrument.
To evaluate $\Pwn$ for EoR-Spec, we adopted the sensitivity estimate reported in Table 1 of \citet{CCATprime_collab}, expressed as a noise-equivalent intensity\footnote{The NEI is defined as the root-mean-square (rms) intensity noise per unit solid angle accumulated in 1 s of integration time, averaged over the instrument's field of view.} (NEI) of 98000 Jy sr$^{-1}\sqrt{\mathrm{s}}$ for a resolving power $R=100$. This value represents a weighted average across the top three weather quartiles, assuming two EoR-Spec modules observe concurrently and that, on average, 80\% of the detectors are operational.

For simplicity, we first computed $\Pwn$ assuming sub-Nyquist sampling with $\eta_\perp=\eta_\parallel=1$ (i.e., one voxel per resolution element). If the survey volume is observed uniformly, the white-noise power spectrum is given by
\begin{equation}
    P_\mathrm{WN}=\sigma^2_\mathrm{vox}\,V_\mathrm{vox} =\frac{\mathrm{NEI}^2}{t_\mathrm{vox}}\,V_\mathrm{vox}\;,
\end{equation}
where $\sigma_\mathrm{vox}$ is the rms noise in a resolution element of comoving volume $V_\mathrm{vox}$, and $t_\mathrm{vox}$ is the average observing time per voxel per detector.

Assuming a scan strategy that uniformly covers the survey area and allocates equal integration time to each spectral-resolution channel, the average integration time per voxel is
\begin{equation}
    t_\mathrm{vox}=\frac{t_\mathrm{surv}}{N_\mathrm{vox}}\;,
\end{equation}
where $N_\mathrm{vox}=N_\mathrm{pix}\,N_\nu$ is the total number of voxels with
\begin{equation}
    N_\mathrm{pix}=
    \frac{\Omega_\mathrm{surv}}{\Omega_\mathrm{beam}}\;, \ \ \ \text{and} \ \ \ 
    N_\nu=
    R\,\ln\frac{\nu_\mathrm{max}}{\nu_\mathrm{min}}\;,
\end{equation}
the number of pixels on the sky and the number of frequency channels, respectively.
For our 16 sq. deg. survey with $R=100$, we obtain $V_\mathrm{vox} \simeq 14.3 \, h^{-3}\, \mathrm{Mpc}^3$, $N_\mathrm{pix} \simeq 410^2$, and $N_\nu \simeq 30$ (in the high-frequency band), leading to:
\begin{equation}
    \Pwn\simeq 2.4\times 10^{10} \,\frac{8000 \,\mathrm{hours}}{t_\mathrm{surv}}\,h^{-3}\,\mathrm{Mpc}^3 \,\mathrm{Jy}^2 \,\mathrm{sr}^{-2}\;.
\end{equation}

Adopting a finer sampling scheme with $\eta_\perp > 1$ and $\eta_\parallel > 1$, the total observing time is divided among more (smaller) voxels, which increases the rms noise per voxel by a factor of $\eta_\perp^2 \eta_\parallel$. The voxel volume decreases by the same factor, however, and leaves $\Pwn$ unchanged. It is important to note, though, that this finer sampling introduces correlations in the noise between neighbouring voxels, as multiple measurements contribute to each resolution element.

In the following sections, we extend our analysis to include futuristic instruments characterised by a resolving power of $R=500$. It is therefore essential to assess how instrumental noise scales with $R$. According to the radiometer equation, the NEI scales as $\mathrm{NEI} \propto \Delta\nu^{-1/2} \propto R^{1/2}$, while the voxel volume behaves as $V_\mathrm{vox} \propto \Delta\nu \propto R^{-1}$, and the number of frequency channels grows as $N_\nu \propto R$. Consequently, for a fixed total survey duration $t_\mathrm{surv}$, the noise power spectrum scales as $\Pwn \propto R$. It is possible to maintain a constant $\Pwn$ by increasing $t_\mathrm{surv}$ proportionally to $R$, however.

\subsection{Map making, foregrounds, and interlopers}
\label{sec:foregrounds}
Foreground contamination constitutes a major challenge for LIM studies as it superimposes prominent fluctuations to the target signal.
For each experimental setup, the contamination needs be characterised and, if possible, isolated within the data analysis pipeline.

For \cii experiments, the strongest contaminants are atmospheric noise, the cosmic infrared background (CIB, i.e. the integrated continuum emission from cosmic dust in galaxies), and redshifted CO rotational lines emitted by foreground galaxies. 
Removing or mitigating the impact of these contaminants possibly introduces systematic effects in the measured summary statistics.
For instance, filtering out atmospheric noise during the map-making process can lead to the suppression of the final PS on the largest scales \citep[e.g.][]{Lunde+24}. Although this systematic effect can be corrected by estimating the pipeline transfer function, the suppression becomes rather extreme at low $k_\parallel$.
Additional systematic effects on large scales might be introduced by the corrections for continuum emission.
The CIB is highly dominant in terms of intensity but has a smooth dependence on frequency which makes its separation from the highly fluctuating \cii signal doable using methods that have been originally developed for the 21 cm line. 
In general, continuum foregrounds mostly affect a few Fourier modes perpendicular to the line of sight with the lowest wavenumbers, i.e. with $k_\parallel \simeq  0$  \citep[e.g][]{Switzer+19, Zhou+23}. 
This source of contamination can therefore simply be removed by discarding these modes.
All these considerations suggest that an approximate method to account for foreground contamination in our forecasts is to only consider Fourier modes above a minimum $k_\parallel$. In what follows, we only use modes with $k_\parallel\geq k_\mathrm{f}^\parallel$ (i.e. we discard those with $k_\parallel=0$) which is equivalent to setting $k\geq k_\mathrm{f}^\parallel$ independently of the angular size of the survey (i.e. increasing $\Omega_\mathrm{surv}$ will reduce $\sigma_{P_0}$ because $V_\mathrm{surv}$ and $N_\mathrm{m}(k)$ will grow but will not extend the power-spectrum analysis to smaller values of $k$ corresponding to larger transverse length scales).

Finally, we briefly discuss line interlopers which can also significantly alter the \cii PS. 
Many different approaches have been proposed to correct for this contaminant. For instance, acting at the map level, one could mask the voxels that should contain CO emission from galaxies that have been detected in external surveys \citep{Yue+15, Sun+18, Bethermin+22, Karoumpis+24}.
While targeted masking has been proven to be successful in mitigating the contamination at the highest frequencies, it also reduces  $V_\mathrm{surv}$ (thus increasing the statistical errors on the PS) and convolves the expected signal with a complicated window function which induces correlations between the measurements in different $k$-bins.
Alternatively, working at the PS level, the contamination from interlopers could be characterised by cross-correlating the LIM data with galaxy catalogues or with intensity maps at different frequencies \citep{Wolz+16,Schaan-White_21,Keenan+22,Roy-Battaglia_24, Bernal-Baleato_25}. 
Lastly, without requiring any external input, one could use the technique of `spectral line de-confusion' which is based on the fact that sources at different redshifts than the target lines will be mapped to the wrong comoving coordinates so that their PS will be highly anisotropic along the $k_\parallel$ and $k_\perp$ directions \citep{Visbal-Loeb-10, Lidz-Taylor-16, Cheng+16}.

Current estimates on the level of contamination depend on assumptions about the CO spectral line energy distribution (SLED; i.e. the relative intensities of the different rotational transitions) in the interloper galaxies. \citet{Roy-Battaglia-23} found that CO interlopers generate a strong bias
in the PS at 410 GHz, while several other authors
concluded that contamination is severe only below
350 GHz and that fewer than 10\% of the voxels need to be masked at 410 GHz \citep{Yue+15, Bethermin+22, Karoumpis+24}. 
Based on this second set of results, we did not modify our forecasts to account for interloper contamination.

\begin{table}
    \caption[]{Parameters derived from the halo model for the LIM PS at $z\simeq3.6$ for different input \cii LF.}
    \label{tab:bias-sn}
    \centering
    \begin{tabular}{lcccc}
        \hline
        \hline
        \noalign{\smallskip}
        LF model & $\alpha$ & $\bar{I}_\nu$ & $b$& $\bar{n}_\mathrm{eff}^{-1}$\\
        & & $(10^3\,h^{2} \ \mathrm{Jy})$ & & $(10^2\,h^{-3}$ Mpc$^3)$\\
        \noalign{\smallskip}
        \hline
        \noalign{\smallskip}
        Optimistic & $-1.1$ & 14.52& 3.53 & 2.66 \\  
        \noalign{\smallskip}
        \hline
        \noalign{\smallskip}
        Pessimistic & $-1.9$ & 5.34& 2.71 & 0.88\\
        Pessimistic & $-1.7$ & 3.94& 3.10 & 1.37\\  
        Pessimistic & $-1.4$ & 3.16& 3.35 & 1.90\\  
        Pessimistic & $-1.1$ & 2.73& 3.48 & 2.39\\  
        Pessimistic & $-0.8$ & 2.45& 3.58 & 2.88\\  
        Pessimistic & $-0.5$ & 2.26& 3.65 & 3.30\\  
        \noalign{\smallskip}
        \hline
    \end{tabular}
\end{table}

\subsection{Clustering and shot-noise amplitudes}
\label{sec:clust_shot-noise}
Our initial goal is to make predictions about the LIM PS that will be detected with EoR-Spec at $z\simeq 3.6$ based on the halo model presented in Sect.~\ref{sec:halomodel} and the abundance-matching technique described in Sect.~\ref{sec:AM}.
In order to achieve this, however, we have to face the fact that the ALPINE estimates for the LF are only available in the redshift interval $4<z<6$, meaning that the abundance matching can only be performed at $z\simeq 5$. 
Since both observations and simulations suggest that the \cii LF evolves rather rapidly with time \citep[e.g.][]{yan_20, Khatri+24_marigold}, assuming that it remains unchanged within the $\sim550$ Myr intervening between $z=5$ and $3.6$ seems implausible.
The \textsc{Marigold} simulations offer a way out of this dilemma. Fig. 2 in \citet{Khatri+24_marigold} shows that the CLF of the simulated \cii emitters does not change much between redshift 5 and 3. This is also evident in our Fig.~\ref{fig:HAM_CDM}, where we compare the relation $\mathcal{L}(M)$ extracted from the simulations at $z=5$ and 4.
We thus proceed by assuming that the function $\mathcal{L}(M)$ determined from the ALPINE data (see Fig.~\ref{fig:HAM_CDM}) can be reliably used to compute the LIM PS at $z\simeq 3.6$ when combined with the evolved halo mass function and halo bias.

In Table~\ref{tab:bias-sn}, we report the mean \cii intensity, linear bias and effective volume per emitter obtained at $z\simeq3.6$ for different models of the \cii LF (at $z=5$).
In our pessimistic case, due to the opposite trends of $\bar{I}_\nu$ and $b$, the clustering signal ($\propto \bar{I}_\nu^2\,b^2$), does not vary much with $\alpha$. It is the highest for $\alpha=-1.9$ and the lowest for $\alpha=-0.5$, but it only changes by a factor of 3 overall.
The shot-noise term ($\propto \bar{I}_\nu^2 \,\bar{n}_\mathrm{eff}^{-1}$) also decreases with $\alpha$ and varies even less, with an overall change by a factor of 1.5 when $\alpha$ spans from $-1.9$ to $-0.5$.
Obviously, our optimistic and pessimistic predictions differ much more and their ratio at fixed $\alpha$ is driven by $\bar{I}_\nu$. For $\alpha=-1.1$, both the clustering and shot-noise amplitudes deviate approximately by a factor of 25.

It is interesting to note that the effective number of galaxies per voxel $\bar{n}_\mathrm{eff}\,V_\mathrm{vox}$ \citep[sometimes called sparsity,][]{Schaan-White_21} is always well below unity for both $R=100$ and 500. This means that only a small fraction of voxels contain a bright galaxy at the
redshift of interest. In particular, voxels with more than one such galaxy are extremely rare.
This opens interesting perspectives for mitigating
the interloper contamination \citep[e.g.][]{Cheng+2020}.

\begin{figure*}
\centering
\includegraphics[width=0.45\textwidth]{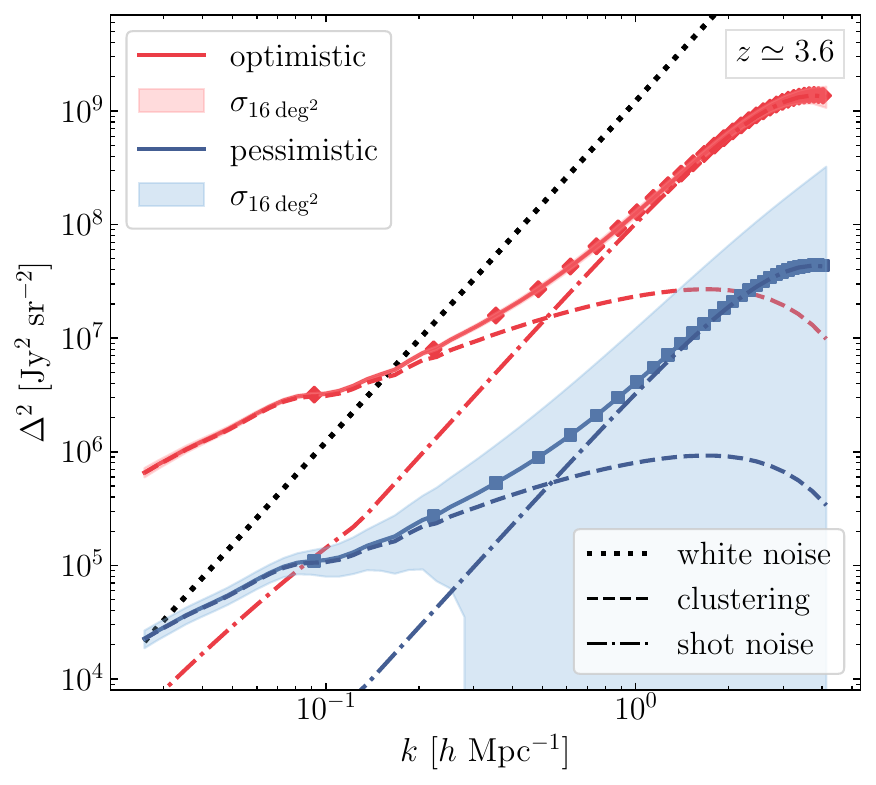} \hspace{0.5cm} \includegraphics[width=0.45\textwidth]{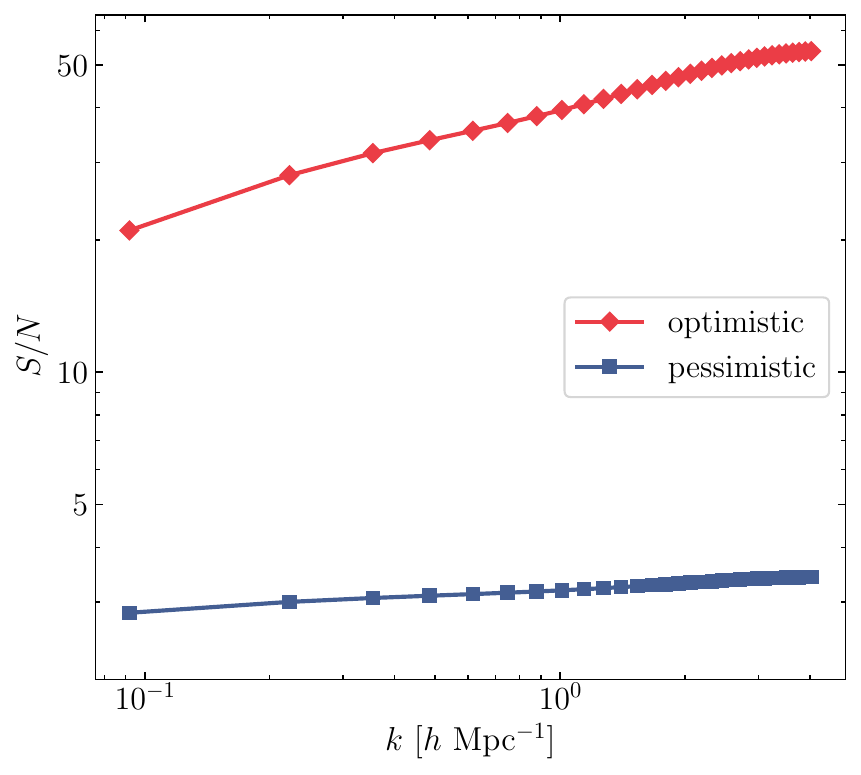} 
\caption{Left: Function $\Delta^2(k,z\simeq 3.6)$ for our optimistic and pessimistic cases (solid lines) and its statistical uncertainty (shaded regions) estimated for a 16 sq. deg. survey with $R=100$.
The dotted line shows the white-noise spectrum for EoR-Spec, and the dashed and dot-dashed lines refer to the clustering and shot-noise components, respectively.
Right: Cumulative S/N for the spectra shown in the left panel. In both panels, the markers indicate the centre of our $k$-bins.}
\label{fig:powerspectrum_rectangular}
\end{figure*}

\subsection{Results}
\label{sec:PSresults}

Our results for the PS are presented in the left panels of Fig.~\ref{fig:powerspectrum_rectangular} 
for $R=100$, and Fig.~\ref{fig:powerspectrum_R500} for $R=500$.
Shown is the contribution to the variance of the specific intensity per unit log interval in $k$
\begin{equation}
    \Delta^2(k,z)=\frac{\der \sigma^2_{I_\nu}}{\der \ln k}=\frac{k^3}{2\pi^2}\,P_0(k,z)\;,
\end{equation}
(solid curves) together with its statistical error derived from Eq.~(\ref{powerspectrum_error}) assuming $\Delta k=5k_\mathrm{f}^\parallel$
(shaded areas). Red and blue colours correspond to our optimistic and pessimistic LFs, respectively, both with $\alpha=-1.1$. 
While the solid curves trace the theoretical predictions continuously in $k$, the overlaid symbols show the results obtained with our actual binning scheme.

In all cases, we adopted a fiducial value of $\sigma=3\,h^{-1}$ Mpc.
This should be regarded as a rough estimate, given the absence of clustering measurements at these redshifts.
It is motivated by the following considerations:
(i) the line-of-sight pairwise velocity dispersion measured by the 2dF Galaxy Redshift Survey (2dF GRS) for star-forming galaxies in the local Universe is $\sigma_{12}\sim 420$ km s$^{-1}$ \citep{Madgwick+03}, which corresponds to a damping parameter $\sigma_{12}/\sqrt{2}\sim 300 $ km s$^{-1}$, or equivalently $\sigma\sim 3\,h^{-1}$ Mpc;
(ii) the velocity dispersion of DM haloes
is expected to scale with mass and redshift as $M^{1/3}\,H(z)^{1/3}$;
(iii) our LIM signal at $z=3.6$ is dominated by haloes of mass $M=$ a few$\,\times \,10^{11}\,h^{-1}$ M$_\sun$, corresponding to a one-dimensional velocity dispersion of $\sim160$ km s$^{-1}$ which is slightly higher than that of the haloes hosting the 2dF GRS star-forming galaxies at $z=0$ ($10^{12}\,h^{-1}$ M$_\sun$ and $\sim130$ km s$^{-1}$);
(iv) at high redshift, the large-scale structure of the Universe is less evolved resulting in smaller relative velocities between distinct haloes; and
(v) the intrinsic velocity dispersion of the \cii emitting gas---driven by both ordered rotation and turbulent motions within galaxies
\cite[see e.g.][]{Kohandel+20}---must also be taken into account. Taken together, these considerations suggest that $\sigma$ should lie within the range $1-3\,h^{-1}$ Mpc, and we adopted the upper end of this interval for our analysis.

The optimistic and pessimistic LFs generate power spectra with similar shapes that, however, differ in amplitude by a factor of $\sim25$. This gap approximately encompasses the range spanned by the different predictions that have appeared in the literature
\citep{Silva+15,Serra+16,Dumitru+19,chung_20,Padmanabhan22_oct,Kannan+22,karoumpis_22,Sun+23,Clarke+24}.
The individual contributions from the clustering and shot-noise components are indicated with dashed and dot-dashed lines, respectively.
In a concrete experimental setting, the white-noise spectrum $P_\mathrm{WN}$ must be subtracted from the measured signal---or incorporated into the model used to fit the data---in order to isolate $P_0$. For this reason, we also include the white noise PS in the figure (indicated by a dotted line). 
It is worth noting that $\Delta^2$ exceeds the white-noise level at only one data point (two in fig.~\ref{fig:powerspectrum_R500}), even in the optimistic scenario. This underscores the critical importance of accurately characterising the white noise associated with the instrument in order to reliably isolate the LIM signal.

The right panel of Fig.~\ref{fig:powerspectrum_rectangular} displays the cumulative S/N for $\Delta^2$ as a function of $k$, based on our binning strategy. In all cases, a detection of the LIM signal should be achievable: the cumulative S/N reaches values of approximately 3.4 in the pessimistic case and up to 54 in the optimistic scenario. These findings are broadly consistent with the recent forecasts for EoR-Spec presented by \citet{karoumpis_22} and \citet{Clarke+24}.

\begin{figure*}
\centering
\includegraphics[width=0.42\textwidth]{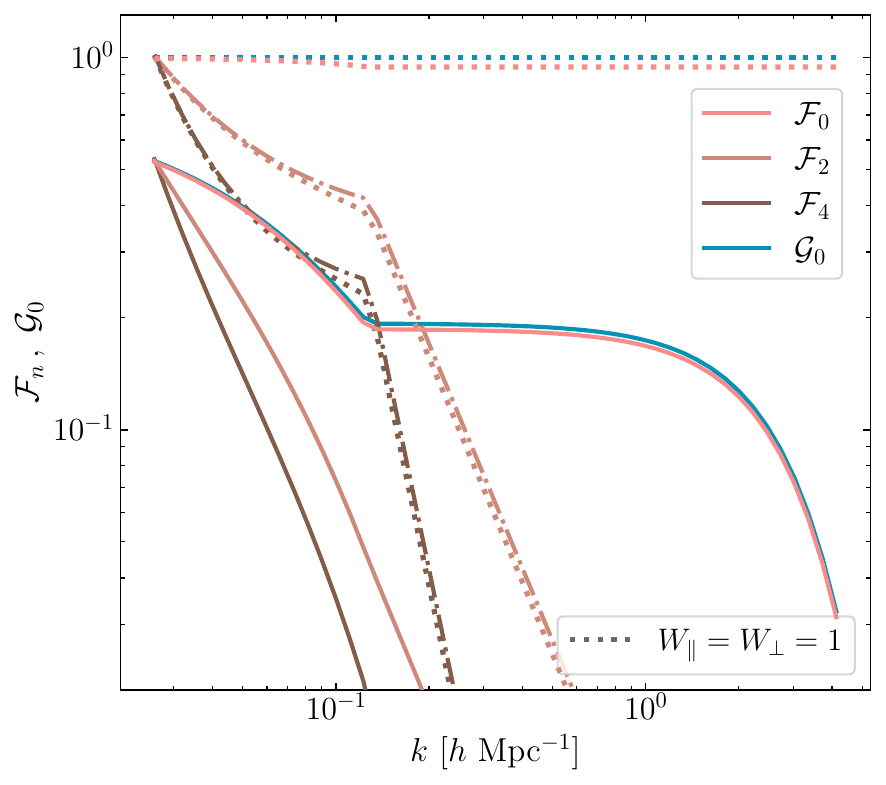} \hspace{1cm}
\includegraphics[width=0.42\textwidth]{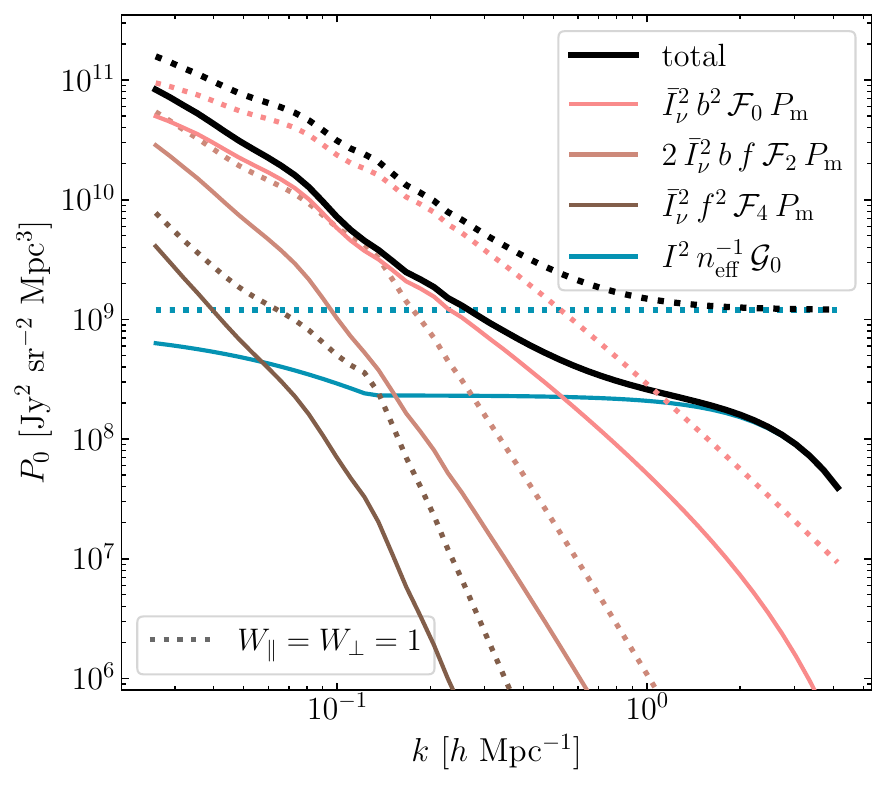}\\
\includegraphics[width=0.42\textwidth]{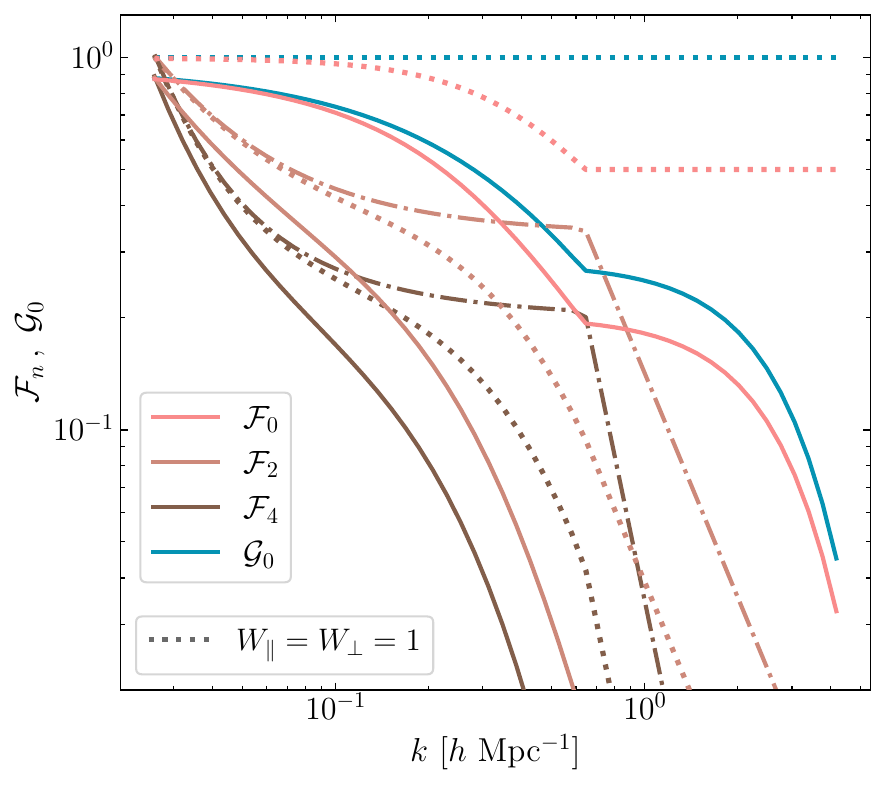} \hspace{1cm}
\includegraphics[width=0.42\textwidth]{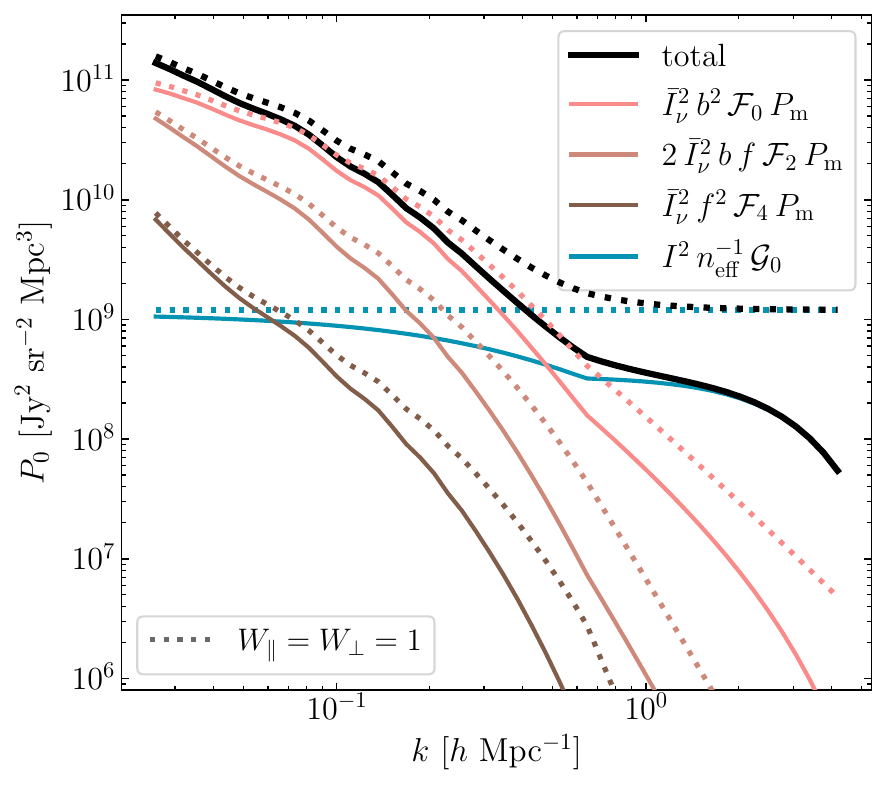}
\caption{
Left: Functions $\mathcal{F}_0$, $\mathcal{F}_2$, $\mathcal{F}_4$, and $\mathcal{G}_0$ defined in Eqs.~(\ref{eq:Fndef}) and (\ref{eq:G0def}) for our observational setup, assuming $\sigma = 3\,h^{-1}\,\mathrm{Mpc}$ and a spectral resolving power of $R = 100$ (top panel) or $R = 500$ (bottom panel). The dash-dotted lines represent the analytic approximations given in Eqs.~(\ref{eq:F2_analitic}) and (\ref{eq:F4_analitic}), with colours matching those used for $\mathcal{F}_2$ and $\mathcal{F}_4$, respectively.
Right: Individual components of the LIM power spectrum entering Eq.~(\ref{eq:finalP0}) for our pessimistic luminosity function with $\alpha = -1.1$ (similar results are found for other LF assumptions).}
\label{fig:sigmapv-effect}
\end{figure*}

\subsubsection{Redshift-space distortions and resolution effects}
The power spectra displayed in Fig.~\ref{fig:powerspectrum_rectangular} present some characteristic features at both small and large scales.
In order to explain their origin, we discuss the impact that redshift-space distortions and instrumental effects have on $\Delta^2$. 
By inserting Eqs.~(\ref{anisotropic_power_spectrum}),~(\ref{eq:Pclust}),~(\ref{shotnoise_component}) and ~(\ref{eq:newsmoothing}) in Eq.~(\ref{observedPk_CCAT}), we obtain
\begin{multline}
  P_0(k,z)=\bar{I}_\nu^2(z)\,\left\{\left[b^2(z)\,\mathcal{F}_0(k,z)+2\,b(z)f(z)\,\mathcal{F}_2(k,z)\right.\right.\\\left.\left.+f^2(z)\,\mathcal{F}_4(k,z)\right]\,P_\mathrm{m}(k,z)+\bar{n}^{-1}_\mathrm{eff}(z)\,\mathcal{G}_0(k,z)\right\}\;,
  \label{eq:finalP0}
\end{multline}
with
\begin{equation}
    \label{eq:Fndef}
    \mathcal{F}_n(k,z)=\langle  \mu^n\,\mathcal{D}(k,\mu,z)\, W_\perp(k,\mu)\,W_\parallel(k,\mu) \rangle_\mu
\end{equation}
and
\begin{equation}
    \label{eq:G0def}
    \mathcal{G}_0(k,z)=\langle W_\perp(k,\mu)\,W_\parallel(k,\mu)\ \rangle_\mu\;.
\end{equation}

In the absence of instrumental effects (i.e. $W_\perp=W_\parallel=1$), non-linear
redshift-space distortions (i.e. $\sigma=0$ or $\mathcal{D}=1$), and when averaging over the full range of $\mu\in [0,1]$, the functions
$\mathcal{F}_0, \mathcal{F}_2, \mathcal{F}_4$ and $\mathcal{G}_0$ take on their `classical' values of $1, 1/3, 1/5$, and 1, respectively. When $\mu$ is restricted to the interval $k_\mathrm{f}^\parallel/k\leq \mu\leq k_\mathrm{max}^\parallel/k$, $\mathcal{F}_0$ and $\mathcal{G}_0$ remain unchanged; however, $\mathcal{F}_2$ and $\mathcal{F}_4$ are significantly modified, becoming
\begin{align}
    \mathcal{F}_2=\begin{cases}
    \displaystyle{\frac{k^3-(k_\mathrm{f}^\parallel)^3}{3k^2(k-k_\mathrm{f}^\parallel)}}\;,& \text{if $k_\mathrm{f}^\parallel\leq k\leq k_\mathrm{max}^\parallel$\;,}\\    
    \displaystyle{\frac{(k_\mathrm{max}^\parallel)^3- (k_\mathrm{f}^\parallel)^3}{3k^2(k_\mathrm{max}^\parallel-k_\mathrm{f}^\parallel)}}\;,& \text{if $k> k_\mathrm{max}^\parallel$\;,}
    \end{cases}
    \label{eq:F2_analitic}
\end{align}
\begin{align}
    \mathcal{F}_4=\begin{cases}
    \displaystyle{\frac{k^5-(k_\mathrm{f}^\parallel)^5}{5k^4(k-k_\mathrm{f}^\parallel)}}\;,& \text{if $k_\mathrm{f}^\parallel\leq k\leq k_\mathrm{max}^\parallel$\;,}\\   
   \displaystyle{\frac{(k_\mathrm{max}^\parallel)^5-(k_\mathrm{f}^\parallel)^5}
   {5k^4 (k_\mathrm{max}^\parallel-k_\mathrm{f}^\parallel)}}\;,&  \text{if $k> k_\mathrm{max}^\parallel$\;,}
    \end{cases}
    \label{eq:F4_analitic}
\end{align}
and decline rapidly as a function of $k$.
In the low-$k$ regime, this behaviour arises because $\mu_\mathrm{max}$ is fixed to 1, while $\mu_\mathrm{min}$ progressively approaches 0 as $k$ increases. This shift reduces the average values of $\mu^2$ and $\mu^4$ from unity at $k=k_\mathrm{f}^\parallel$ to values approaching the classical ones at $k=k_\mathrm{max}^\parallel$.
Conversely, in the high-$k$ regime, where $\mu_\mathrm{max}<1$, both the upper and lower bounds of $\mu$ decrease with increasing $k$. As a result, the functions exhibit a more rapid decline in this regime.

Activating non-linear redshift-space distortions further suppresses $\mathcal{F}_2$ and $\mathcal{F}_4$, while also causing $\mathcal{F}_0$ to deviate from unity (see the dotted curves in the left panel of Fig.~\ref{fig:sigmapv-effect}).
The suppression of the clustering signal due to the incoherent redshift-space distortions remains mild for the EoR-Spec resolving power of $R=100$, with the damping factor $\mathcal{D}$ confined in the range $0.86 \leq \mathcal{D}< 1$ (assuming $\sigma=3\,h^{-1}$ Mpc) since our setup does not sample Fourier modes with $k_\parallel>k^\parallel_\mathrm{max}\simeq 0.13\,h \ \mathrm{Mpc}^{-1}$. In contrast, for an hypothetical future spectrograph with $R=500$, the damping can be significantly stronger, with $\mathcal{D}$ reaching values as low as 0.11 when
$k_\parallel=k^\parallel_\mathrm{max}\simeq 0.64\,h \ \mathrm{Mpc}^{-1}$.
The function $\mathcal{F}_0$ (pink dotted lines) gives the effective damping factor due to the incoherent small-scale motions as a function of $k$.

Finally, when instrumental effects are taken into account, all the functions are affected in two key ways:
(i) they experience significant damping due to the limited spectral resolution, as captured by $W_\parallel$; and (ii)
they are strongly suppressed at $k\gtrsim \sigma_\perp^{-1}$ owing to the finite angular resolution of the observations, encoded in $W_\perp$.
The resolving power of the spectrograph plays a critical role in modulating these effects.
For $R=100$, both $\mathcal{F}_0$ and $\mathcal{G}_0$ are reduced by a factor of approximately $5$ at $k\simeq 0.2\,h$ Mpc$^{-1}$, significantly lowering the amplitude of the observed PS.
Moreover, since the FWHM of the Lorentzian line profile exceeds $8\sigma$, the resulting exponential damping becomes so severe that the observed signal carries virtually no sensitivity to $\sigma$, making it effectively impossible to place relevant constraints on this parameter at the given spectral resolution.
The situation improves substantially for $R=500$, where the suppression due to the spectral resolution is less severe and $\Delta_\parallel^\mathrm{FWHM}\simeq 1.6\,\sigma$. In this case, although the damping of the clustering signal due to the incoherent redshift-space distortions remains subdominant, it cannot be neglected entirely, and setting constraints on $\sigma$ becomes possible.

To mitigate the suppression caused by $W_\parallel$, we could use a different clustering statistic. 
One possible approach, involves reducing the value of $\mu_\mathrm{max}$ in Eq.~(\ref{observedPk_CCAT}) to restrict the analysis to Fourier modes with small $|k_\parallel|$.
For example, one could impose $\mu_\mathrm{max}=\min (1,k^\parallel_\mathrm{max})$, where $k^\parallel_\mathrm{max}\ll \pi/\Delta_\parallel^\mathrm{FWHM}$. 
While this strategy enhances the signal by limiting the contribution from modes strongly affected by spectral resolution, it also leads to increased noise. This is because the averaging would be performed over a smaller subset of Fourier modes, and $N_\mathrm{m}(k)$ in Eq.~(\ref{eq:nmodes}) would need to be calculated with  $k^\parallel_\mathrm{max}$ replacing $\pi/\Delta_\parallel^\mathrm{FWHM}$.
In practical terms, this amounts to analysing  line-intensity maps with intentionally degraded resolution along the line of sight. The optimal choice of $k^\parallel_\mathrm{max}$ could be determined by maximising the overall S/N of the measurement. However, we find that this strategy yields only modest gains in cumulative S/N (e.g. around 7\% for $k^\parallel_\mathrm{max}=0.2\,h$ Mpc$^{-1}$ and $R=500$) and does not lead to tighter constraints on the model parameters. For this reason, we retain the full range range of Fourier modes in the analysis presented throughout the remainder of the paper.

\subsubsection{Sensitivity to model parameters}
\label{Sec:sensitivity}
Eq.~(\ref{eq:finalP0}) can be employed to constrain model parameters through template fitting. For a fixed cosmological model, theory of gravity (which determines the growth rate $f$), and observational setup, the functions $\mathcal{F}_0, \mathcal{F}_2, \mathcal{F}_4$ and $\mathcal{G}_0$ depend solely on $\sigma$. 
This allows the parameters $\bar{I}_\nu, b, \bar{n}_\mathrm{eff}^{-1}$ and $\sigma$ to be varied in order to fit the observed data. 
The dominant clustering contribution to the power spectrum---proportional to $\mathcal{F}_0$---is only sensitive to the combination $\bar{I}_\nu^2\,b^2$ and to $\sigma$. Additional information on $\bar{I}_\nu^2\,b$ and $\bar{I}_\nu^2$ can, in principle, be extracted from the sub-dominant components proportional to $\mathcal{F}_2$ and $\mathcal{F}_4$, respectively. However, these terms can only be meaningfully constrained if they are sufficiently large compared to the statistical uncertainties of the measurements.
Moreover, constraining three parameters in this regime requires multiple well-measured data points.
On small scales, where the shot-noise component dominates, the power spectrum is primarily sensitive to the product $\bar{I}_\nu^2\,\bar{n}_\mathrm{eff}^{-1}$.

The solid curves in the right-hand panels of Fig.~\ref{fig:sigmapv-effect} show the individual contributions to the PS appearing in Eq.~(\ref{eq:finalP0}) for our pessimistic scenario characterised by $\alpha=-1.1$ and $\sigma=3\,h^{-1}$ Mpc. 
At the centre of our first bin and for $R=100$, the components proportional to ${\mathcal F}_0$, ${\mathcal F}_2$, ${\mathcal F}_4$ and ${\mathcal G}_0$ contribute approximately 80.2\%, 15.3\%, 1.1\% and 3.4\% of the total signal, respectively. In the second bin, these value change to 81.7\%, 3.6\%, 0.1\%, and 14.6\%.
As expected, the coherent large-scale flows---encoded in the terms proportional to $f$ and $f^2$---only modestly enhance the clustering signal for highly biased tracers such as \cii emitters. 
Nonetheless, the fact that shot noise and multiple clustering terms contribute at comparable levels highlights the necessity of employing statistical inference to disentangle and constrain the individual components of the signal.

\section{Bayesian inference}\label{sec:bayesian_inference_results}

In this section, we assess what information can be extracted about the population of \cii emitters from the measurements of the LIM PS.
The most direct approach is to fit Eq.~(\ref{eq:finalP0}) to the data using $\bar{I}_\nu, b, \sigma$ and $\bar{n}_\mathrm{eff}^{-1}$
as tunable parameters while keeping fixed the cosmological parameters (and thus $f$).
This procedure allows us to determine information about the \cii LF without assuming its functional form and without relying on abundance matching.
In fact, Eqs.~(\ref{mean_specific_intensity}) and (\ref{eq:neff}) show that $\bar{I}_\nu$ is proportional to the first moment of the LF ($\bar{\rho}_L$) and $\bar{\rho}_L^2\,\bar{n}_\mathrm{eff}^{-1}$ gives exactly the second moment.
We term this approach minimal modelling and we pursue it in Sects.~\ref{sec:nonparam} and \ref{sec:moments}.

Alternatively, one could pick a functional form for the LF and set constraints on its free parameters and $\sigma$ from the LIM PS. In this case, $b$ is a function of the LF parameters which is evaluated via abundance matching and the halo model using Eq.~(\ref{eq:effectivebas}).
This analysis is presented in Sect.~\ref{sec:param}.

We performed Bayesian inference of the model parameters $\boldsymbol{\theta}$ given some mock observations $\mathbf{D}\equiv\{D_i\}$ representing the LIM power-spectrum monopole in different $k$-intervals and/or the LF observed in luminosity bins.
We assumed Gaussian independent errors and wrote the likelihood function as
 \begin{equation}
        \mathcal{L}(\boldsymbol{\theta}|\mathbf{D}) \propto \exp\left\{-\frac{1}{2} \sum_i \frac{\left[D_i-M_i(\boldsymbol{\theta}) \right]^2}{\sigma_i^2(\boldsymbol{\theta})}\right\} \; ,
        \label{gaussian_likelihood}
    \end{equation}
where $M_i$ denotes the model predictions in a given bin and $\sigma_i$ is the corresponding statistical errors.
We sampled the posterior distribution of $\boldsymbol{\theta}$ with the \textsc{emcee} code \citep{emcee} which implements the affine-invariant ensemble sampler for Markov chain Monte Carlo (MCMC) by \citet{Goodman-Weare-10}.
Given the current limited knowledge of the \cii LF at high redshift (Sect.~\ref{sec:LF}), we repeated our analysis several times with different mock data. On the one hand, in our optimistic case, we generated the data based on the Y20 LF. On the other hand, in our pessimistic case, we used our own fit to the LF of the targeted ALPINE detections with $\alpha=-1.1$. For one particular application, we also considered a steeper faint end,  with $\alpha=-1.9$.
In all cases, we assumed that $\sigma=3\,h^{-1}$ Mpc. 

Table~\ref{tab:surveys} summarises the survey configurations considered in our analysis (labelled A to D).
As a baseline (A), we adopted the 16 sq. deg. survey with DSS sensitivity introduced in Sect.~\ref{sec:survey}.  This reflects the current state of the art, although it already assumes twice the sky area of the planned DSS survey. 
Building on this, we considered three increasingly ambitious LIM scenarios (B through D), which feature wider sky coverage, enhanced sensitivity, or both, while keeping the spectral resolving power fixed at $R=100$.
It is important to note that within our modelling framework, variations in $\Omega_\mathrm{surv}$ or $P_\mathrm{WN}$ only affect the statistical errors on $\Delta^2$ and do not alter the underlying signal or the range of accessible wavenumbers.

\begin{table}[t]
    \caption[]{Characteristics of the abstract surveys.}
    \label{tab:surveys}
    \centering
    \begin{tabular}
    {l@{\hskip 5pt}c@{\hskip 5pt}c@{\hskip 5pt}c@{\hskip 0pt}c}
        \hline
        \hline
        \noalign{\smallskip}
        Name & $\Omega_\mathrm{surv}$ & $R$ &$P_\mathrm{WN}$& Comment\\
        & sq. deg. & & Jy$^2$ sr$^{-2}$ Mpc$^3$ &\\
        \noalign{\smallskip}
        \hline
        \noalign{\smallskip}
        A & 16 & 100 &$7.8\times 10^{10}$& Baseline\\
        B & 160 & 100 & $7.8\times 10^{10}$ &  Wider $(10\times)$\\
        C & 16 & 100 &$7.8\times 10^{9}$&  More sensitive $(\sqrt{10}\times)$\\
        D & 160 & 100 &$7.8\times 10^{9}$ & Wider and more sensitive\\
        \noalign{\smallskip}
        \hline
    \end{tabular}
\end{table}

Finally, to assess the impact of increased spectral resolution, we also considered variants of each configuration carried out with a futuristic instrument operating at $R=500$ (denoted A$^+$ through D$^+$). To ensure a consistent comparison and preserve the white-noise power spectrum level, the total survey duration was scaled up by a factor of five.

\subsection{Minimal modelling}
\label{sec:nonparam}
Although it would be possible in principle to use $\boldsymbol{\theta}=\{\bar{I}_\nu, b, \sigma, \bar{n}_\mathrm{eff}^{-1}\}$ in MCMC sampling, this would lead to a very inefficient exploration of parameter space, because the model parameters are highly correlated. In order to minimise degeneracies and make MCMC sampling much more efficient, we re-parametrise the model using
$\boldsymbol{\theta}=
\{ \bar{I}_\nu^2\,b^2,b, \bar{I}_\nu^2/\bar{n}_\mathrm{eff},\sigma\}$.

We adopted independent uniform priors for all parameters, with the corresponding ranges listed in the upper section of Table~\ref{tab:priors}. For the clustering and shot-noise amplitudes, we selected broad intervals extending from zero to large values. These choices have little impact on our inference, as the mock data tightly constrain both parameters. The situation is different for $b$ and $\sigma$, which cannot always be determined precisely and whose posteriors may remain prior-dominated depending on the survey configuration.
For these parameters, we imposed theoretically motivated bounds that reflect the expected nature of the LIM signal. We assumed that it is primarily generated by galaxies residing in DM haloes with masses in the range $10^{10}\le M\le 10^{13}\,h^{-1}$ M$_\sun$, ensuring a physically plausible and conservative exploration of parameter space. In particular, we only considered massive, biased haloes, and excluded the possibility of anti-biased tracers by restricting $b>1$. The lower limit on $\sigma$ reflects the velocity dispersion expected for central galaxies in low-mass haloes, while the upper limit accounts not only for satellite galaxies in massive haloes but also includes contributions from internal gas motions and relative velocities between distinct haloes (see also Sect.~\ref{sec:PSresults}).

\begin{figure}
\centering
\includegraphics[width=0.48\textwidth]
{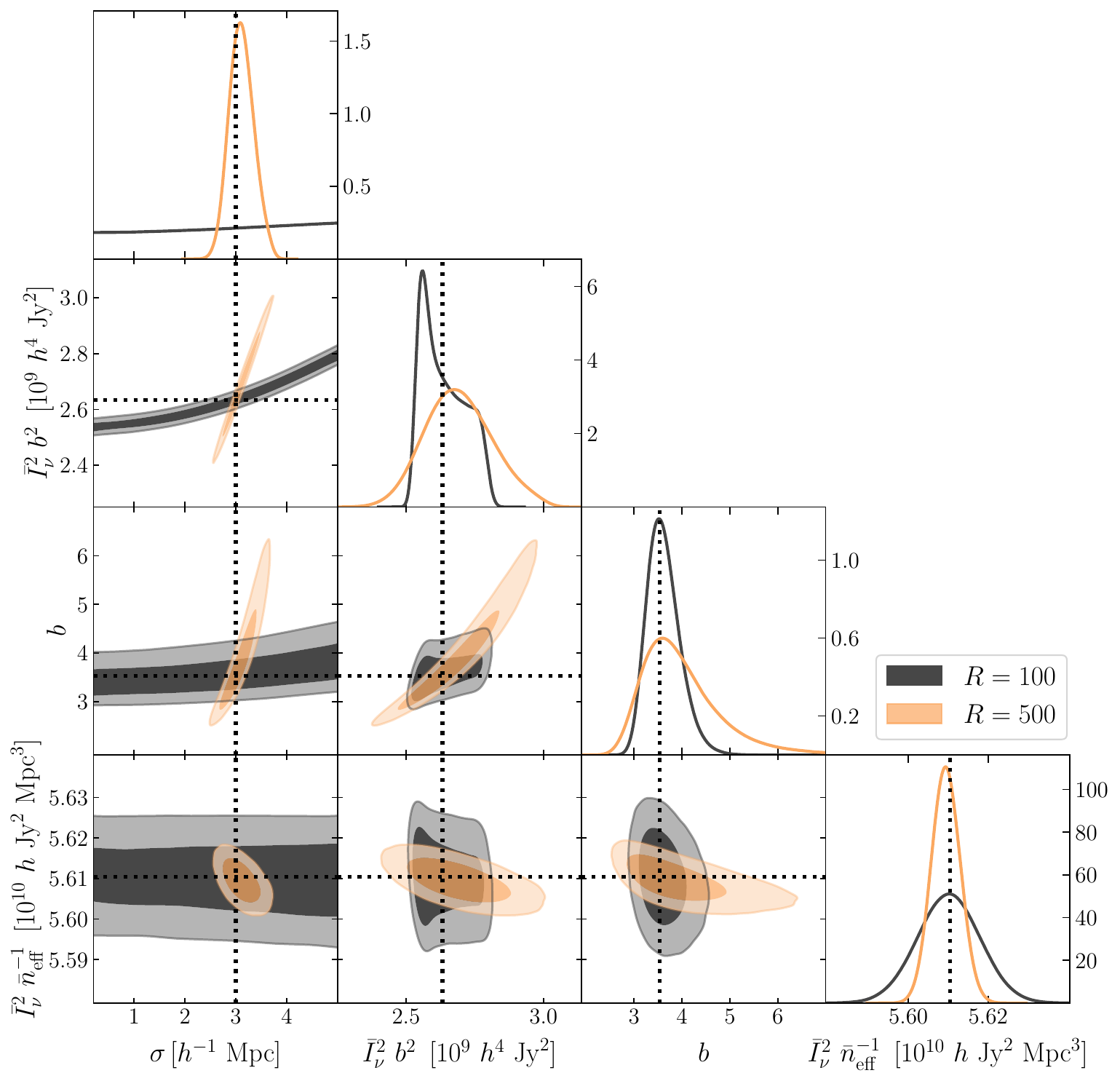} 
\caption{Marginalised posterior distributions of the model parameters we obtained by fitting synthetic data for the LIM PS. The displayed results assume the optimistic \cii LF and refer to survey D (grey) and D$^+$ (orange).
The shaded areas indicate the 68\% (dark) and 95\% (light) highest posterior density (HPD) regions (hereafter, credible regions). The dotted lines highlight the underlying true values.}
\label{fig:fullcorneroptimistic}
\end{figure}

\begin{table}[t]
    \caption[]{Uniform prior probabilities.}
    \label{tab:priors}
    \centering
    \begin{tabular}{lcc}
        \hline
        \hline
        \noalign{\smallskip}
      Parameter & Units & Prior range\\
        \noalign{\smallskip}
        \hline
        \noalign{\smallskip}
        $\bar{I}_\nu^2 \,b^2$ & $h^4$ $\mathrm{Jy}^2$ & $(0, 10^{10})$\\
        $b$ & -- & $(1,7)$\\
        $\bar{I}_\nu^2 \,\bar{n}_\mathrm{eff}^{-1}$ & $h$ $\mathrm{Jy}^2$ Mpc$^{3}$& $(0,10^{11})$\\
        $\sigma$ & $h^{-1}$ Mpc & $(0.2,5)$\\
        \noalign{\smallskip}
        \hline
        $\log_\mathrm{10}[\Psi_*/(\mathrm{Mpc}^{-3}\mathrm{dex}^{-1})]$ & --& $(-6, 0)$ \\
        $\log_\mathrm{10} (L_*/L_\sun)$ & -- & $(5,10)$\\
        $\alpha$ & -- & $(-2,3)$ \\
        $\sigma$ & $h^{-1}$ Mpc & $(0.2,5)$\\
        \noalign{\smallskip}
        \hline
    \end{tabular}
\end{table}

As an example, in Fig.~\ref{fig:fullcorneroptimistic}, we present the marginalised one- and two-dimensional posterior distributions of the model parameters obtained for the D and D$^+$ surveys using mock data based on our optimistic LF. The leftmost column of the figure shows that, even in this favourable scenario, it is not possible to place significant constraints on the damping parameter $\sigma$ when adopting a resolving power of $R = 100$. At this resolution, the spectral smoothing induced by the instrumental response function $W_\parallel$ dominates over the physical damping from incoherent redshift-space distortions, rendering variations in $\sigma$ effectively unobservable. As a result, the posterior distribution for $\sigma$ remains flat across its prior range, and its degeneracies with other parameters are largely suppressed.
In contrast, when the resolving power is increased to $R = 500$, the damping signature of redshift-space distortions becomes distinguishable from the instrumental effects, which enables us to determine $\sigma$. This gain in sensitivity comes with an important caveat, however: The degeneracy between $\sigma$ and the clustering amplitude parameters, in particular $\bar{I}_\nu^2 b^2$, becomes significantly more pronounced. 
This is evident in the shape and orientation of the credible regions in the $\bar{I}_\nu^2 b^2$--$\sigma$ plane. At $R = 100$, the contours follow the approximate relation $\bar{I}_\nu^2\,b^2 = 0.097\,(\sigma/\sigma_\mathrm{true})^{1.9} + 2.54$ (in units
of $10^9\,h^4$ Jy$^2$), while for $R = 500$, the degeneracy tightens along a steeper trajectory, $\bar{I}_\nu^2\,b^2 =1.53\,(\sigma/\sigma_\mathrm{true}) + 1.1$.
This change reflects the fact that, at high spectral resolution, the observed shape of the power spectrum becomes more sensitive to $\sigma$, but in a way that strongly couples to the amplitude of clustering. As a result, uncertainties in $\sigma$ propagate more directly into $\bar{I}_\nu^2 b^2$, elongating the credible regions along a near-linear degeneracy direction.
These findings highlight a subtle but important trade-off: While higher spectral resolution enables the measurement of physical parameters like $\sigma$, it can also exacerbate parameter degeneracies, particularly when the number of observables remains limited. This suggests that complementary information or additional observables may be required to fully disentangle the contributions of $\sigma$, $b$, and $\bar{I}_\nu$ to the LIM signal.

Based on these considerations, we hereafter present results marginalised over $\sigma$, unless explicitly stated otherwise. In this section, we focus on the results obtained for $R = 100$, while the corresponding analysis for surveys with $R = 500$ is provided in Appendix~\ref{Sect:R500}.

\begin{figure*}[h!]
\centering
\includegraphics[width=0.48\textwidth]{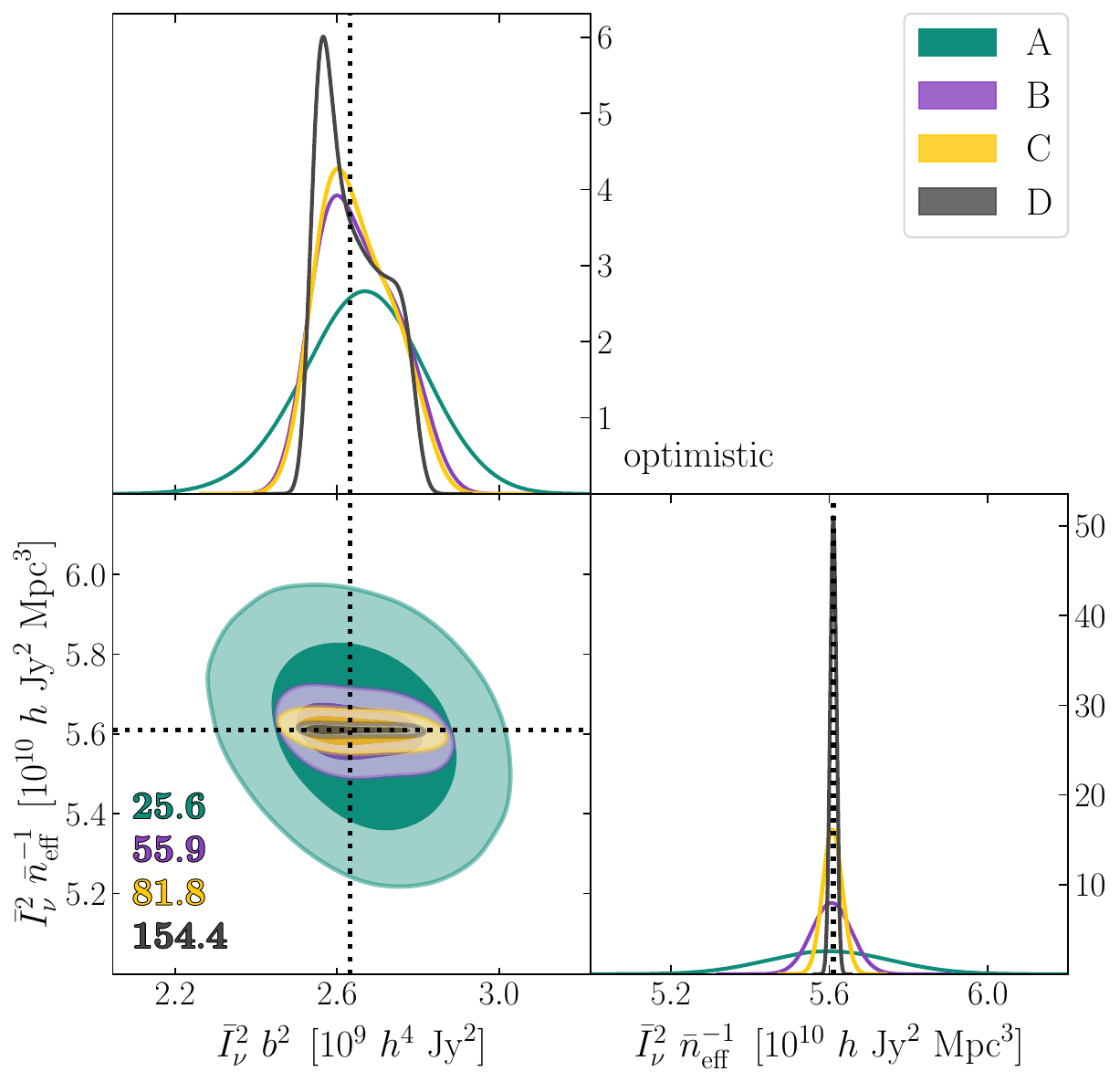} \hspace{0.5cm}
\includegraphics[width=0.48\textwidth]{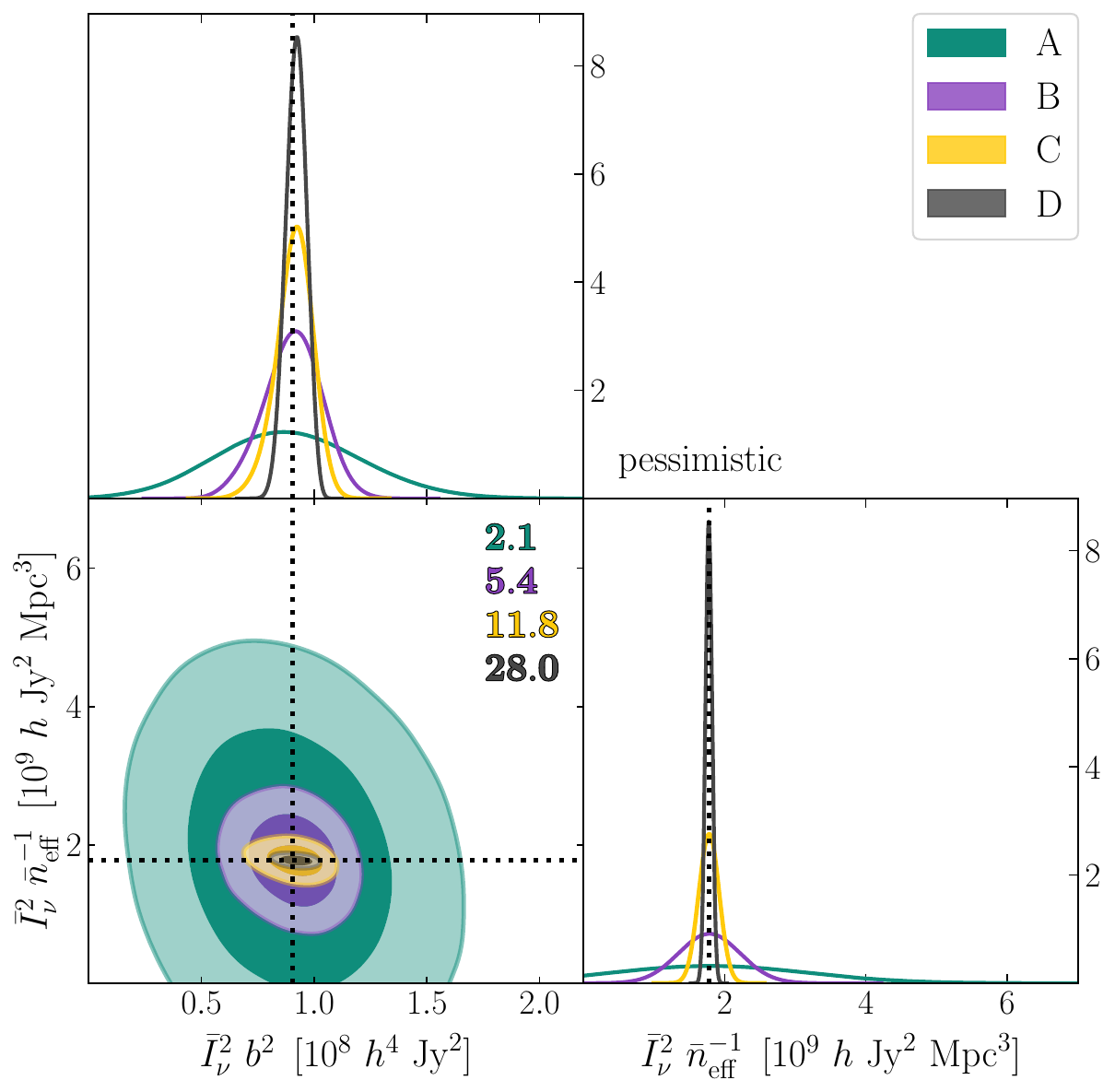}
\caption{Marginalised posterior distributions of the parameters $\bar{I}_\nu^2\,b^2$ and $\bar{I}_\nu^2\,\bar{n}_\mathrm{eff}^{-1}$ for the different surveys listed in Table~\ref{tab:surveys}. The left and right panels refer to the optimistic and pessimistic cases, respectively. 
Shown are the 68\% and 95\% credible regions (shaded) and the underlying true values (dotted). Also indicated is the figure of merit defined in Eq.~(\ref{eq:fom}).}
\label{fig:constraints_momentsLF}
\end{figure*}

In Fig.~\ref{fig:constraints_momentsLF}, we zoom into the marginalised joint posterior distribution for the clustering and shot-noise amplitudes $\bar{I}_\nu^2\,b^2$ and $\bar{I}_\nu^2/\bar{n}_\mathrm{eff}$. Here, we overplot the results obtained for the four different surveys described in Table~\ref{tab:surveys}. 
The left and right panels refer to different mock data generated using the optimistic and pessimistic LF, respectively. The axes ranges are different in the two panels.
The first thing that one spots is that the peak of the marginalised posterior for $\bar{I}_\nu^2\,b^2$ is shifted from the true value  when the optimistic LF is used. This is a projection (or prior-volume) effect which arises because of the degeneracy between $\bar{I}_\nu^2\,b^2$ and $\sigma$.
The projection of the banana shaped region (see e.g.
Fig.~\ref{fig:fullcorneroptimistic}) generates the shifted peak of the marginalised posterior. The peak of the likelihood lies at the true value, however.

The most important thing we learn from Fig.~\ref{fig:constraints_momentsLF} is how the parameter constraints respond to survey and instrumentation improvements. 
In order to more easily compare the constraining power of the different surveys, we introduce a figure of merit defined as 
\begin{equation}
\mathrm{FoM}=\left[(\det \Sigma_{n})^{-1/2}\,\displaystyle{\prod_{i=1}^n \theta_i^\mathrm{true}}\right]^{1/n}\;,
\label{eq:fom}
\end{equation}
with $n=2$, in this case, where the symbols $\theta_i^\mathrm{true}$ indicate the actual values of the model parameters that have been used to generate the mock data and $\Sigma_{n}$ denotes the corresponding minor of the covariance matrix extracted from the MCMC chains. This dimensionless quantity is a measure of tightness of the posterior probability: the higher is FoM, the stronger are the constraints on the model parameters. For non-correlated variables, it gives the geometric average of their S/N.
It turns out that increasing the sensitivity of the survey is more beneficial than increasing its area.
With respect to survey A, the FoM increases by a factor of 2.2 (2.6) in the optimistic (pessimistic) case for survey B and of 3.2 (5.6) for survey C. The corresponding figure for survey D is 6.0 (13.3).

\begin{figure}
\centering
\includegraphics[width=0.37\textwidth]{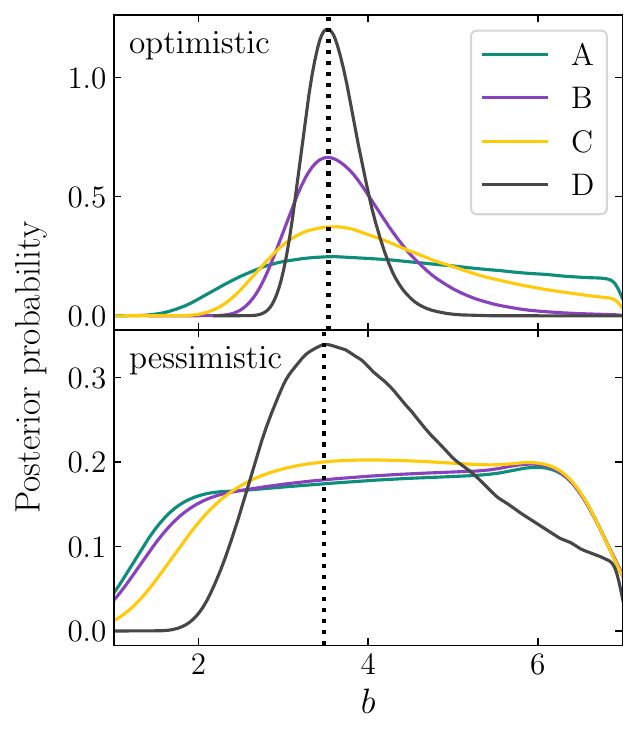} 
\caption{Marginalised posterior distributions for the linear bias parameter of the \cii emitters.
The dotted lines indicate the true values.}
\label{fig:bposterior}
\end{figure}

\begin{figure*}
\centering
\includegraphics[width=0.35\textwidth]{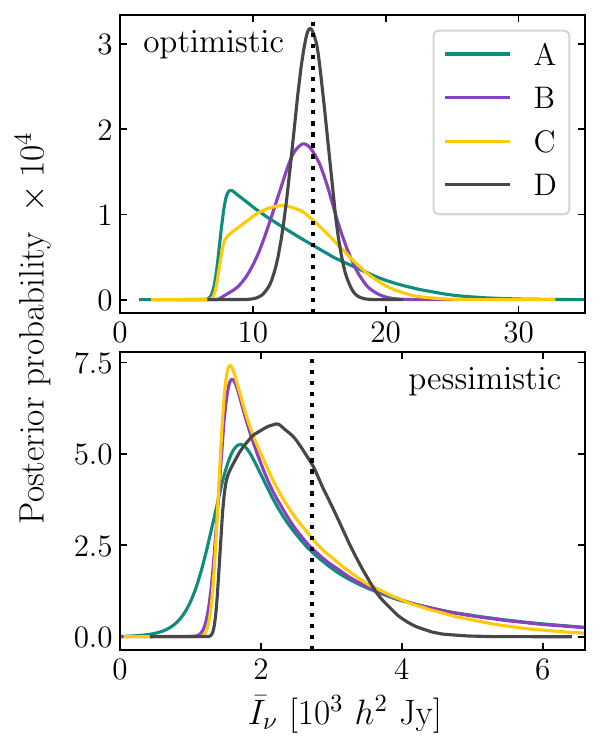} \hspace{1.5cm}
\includegraphics[width=0.35\textwidth]{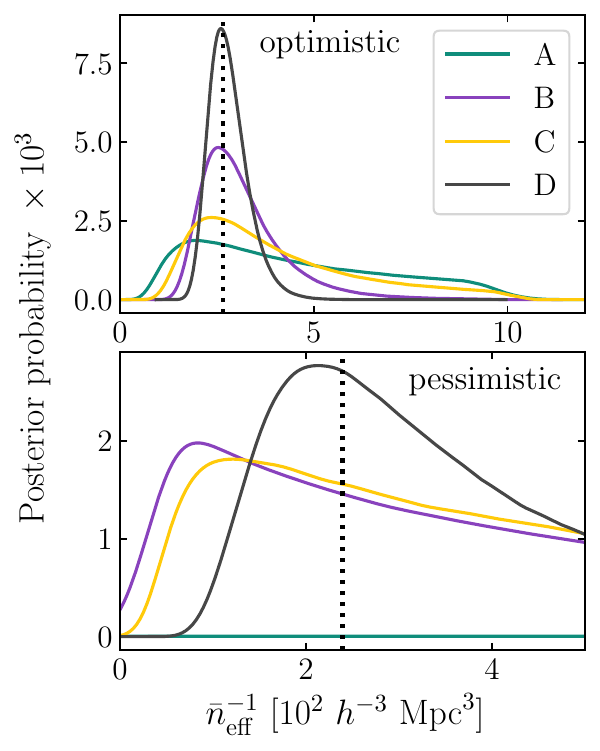} 
\caption{Marginalised posterior distributions for the mean LIM intensity (left) and the effective volume per \cii emitter (right). 
The dotted lines indicate the true values.}
\label{fig:Iposterior}
\end{figure*}

Fig.~\ref{fig:bposterior} shows the marginalised posterior distribution of the linear bias parameter which provides information about the DM haloes hosting the \cii emitters.
Survey A is incapable of setting any useful constraints on $b$. In the optimistic case, all the other configurations are sufficient to provide a measurement with a S/N greater than one. In particular, a larger survey area (B) gives tighter constraints than a more sensitive survey (C). Conversely, in the pessimistic case, survey D is needed to measure $b$.

In the left panel of Fig.~\ref{fig:Iposterior}, we present the marginalised posterior distributions for the mean LIM signal, $\bar{I}_\nu$, which is treated as a derived parameter in our analysis. Since our inference is based on independent uniform priors for the primary model parameters, the resulting effective prior on $\bar{I}_\nu$ is non-trivial. It features a sharp cutoff at very small intensities, followed by a pronounced peak and a long tail extending toward higher values (see Appendix~\ref{sec:I-prior}). 
For the optimistic luminosity function, we find that both surveys B and D yield relatively tight posterior distributions for $\bar{I}_\nu$, demonstrating the capability of these configurations to constrain the mean LIM intensity. In the more challenging scenario based on the pessimistic luminosity function, however, only survey D shows a marked suppression of the extended high-intensity tail, indicating a genuine gain in constraining power. 

Similar conclusions emerge from the analysis of the marginalised posterior distribution of the derived parameter $\bar{n}_\mathrm{eff}^{-1}$, shown in the right panel of Fig.~\ref{fig:Iposterior}. While the shot-noise amplitude, $\bar{I}_\nu^2\,\bar{n}_\mathrm{eff}^{-1}$, is relatively well constrained across all test surveys---with the exception of Survey A under the pessimistic scenario (see Fig.~\ref{fig:constraints_momentsLF})---disentangling and independently constraining the amplitude $\bar{I}_\nu$ and the effective volume $\bar{n}_\mathrm{eff}^{-1}$ of the emitters remains significantly more challenging.

\begin{figure}
\centering
\includegraphics[width=0.43\textwidth]{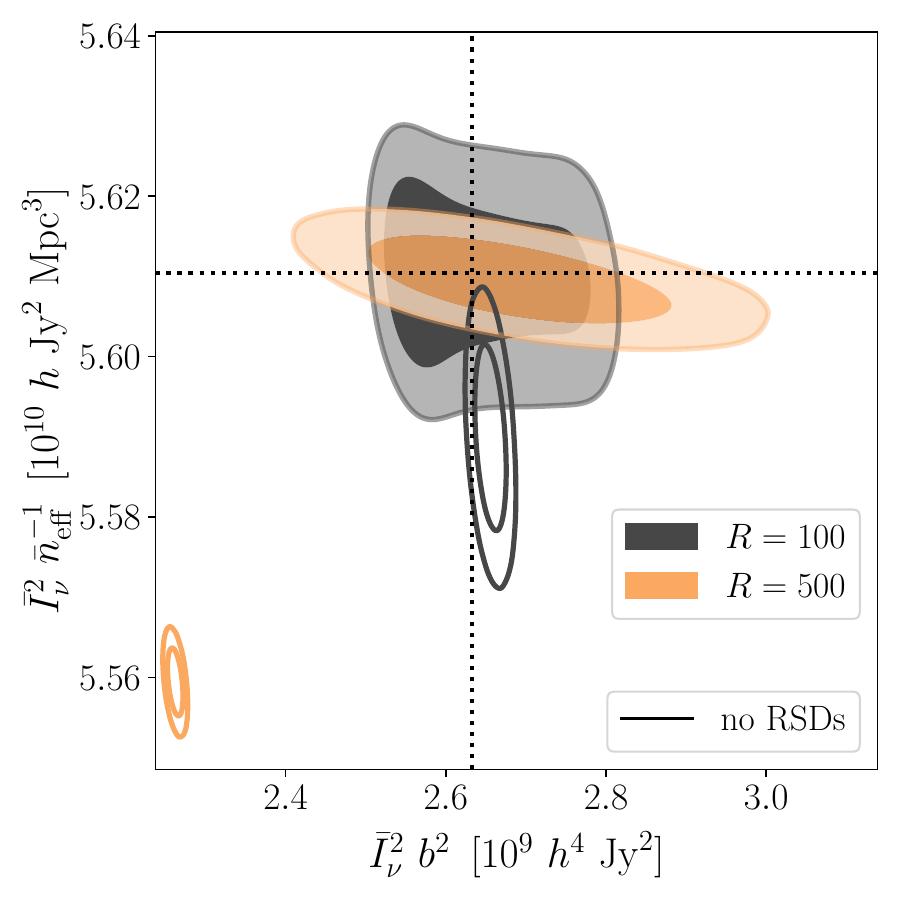} 
\caption{Joint posterior distributions of the clustering and shot-noise amplitudes for the D (grey) and D$^+$ (orange) survey configurations. Shaded regions denote the 68\% and 95\% credible regions obtained using the full four-parameter model in Eq.~(\ref{eq:finalP0}), which includes redshift-space distortions (RSDs), after marginalising over the parameters $b$ and $\sigma$. Unfilled contours show the corresponding credible regions derived from a simplified two-parameter model that neglects RSDs.}
\label{fig:RSDs}
\end{figure}

In Sect.~\ref{Sec:sensitivity}, we have demonstrated that redshift-space distortions introduce only per-cent level correction to the PS and that their impact rapidly diminishes with increasing wavenumber.
Armed with this knowledge, one might be tempted to simplify the model for the LIM PS by neglecting the terms proportional to $f$ in Eq.~(\ref{eq:finalP0}) and by setting $\mathcal{D}=1$.
The consequences of such a simplification are illustrated in Fig.~\ref{fig:RSDs} for the D and D$^+$ surveys using the optimistic LF. We compare the marginalised posterior distributions in the $\bar{I}_\nu^2\,b^2$--$\bar{I}_\nu^2/\bar{n}_\mathrm{eff}$ plane obtained by (i) using the full model, which includes RSDs and marginalises over both $b$ and $\sigma$ (shaded areas) and (ii) adopting a simplified model that neglects RSDs entirely (solid lines).
It is evident that excluding RSDs from the analysis introduces significant biases in parameter estimation, particularly for high-resolution observations with $R=500$.
Furthermore, the simplified model underestimates the uncertainty on $\bar{I}_\nu^2\,b^2$,  highlighting that accurate modelling of RSDs remains essential, even when their apparent impact on the PS amplitude is modest.

\begin{figure*}[h!]
\centering
\includegraphics[width=0.46\textwidth]{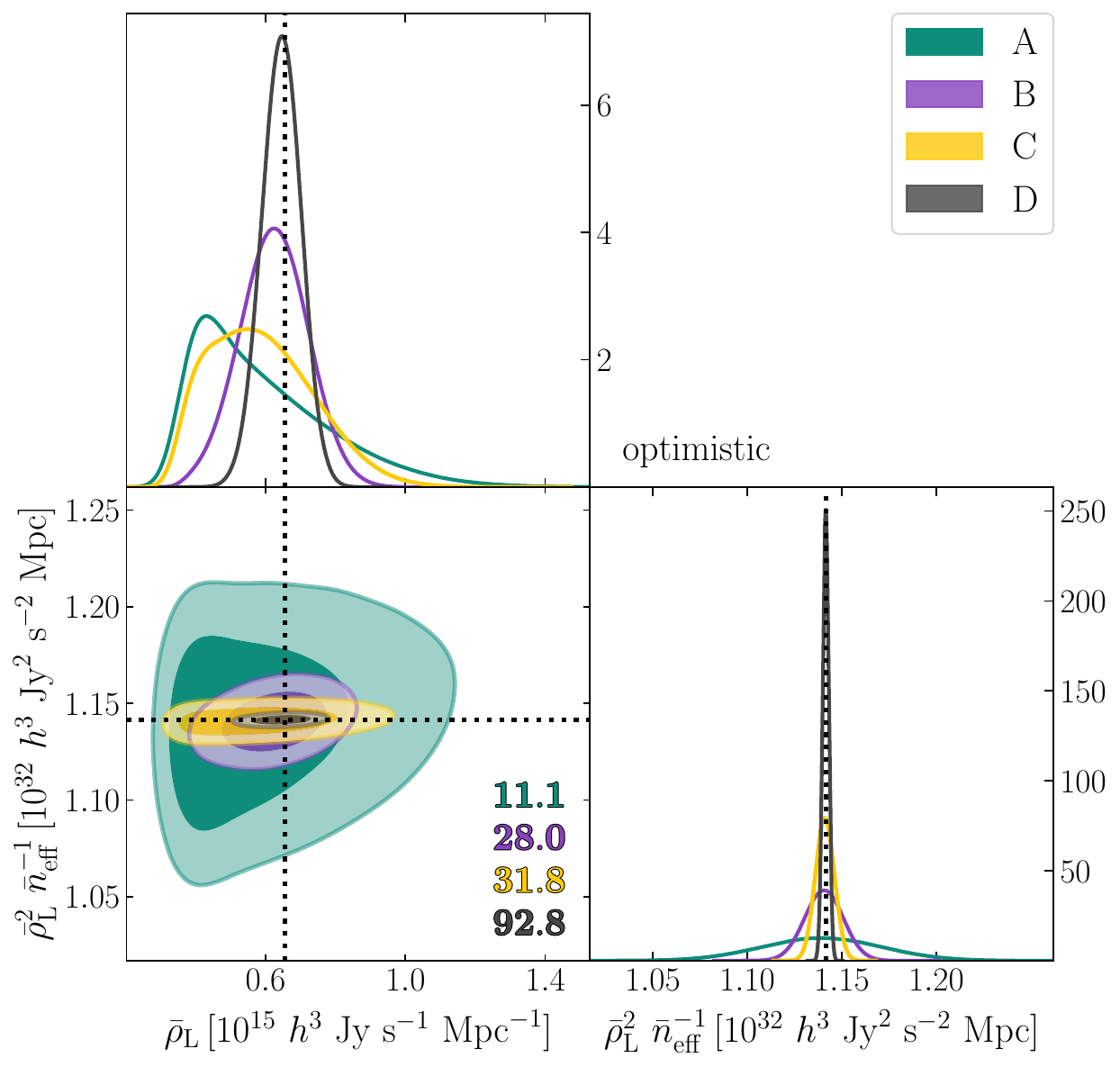} \hspace{0.5cm}
\includegraphics[width=0.46\textwidth]{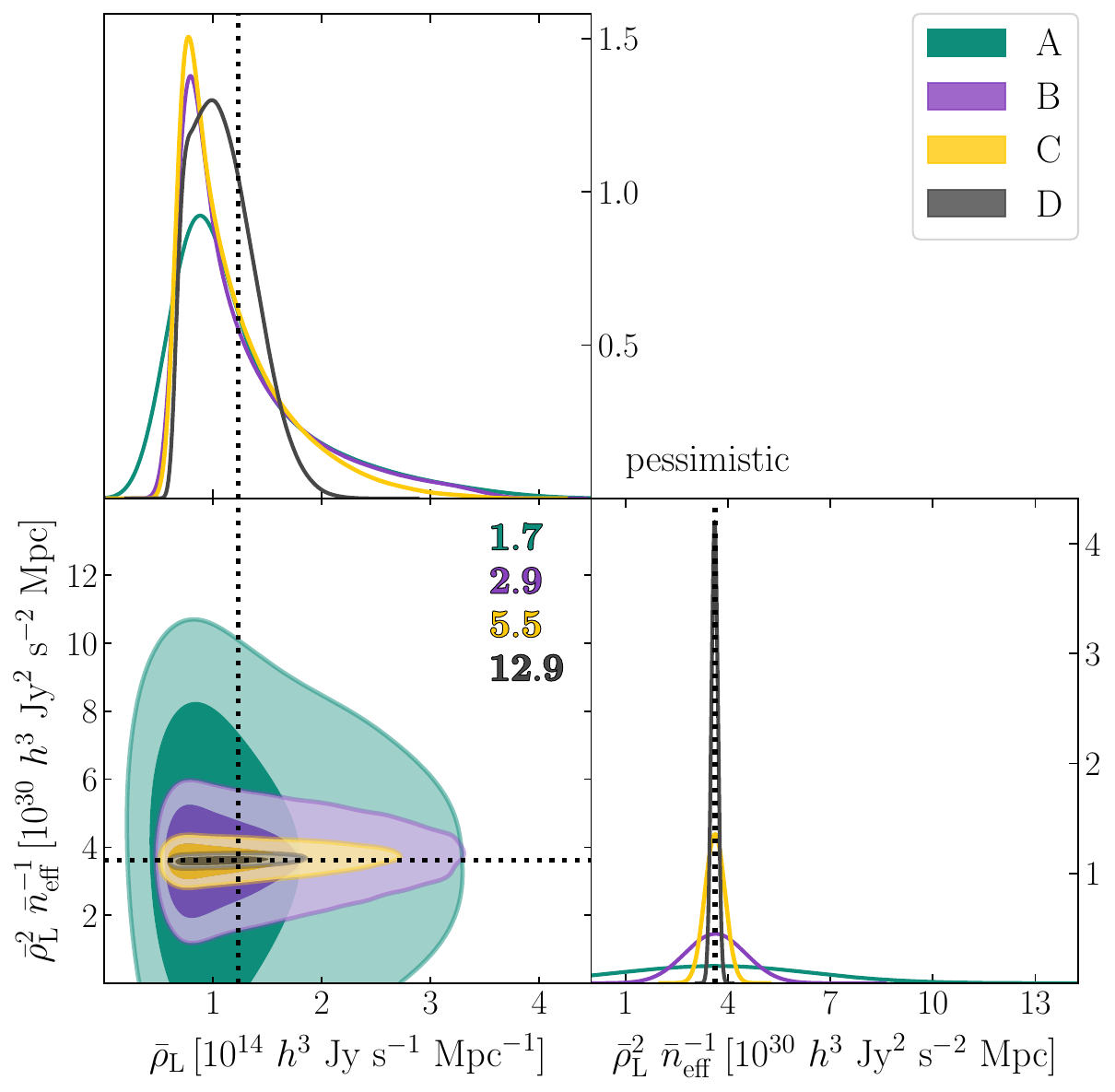}
\caption{As in Fig.~\ref{fig:constraints_momentsLF}, but for the derived variables that give the first two moments of the LF.}
\label{fig:constraints_derived_momentsLF}
\end{figure*}

\subsection{Moments of the luminosity function}
\label{sec:moments}
The LIM PS is sensitive to the first two moments of the LF, $\bar{\rho}_L$ and 
$\bar{\rho}_L^2\,\bar{n}_\mathrm{eff}^{-1}$.
It is thus interesting to investigate what constraints can be set on these quantities. 
Eq.~(\ref{mean_specific_intensity}) shows that $\bar{\rho}_L$ can be obtained rescaling $\bar{I}_\nu$ by a (cosmology dependent) constant factor.
In  Fig.~\ref{fig:constraints_derived_momentsLF}, we plot the joint marginalised posterior distribution of $\bar{\rho}_L$ and $\bar{\rho}_L^2\,\bar{n}_\mathrm{eff}^{-1}$ by treating them as derived variables in our MCMC chains.
The two parameters turn out to be nearly uncorrelated.

For the optimistic case (left panel), the second moment of the luminosity function is very precisely and accurately measured, while the first moment, $\bar{\rho}_\mathrm{L}$, is also well constrained, particularly in surveys B and D.
These two surveys yield nearly unbiased estimates, with survey D achieving a high S/N of 11.7 and a posterior distribution that peaks close to the true value.
In the pessimistic case (right panel), the second moment remains tightly constrained (except for survey A), but the posterior for $\bar{\rho}_\mathrm{L}$ becomes broader and moderately biased, reflecting the degeneracies with $b$ and $\sigma$, which are poorly constrained (not shown in the figure).
Nonetheless, survey D still achieves a detection with $S/N \simeq 4.3$, and the peak of the posterior is shifted by only $0.84$ standard deviations from the true value.
We conclude that, while the LIM PS can robustly constrain the second moment of the LF across a wide range of scenarios, the first moment can also be reliably inferred even though mild biases may arise under pessimistic conditions due to parameter degeneracies.

\begin{figure}
\centering
\includegraphics[width=0.43\textwidth]{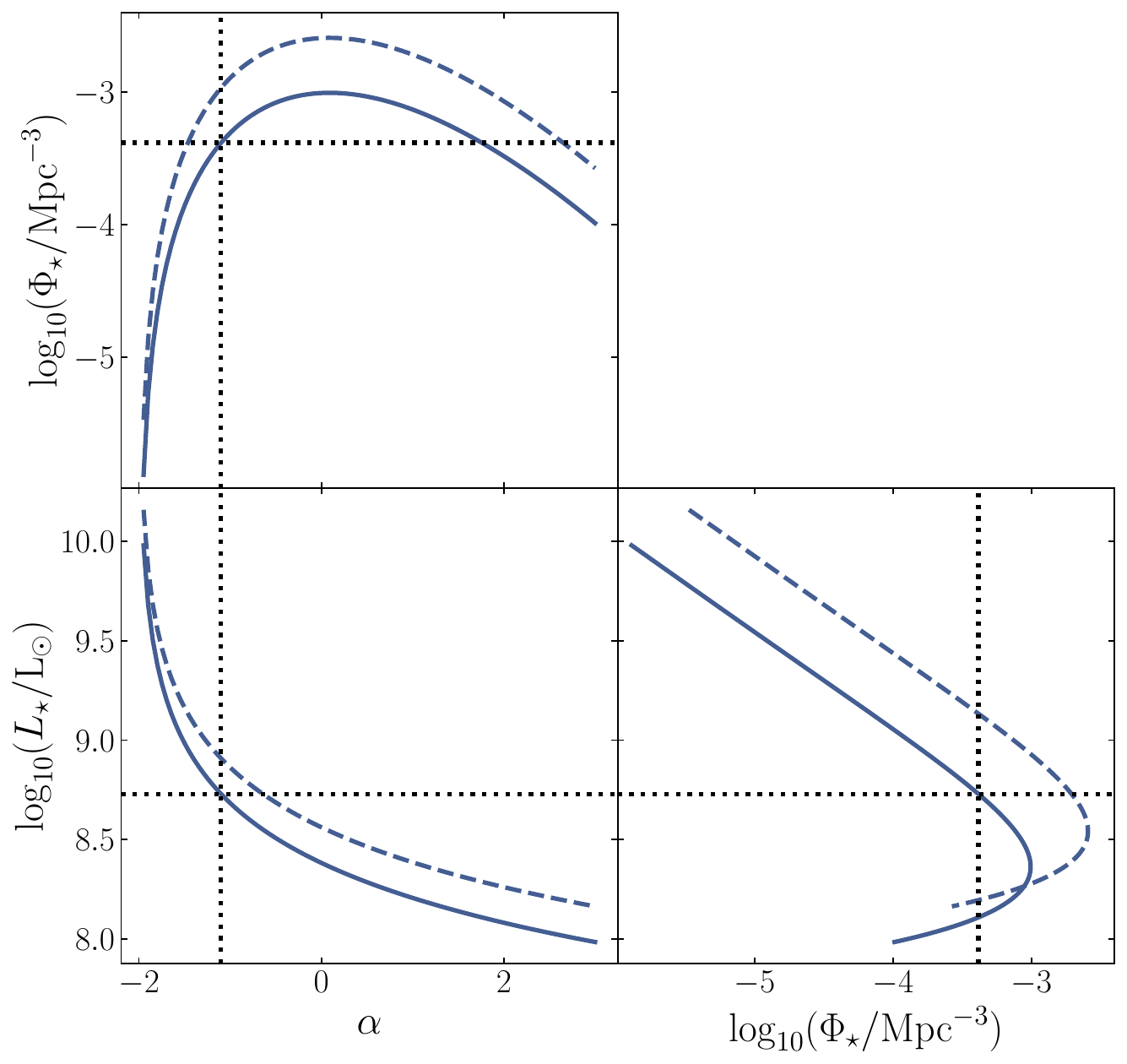} 
\caption{Triplets of the Schechter function parameters $(\Phi_*, L_*,$ and $\alpha)$ that give exactly the same
values of the first two moments as our pessimistic LF at $z=5$ (solid) and $3.6$ (dashed).}
\label{fig:degLF}
\end{figure}

If one is ready to assume that the LF has a particular functional form, then the constraints on the moments can be turned into constraints on the parameters.
These will be degenerate if the model for the LF contains more than two parameters.
For instance, assuming a Schechter function gives
\begin{align}
    \bar{\rho}_L&=\Gamma(\alpha+2)\,\Phi_*\,L_*\;,
    \\ \bar{\rho}_L^2\,\bar{n}_\mathrm{eff}^{-1}&=\Gamma(\alpha+3)\,
    \Phi_*\,L_*^2\;,
\end{align}
or, equivalently,
\begin{align}
    \frac{\bar{\rho}_L^2\,\bar{n}_\mathrm{eff}^{-1}}{\bar{\rho}_L}&=
    (\alpha+2)\,L_*\;,
    \\ \frac{\bar{\rho}_L^2}{\bar{\rho}_L^2\,\bar{n}_\mathrm{eff}^{-1}}&
    =\frac{\Gamma(\alpha+2)}{\alpha+2}\,\Phi_*\;,
\end{align}
where we have used the relation $\Gamma(1+x)=x\,\Gamma(x)$.
Fig.~\ref{fig:degLF} shows different projections of the degeneracy locus of the LF parameters corresponding to the actual first two momenta of our pessimistic case with $\alpha=-1.1$ at $z=5$ (solid) and 3.6 (dashed).
Uncertain constraints on the moments will thus be remapped to posterior distributions with support that elongates along these complex curves.

\subsection{Parameters of the luminosity function}
\label{sec:param}

A complementary approach, which we adopt in this section, is to constrain a parametric representation of the LF directly from the LIM PS. Specifically, we assume that the LF can be accurately described by a Schechter function and derive the joint posterior distribution of its three parameters, starting from the independent uniform prior distributions listed in the bottom part of Table~\ref{tab:priors}.

By construction, our implementation of this approach is not equivalent to the concept discussed at the end of Sect.~\ref{sec:moments}.
Indeed, there, we showed that the LIM PS can be used to set constraints of the parameters of the LF at $z=3.6$.
Conversely, here, to be consistent with the generation of our mock data presented in Sect.~\ref{sec:clust_shot-noise}, we perform the abundance matching at $z=5$ and we model the power spectra at $z\simeq 3.6$ by assuming that the function $\mathcal{L}(M)$ does not evolve in between. We therefore effectively set constraints on the LF at $z=5$, while in this model, the LF at $z=3.6$ is not even necessarily well described by a Schechter function.\footnote{With our assumptions, due to the evolution of the halo mass function between redshift 5 and 3.6, the first and second moments of the LF increase by a factor of a few, which is in the same ballpark of the variations seen in the \textsc{Marigold} simulations.}

An advantage of this framework is that it enables a joint analysis of the LIM PS at $z=3.6$ and the ALPINE LF data at $z=5$. Because the mock data were generated assuming the pessimistic LF scenario and no redshift evolution in $\mathcal{L}(M)$, however, the resulting constraints are intrinsically conservative. They should be interpreted as lower bounds on the performance of this method, with real data expected to yield tighter constraints as measurements improve.

\begin{figure*}
\centering
\includegraphics[width=0.47\textwidth]{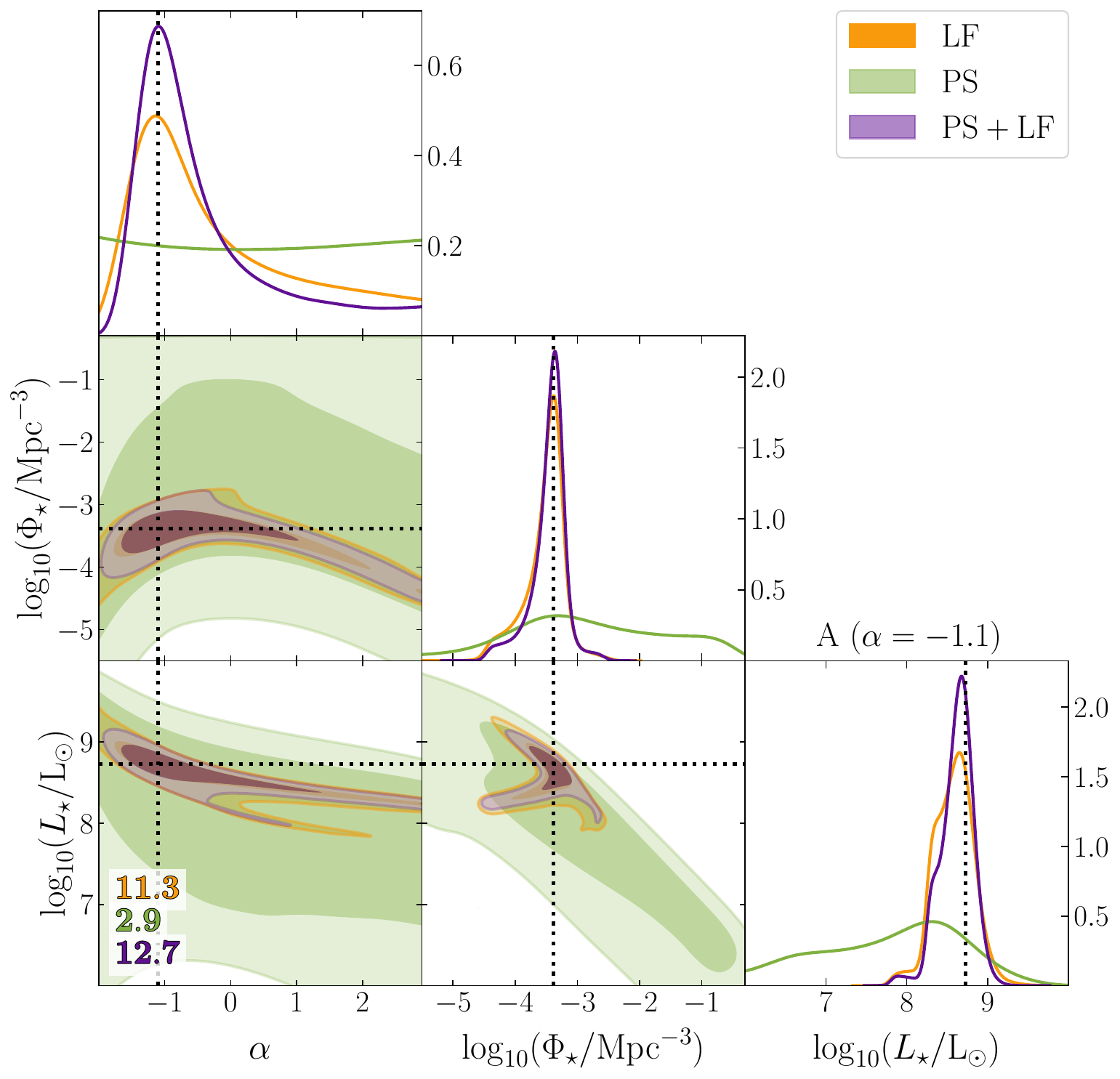} 
\hspace{0.3cm}
\includegraphics[width=0.47\textwidth]{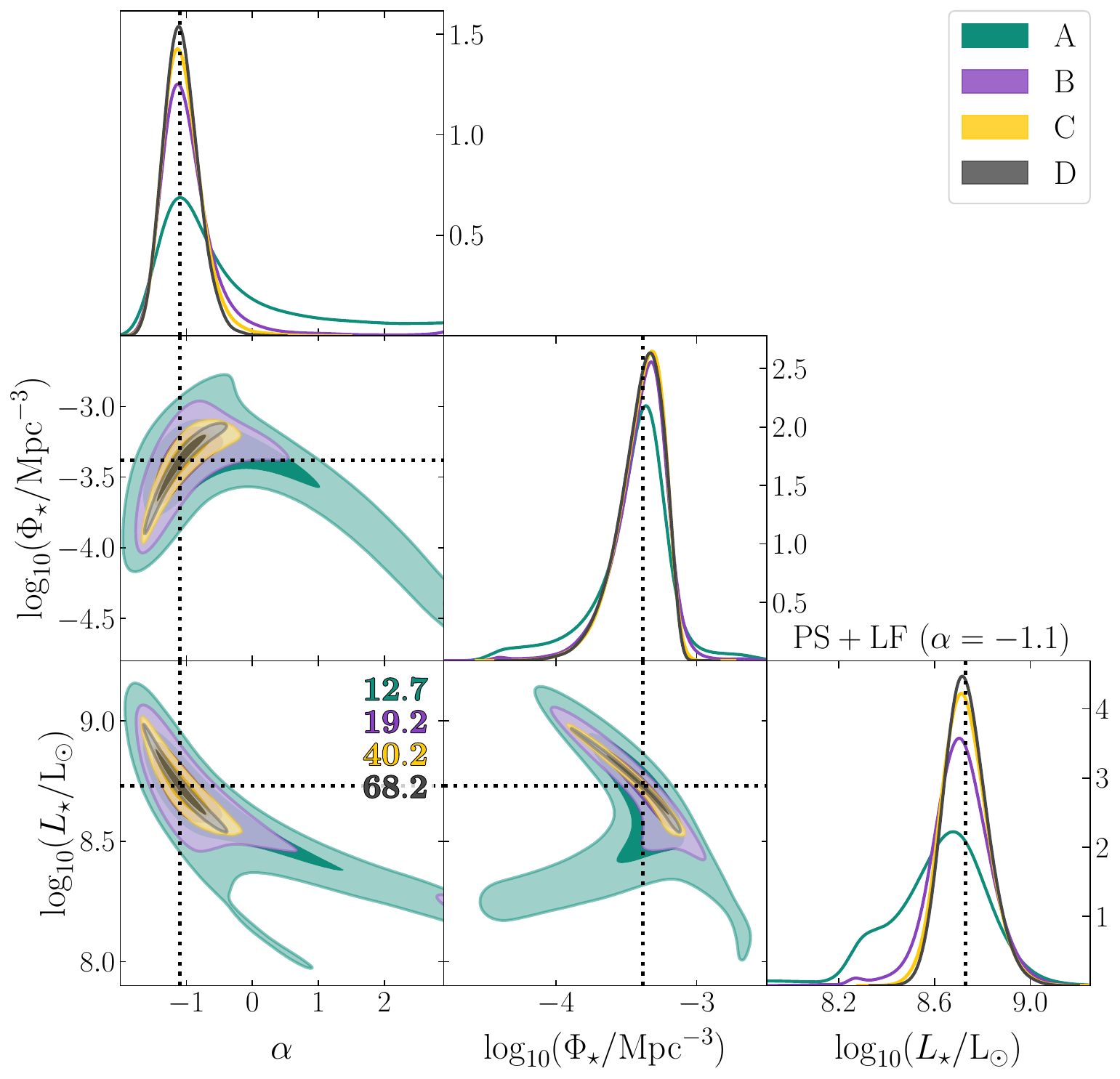}
\caption{
Left: Marginalised posterior distributions of the LF parameters obtained by fitting the LIM PS (green), the number counts of the targeted ALPINE survey (orange), and the combination of the two datasets (violet). The shaded areas indicate the 68\% and 95\% credible regions.
Right: As in the left panel, but for the fit of the combined datasets and for different LIM surveys.}
\label{fig:constraintsLF}
\end{figure*}

The marginalised posterior distribution of the model parameters given the LIM PS for survey A  is represented in the left panel of Fig.~\ref{fig:constraintsLF} using green tones.
It is evident that $\Delta^2(k)$ does not constrain $\alpha$ and that all the LF parameters are strongly correlated.
The contours of the posterior probability elongate along the solid degeneracy lines presented in Fig.~\ref{fig:degLF} but are, of course, broader as the moments of the LF are measured with an uncertainty.
A careful inspection reveals another small difference: the contours in the $\{\Phi_*,L_*\}$ plane close at low $\Phi_*$ (corresponding to $\alpha$ approaching $-2$) while the corresponding lines in Fig.~\ref{fig:degLF} are unlimited.
This happens because, in this region of parameter space, the contribution to $\bar{\rho}_L$ from emitters with $L\ll L_*$ is non-negligible but our halo model only considers haloes with $M>10^6\,h^{-1}$ M$_\sun$ and thus truncates the LF at the extreme faint-end ($L\lesssim 10$ L$_\sun$) underestimating $\bar{\rho}_L$ with respect to the idealised Schechter function.

For comparison, we fit the LF measurements from the ALPINE targeted detections (see Fig.~\ref{fig:fit_LF}) with the same Schechter function.
The corresponding posterior distribution is displayed with orange tones in the left panel of Fig.~\ref{fig:constraintsLF}.
The LF data better constrain the model parameters than the LIM PS: the orange shaded regions are narrower and the marginalised posterior for $\alpha$ shows a clear peak around the true value.

\begin{figure*}
\centering
\includegraphics[width=0.47\textwidth]
{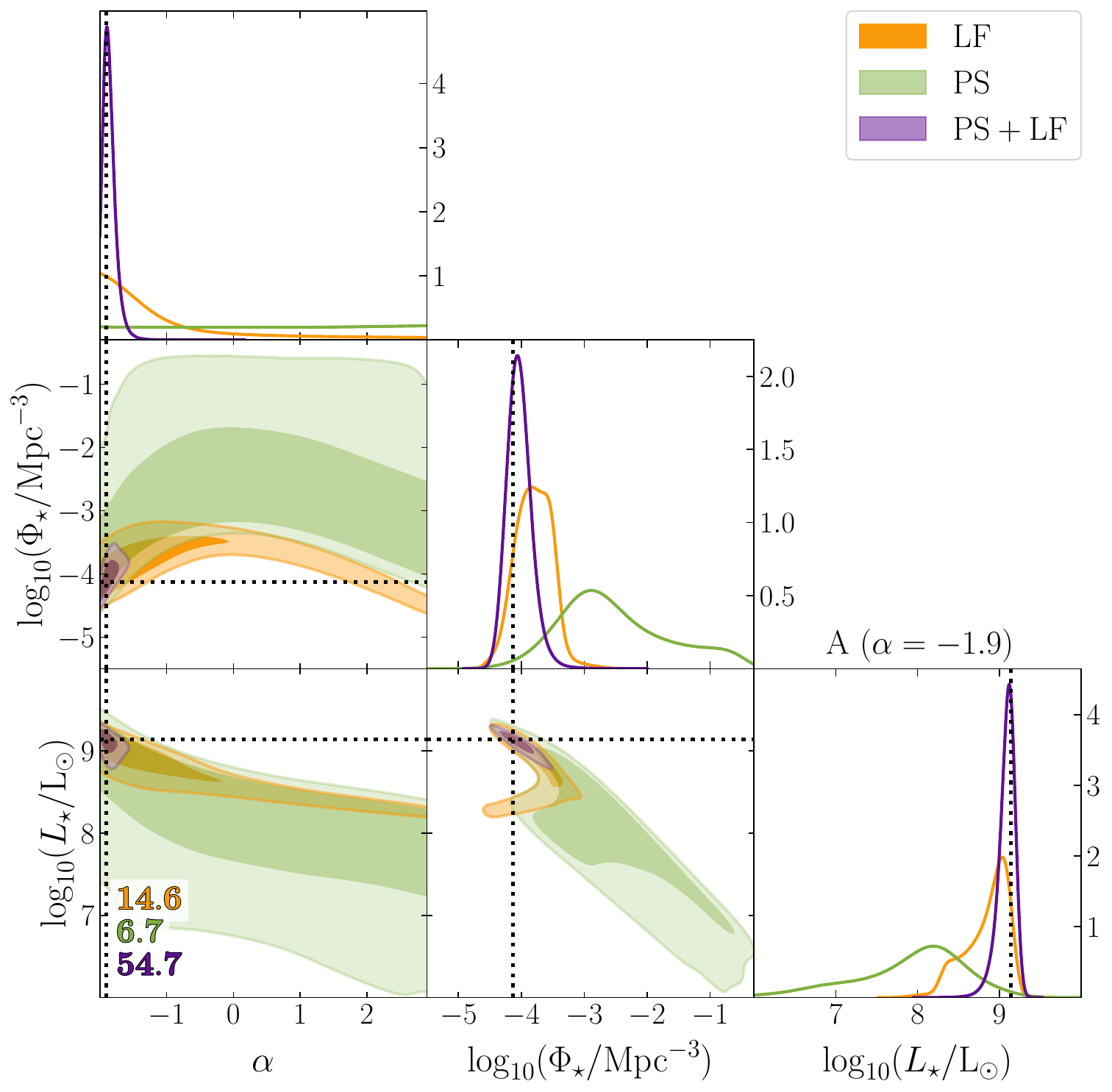}
\hspace{0.3cm}
\includegraphics[width=0.47\textwidth]{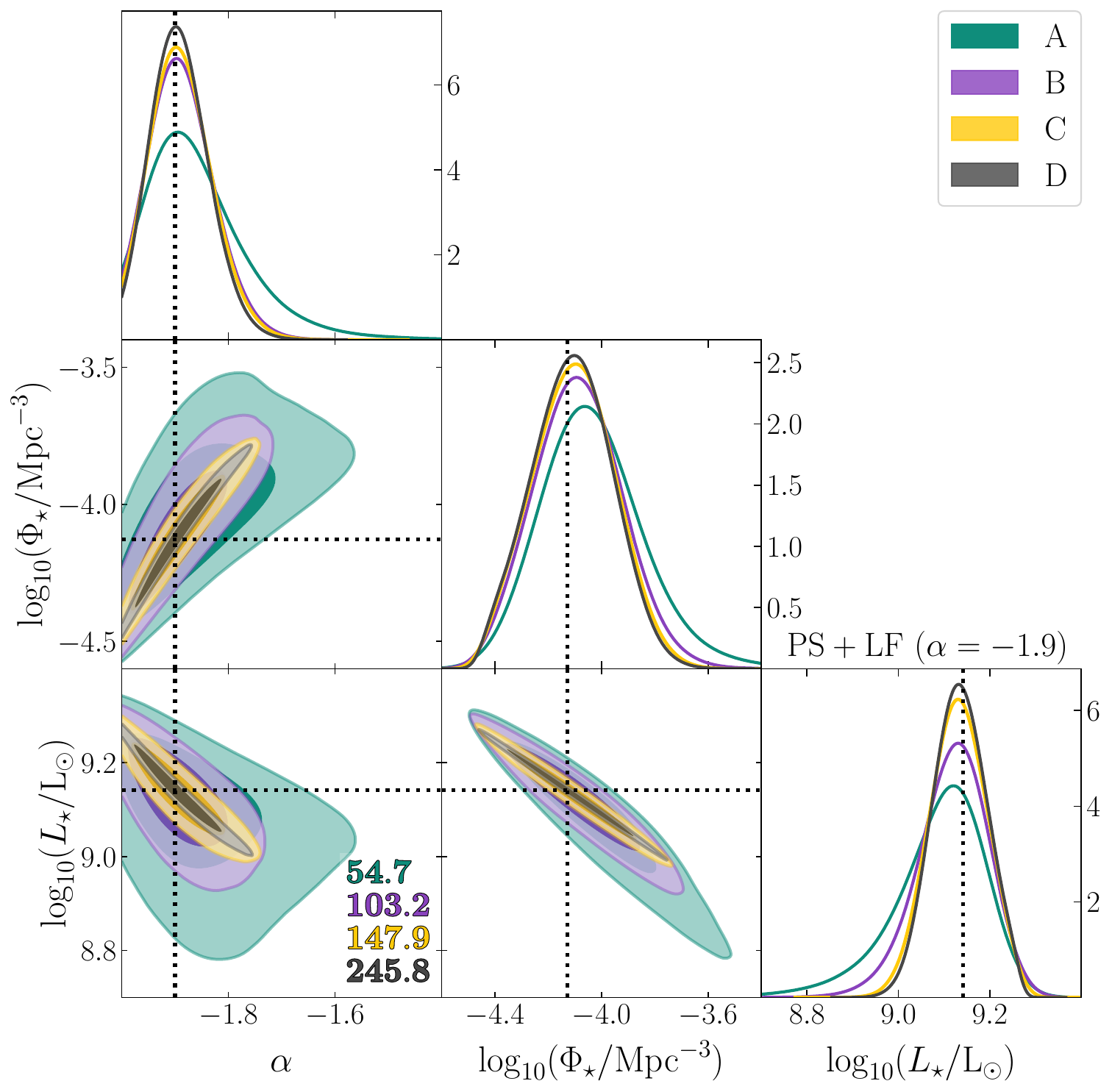} 
\caption{ As in Fig.~\ref{fig:constraintsLF}, but for the pessimistic case with a faint-end slope of $\alpha=-1.9$.}
\label{fig:constraints_parametersLF_joint}
\end{figure*}

Eventually, we fit the LIM and LF data simultaneously. The resulting posterior distribution is shown in Fig.~\ref{fig:constraintsLF} with violet tones in the left panel.
In order to compare the constraining power of the different data with a single number, we introduce a FoM defined analogously to Sect.~\ref{sec:nonparam} but for three parameters. The ALPINE LF provides constraints that are substantially tighter than the LIM PS (the FoM is a factor 3.9 smaller). However, the combination of the two datasets increases the FoM by a factor of 1.1 with respect to the fit to the LF alone.

In the right panel of Fig.~\ref{fig:constraintsLF}, we show the constraints on the Schechter function parameters to the joint LF+PS data for the different surveys. The contours and lines for survey A (teal) coincide with those presented in the left panel (violet) but the plot area is narrower here. Our results show that increasing the sensitivity (survey C) provides a much bigger improvement in the determination of the Schechter parameters with respect to enlarging the survey area (survey B).
The marginalised one-dimensional posterior distributions appear all very similar, however. The improvements mostly come from reducing the importance of the tails.

In Fig.~\ref{fig:constraints_parametersLF_joint}, we repeated the analysis using different mock data representing the pessimistic case with $\alpha=-1.9$. As we have discussed in Sect.~\ref{sec:clust_shot-noise}, a steeper faint-end slope corresponds to stronger clustering and shot-noise amplitudes (see Table~\ref{tab:bias-sn}) which increase the S/N of the PS measurements and thus the FoM of the corresponding fit. 
Since the ALPINE measurements of the LF allow $\alpha=-1.9$ but do not prefer it (see the orange contours in the left panel of Fig.~\ref{fig:constraintsLF}), for our analysis we generated mock LF data that sample a Schechter function with $\alpha=-1.9$ and have the same relative uncertainties as the ALPINE measurements.
In this case, the contours of the posterior distributions given the LF data and given the PS measurements are shifted in $\Phi_*$ and $L_*$ whenever $\alpha$ departs significantly from $-1.9$. 
Since they overlap only around the true values, the joint fit PS+LF has a much higher FoM than the individual ones.
For survey A, the marginalised uncertainties for the individual parameters (i.e., the standard deviations of the one-dimensional posteriors) are 4.9\% for $\log_\mathrm{10}[\Phi_*/\mathrm{Mpc}^{-3}]$, 1.2\% for $\log_\mathrm{10} (L_*/L_\sun)$, and 5.2\% for $\alpha$. For comparison, the corresponding figures for the pessimistic case with $\alpha=-1.1$ displayed in Fig.~\ref{fig:constraintsLF} are 8.6, 2.4 and 100.7\%, respectively.

\section{Summary}\label{sec:conclusions}

Measurements of the \cii LF at high redshift ($z\simeq 3-5$) remain highly uncertain because current observational capabilities are limited. We explored the potential of reconstructing the LF from the LIM PS that will be measured with next-generation instruments.
The first challenge we faced was to reliably predict the expected PS signal. To achieve this goal, we combined empirical constraints from the ALPINE survey with theoretical insights from the \textsc{Marigold} simulations \citep{Khatri+24_marigold}, which include a detailed model of \cii emission from early galaxies.
By analysing the simulations, we drew the following conclusions.
\begin{enumerate}[label=(\roman*)]
    \item Although individual DM haloes typically host multiple \cii emitters, the total \cii luminosity is dominated by the central galaxy (Fig.~\ref{fig:CLF-sims} and Table~\ref{tab:clfparam}).
    \item The abundance-matching technique can be used to statistically connect \cii emitters to haloes. This yields an excellent approximation for the first two moments of the CLF
    (Figs.~\ref{fig:test-am} and \ref{fig:test-am2}).
    \item The halo-occupation properties of \cii emitters evolve very little from $z=5$ to 3.6 (see e.g. the dotted lines in Fig.~\ref{fig:HAM_CDM}). 
\end{enumerate}
Building on the insights gained from the simulations, we used an abundance match of the \cii LF observed by the ALPINE survey in the redshift range $4.4<z<5.9$ to the halo mass function and thereby derived the mean \cii luminosity as a function of halo mass $\mathcal{L}(M)$ (Fig.~\ref{fig:HAM_CDM}). We bracketed the uncertainty on the LF by considering two different scenarios: an optimistic high-normalisation case based on the data compilation of Y20, and a pessimistic low-normalisation case informed solely by the targeted ALPINE detections.
In the latter, the faint-end slope (which remains poorly constrained) was treated as a free parameter.

We then combined the halo model (reviewed in Sect.~\ref{sec:halomodel}) with the function $\mathcal{L}(M)$ to predict the expected LIM PS, for which we incorporated corrections for instrumental and observational effects.
To illustrate the current possibilities, we used the specifications of the EoR-Spec instrument, which is soon to be deployed on FYST, as our reference case.
The resulting PS at $z\simeq3.6$ is shown in Fig.~\ref{fig:powerspectrum_rectangular}.

In the second part of the paper, we presented forecasts for the FYST DSS at $z\simeq3.6$ and extended our predictions to include prospective future surveys featuring a broader sky coverage, an enhanced sensitivity, and/or an increased spectral resolving power.
The main conclusions from our Bayesian analysis are summarised below.
\begin{enumerate}[label=(\roman*)]
\setcounter{enumi}{3}
\item The DSS is expected to constrain both the clustering and shot-noise components of the LIM PS with a S/N of approximately 2 or higher, depending on the true underlying LF (Fig.~\ref{fig:constraints_momentsLF}). 
It does not provide sufficient information to constrain the linear bias parameter of the LIM signal, however (Fig.~\ref{fig:bposterior}). As a result, the first moment of the LF is noticeably biased (Fig.~\ref{fig:constraints_derived_momentsLF}).
In contrast, the second moment is accurately recovered. 
\item Even for more sensitive and wider surveys, the damping effect due to the non-linear redshift-space distortions cannot be disentangled from the overall LIM signal (Fig.~\ref{fig:fullcorneroptimistic}).
This is primarily due to the limited spectral resolving power of the instrument ($R=100$), which induces a strong suppression of the power spectrum along the line of sight that often dominates over the milder damping caused by peculiar velocities (quantified by the parameter $\sigma$).
As a result, the imprint of redshift-space distortions is largely erased, which makes it difficult to isolate this effect from the measured signal. 
This also leads to slightly biased constraints on the clustering amplitude (Fig.~\ref{fig:constraints_momentsLF}). The shot-noise level, on the other hand, is determined with both high precision and accuracy (in particular for surveys C and D).
\item Increasing the resolution to $R = 500$ allows the damping from redshift-space distortions to become distinguishable from instrumental effects, enabling a constraint on $\sigma$. However, this intensifies degeneracies with the clustering amplitude $\bar{I}_\nu^2 b^2$ and reduces precision (Fig.~\ref{fig:fullcorneroptimistic}). Importantly, fitting the data with a model that neglects redshift-space distortions leads to substantial biases in both the clustering and shot-noise amplitudes (Fig.~\ref{fig:RSDs}).
\item Tight and accurate constraints on the first two moments of the LF translate into strong degeneracies when LF models are fitted with more than two free parameters (e.g. the Schechter function; Fig.~\ref{fig:degLF}).
\item To address this limitation, we adopted an alternative approach by modelling the LF as a Schechter function and directly constraining its free parameters through a joint fit to the PS (e.g. from the DSS) and the LF measurements (e.g. from ALPINE).
We found that the overall normalisation, $\Phi_*$, and the characteristic luminosity, $L_*$, are both precisely and accurately determined (Figs.~\ref{fig:constraintsLF} and \ref{fig:constraints_parametersLF_joint}), while the faint-end slope, $\alpha$, remains largely unconstrained unless its true value is close to $-2$.
\item In all scenarios, a survey sensitivity higher by a factor of $\sqrt{10}$ at the same sky coverage yielded substantially tighter constraints than a surveyed area larger by a factor of 10 at fixed sensitivity 
(Figs.~\ref{fig:constraints_momentsLF}, \ref{fig:constraints_derived_momentsLF}, \ref{fig:constraintsLF} and \ref{fig:constraints_parametersLF_joint}).
\end{enumerate}
These results underscore important design trade-offs in LIM survey planning. Gains in one area, such as a higher spectral resolution, can improve the sensitivity to specific physical effects, like the damping scale from redshift-space distortions. However, these improvements can also amplify degeneracies between key parameters, especially when the number of independent observables is limited. This emphasises the need to optimise survey configurations in the light of the specific scientific objectives being pursued, whether they involve precise measurements of clustering, bias, or luminosity function moments. We plan to return to this issue in future work by exploring systematic strategies for designing LIM surveys tailored to distinct science goals.

\begin{acknowledgements}
The authors warmly thank Christos Karoumpis, Ankur Dev, Dominik Riechers and Frank Bertoldi for helpful discussions about the DSS survey and the CCAT-prime project. The authors gratefully acknowledge the Collaborative Research Center 1601 (SFB 1601 sub-project C6) funded by the Deutsche Forschungsgemeinschaft (DFG, German Research Foundation) -- 500700252. They also acknowledge the International Max Planck Research School for Astronomy and Astrophysics (IMPRS A\&A) at the Universities of Bonn and Cologne for supporting EM through a research contract.
EM and PK are members of the IMPRS A\&A, the Bonn Cologne Graduate School (BCGS), and guest researchers at the Max Planck Institute for Radio Astronomy (MPIfR) in Bonn.
CP is grateful to SISSA, the University of Trieste, and IFPU, where part of this work was carried out, for hospitality and support.
\end{acknowledgements}

\bibliographystyle{aa}
\bibliography{biblio}

\clearpage

\begin{appendix}

\onecolumn

\section{Line profile from FPI scans}
\label{sec:SCANS}
In the lossless approximation, the fraction of incident intensity transmitted by a Fabry-Pérot Interferometer (FPI) at a fixed cavity spacing is described by the Airy function:
\begin{align}
    T(\nu) =\frac{1}{1+F\sin^2(\delta/2)} \,,
\end{align}
where the coefficient of finesse\footnote{The actual finesse $\mathcal{F}$ of the interferometer, defined as the ratio between the free spectral range (i.e. the frequency spacing between adjacent transmission peaks) and the full width at half maximum (FWHM) of each peak, is given by $\mathcal{F} = \pi \sqrt{F}/2$.} is 
$F = 4 \mathcal{R} / (1 - \mathcal{R})^2$ and depends on the mirror reflectivity $\mathcal{R}$. The phase shift between successive internally reflected beams is
\begin{align}
    \delta = \frac{4\pi n d \nu}{c} \sqrt{1 - \left(\frac{\sin \theta}{n}\right)^2} \,,
\end{align}
where $d$ is the physical separation of the mirrors, $n$ is the refractive index of the medium between them, $\nu$ is the frequency of the incoming light, and $\theta$ is the external incidence angle. Resonances occur when $\delta = 2\pi m$, with $m$ an integer corresponding to the interference order.

Near a resonance at frequency $\nu_0$, the transmission function is well approximated by a peak-normalised Lorentzian

\begin{align}
    T(\nu) \simeq \frac{1}{1 + \left( \frac{\nu - \nu_0}{\Gamma/2} \right)^2} \,,
\end{align}
where the FWHM is $\Gamma = \nu_0 / (m \mathcal{F})$, yielding a resolving power $R = m \mathcal{F}$.
The corresponding area-normalised line profile is
\begin{align}
    L(\nu;\nu_0) = \frac{1}{\pi} \frac{\Gamma/2}{(\nu - \nu_0)^2 + (\Gamma/2)^2} \,,
\end{align}
which, when translated into comoving radial distance $r$, becomes
\begin{align}
    L(r;r_0) = \frac{1}{\pi} \frac{\Delta_\parallel^\mathrm{FWHM}/2}{(r - r_0)^2 + (\Delta_\parallel^\mathrm{FWHM}/2)^2} \,,
\end{align}
where $\Delta_\parallel^\mathrm{FWHM}$ is the FWHM expressed in comoving distance units.
Its Fourier transform is
\begin{align}
    \tilde{L}(k_\parallel) = e^{i k_\parallel r_0} \, e^{ - |k_\parallel| \Delta_\parallel^\mathrm{FWHM}/2 } \,,
\end{align}
which leads to a window function for the power spectrum:
\begin{align}
    W_\parallel(k_\parallel) = |\tilde{L}(k_\parallel)|^2 = e^{ - |k_\parallel| \Delta_\parallel^\mathrm{FWHM} } \,.
\end{align}
As discussed in Sect.~\ref{sec:resolution}, this exponential suppression significantly damps the clustering signal on small radial scales, particularly in the presence of low resolving power.

If the FPI is scanned over a sequence of discrete frequency steps, which are subsequently combined to construct an intensity map in voxels, then the effective line profile can be approximated as a sum of Lorentzian functions, each centred at a different frequency step. This results in a multi-peaked or broadened profile, depending on the number of steps and their spacing relative to the FWHM of the individual Lorentzian. 

If the system throughput (e.g., atmospheric transmission) or the integration time varies across the scanning process, each Lorentzian should be weighted accordingly. Denoting by $\nu_i$ the centre of the $i$-th scan step and by $w_i$ the corresponding weight, the composite line profile for a voxel centred at $\nu_0$ can be written as
\begin{align}
    T(\nu) &\simeq \sum_{i=1}^N w_i\,L(\nu;\nu_i) \;,
\end{align}
where $L(\nu;\nu_i)$ denotes a Lorentzian profile centred at $\nu_i$. The resulting window function in Fourier space becomes
\begin{align}
    W_\parallel(k_\parallel) = e^{-|k_\parallel| \Delta_\parallel^\mathrm{FWHM}} \,
    \left| \sum_{i=1}^N w_i\,e^{i k_\parallel r_i} \right|^2 \;,
    \label{eq:windowscan}
\end{align}
where $r_i$ is the comoving distance corresponding to $\nu_i$. This structure can lead to a non-trivial modulation of the damping depending on the scanning scheme.

Further analytic insight can be obtained by assuming uniform weights ($w_i = 1/N$), implying negligible variations in system throughput and equal integration time across all steps. If the scan steps are evenly spaced in frequency by $\Delta \nu_\mathrm{step}$, the summation in Eq.~(\ref{eq:windowscan}) produces a comb-like interference pattern that modulates the baseline Lorentzian damping. In the limit of large $N$, the discrete scan approximates a continuous sweep, resulting in a convolution of the Lorentzian with a top-hat function. This leads to the approximate window function:
\begin{align}
    W_\parallel(k_\parallel) = e^{-|k_\parallel|\,\Delta_\parallel^\mathrm{FWHM}} \,
    \left[\frac{\sin(k_\parallel\,\delta_\parallel/2)}{k_\parallel\,\delta_\parallel/2}\right]^2 \;,
    \label{eq:wparappendix}
\end{align}
where $\delta_\parallel$ is the comoving length corresponding to the total scanned frequency range that is mapped to a frequency channel. For finite $N$, the sinc modulation becomes more structured, resulting in additional suppression at large $|k_\parallel|$.

\section{Number of Fourier modes}
\label{sec:Nmodes}
Let us consider a real-valued field defined within a rectangular cuboid of size $(L_\perp, L_\perp, L_\parallel)$ and volume $V = L_\perp^2 L_\parallel$, assuming periodic boundary conditions. The Fourier transform of such a field yields discrete wavevectors
of the form $\mathbf{k} = (i,k_\mathrm{f}^\perp, j,k_\mathrm{f}^\perp, \ell,k_\mathrm{f}^\parallel)$, where $(i, j, \ell) \in \mathbb{Z}^3$ and $k_\mathrm{f}^{\perp/\parallel} = 2\pi / L_{\perp/\parallel}$ defines the fundamental mode in each direction. Basically, each discrete mode occupies a cell of volume $K_\mathrm{f} = (k_\mathrm{f}^{\perp})^2 \, k_\mathrm{f}^\parallel = (2\pi)^3 / V$ in Fourier space.
Due to the Hermitian symmetry of the Fourier transform, only half of the modes carry independent information. The number of independent modes within a thin spherical shell of radius $k$ and thickness $\Delta k \ll k$ can be estimated by dividing the volume of the shell $K_\mathrm{shell}=2\pi \,k^2 \Delta k$ (accounting for the hemisphere) by $K_\mathrm{f}$, yielding the classical result:
\begin{equation}
    N_k=\frac{k^2\,\Delta k}{4\pi^2}\,V\;.
\end{equation}

However, if the parallel component of the wavevector is restricted to $|k_\parallel| \leq k_\parallel^\mathrm{max}$, then the available volume within the shell must be reduced accordingly. Specifically, for $k > k_\parallel^\mathrm{max}$, a spherical cap of volume $2\pi\, k \,(k - k_\parallel^\mathrm{max})\, \Delta k$ must be subtracted from the hemisphere. As a result, the effective shell volume becomes:
\begin{equation}
    K_\mathrm{shell}=
    \begin{cases}
        2\pi\, k^2\,\Delta k\;, &\text{if $k\leq k_\parallel^\mathrm{max}$}\;,\\
        2\pi\,k\,k_\parallel^\mathrm{max}\,\Delta k\;,
        & \text{otherwise}\;.
    \end{cases}
\end{equation}
The number of available modes per bin then reads:
\begin{equation}
N_k=\frac{k\,\min(k,k_\parallel^\mathrm{max}) \, \Delta k}{4\pi^2}\,V\;.    
\end{equation}

\section{Results with $R=500$}
\label{Sect:R500}
In this appendix, we present a set of figures analogous to those shown in Sect.~\ref{sec:nonparam} of the main text, but corresponding to a spectral resolving power of $R = 500$ instead of the baseline value of $R = 100$. This allows us to assess the impact of increased spectral resolution on the various quantities of interest.

\begin{figure}[h!]
\centering
\includegraphics[width=0.4\textwidth]{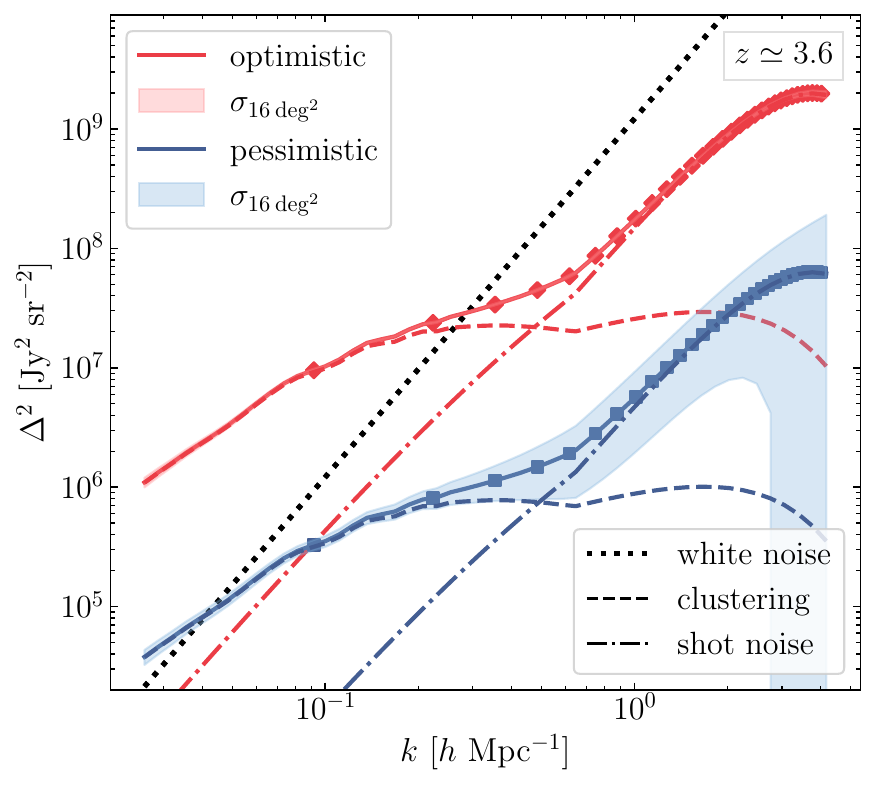} \hspace{0.5cm} \includegraphics[width=0.4\textwidth]{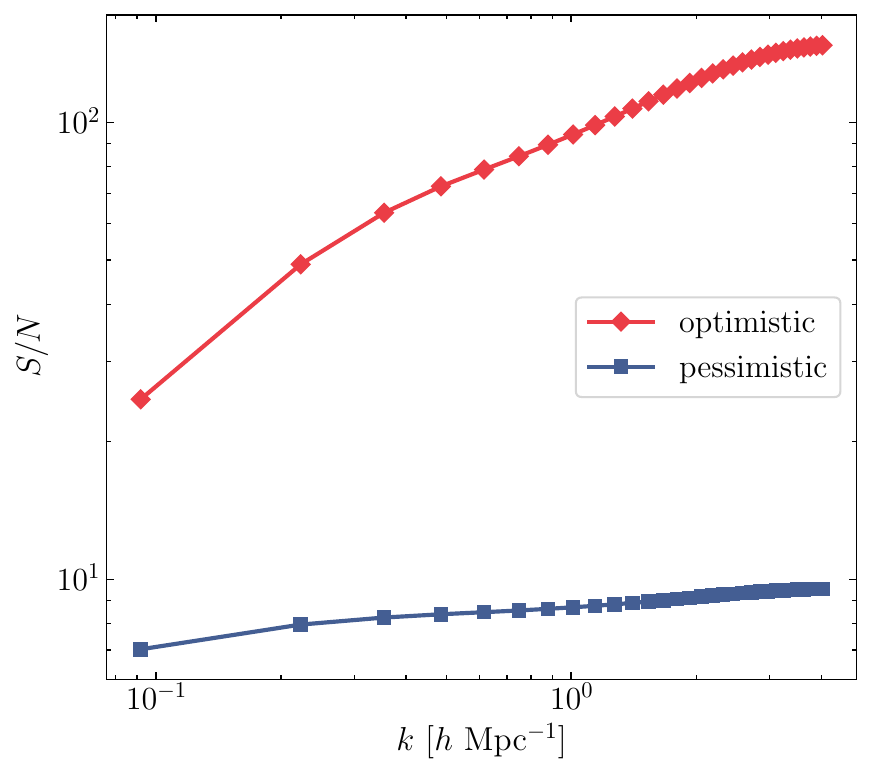} 
\caption{
Same as Fig.~\ref{fig:powerspectrum_rectangular} but for $R=500$.}
\label{fig:powerspectrum_R500}
\end{figure}

\begin{figure}[h!]
\centering
\includegraphics[width=0.4\textwidth]{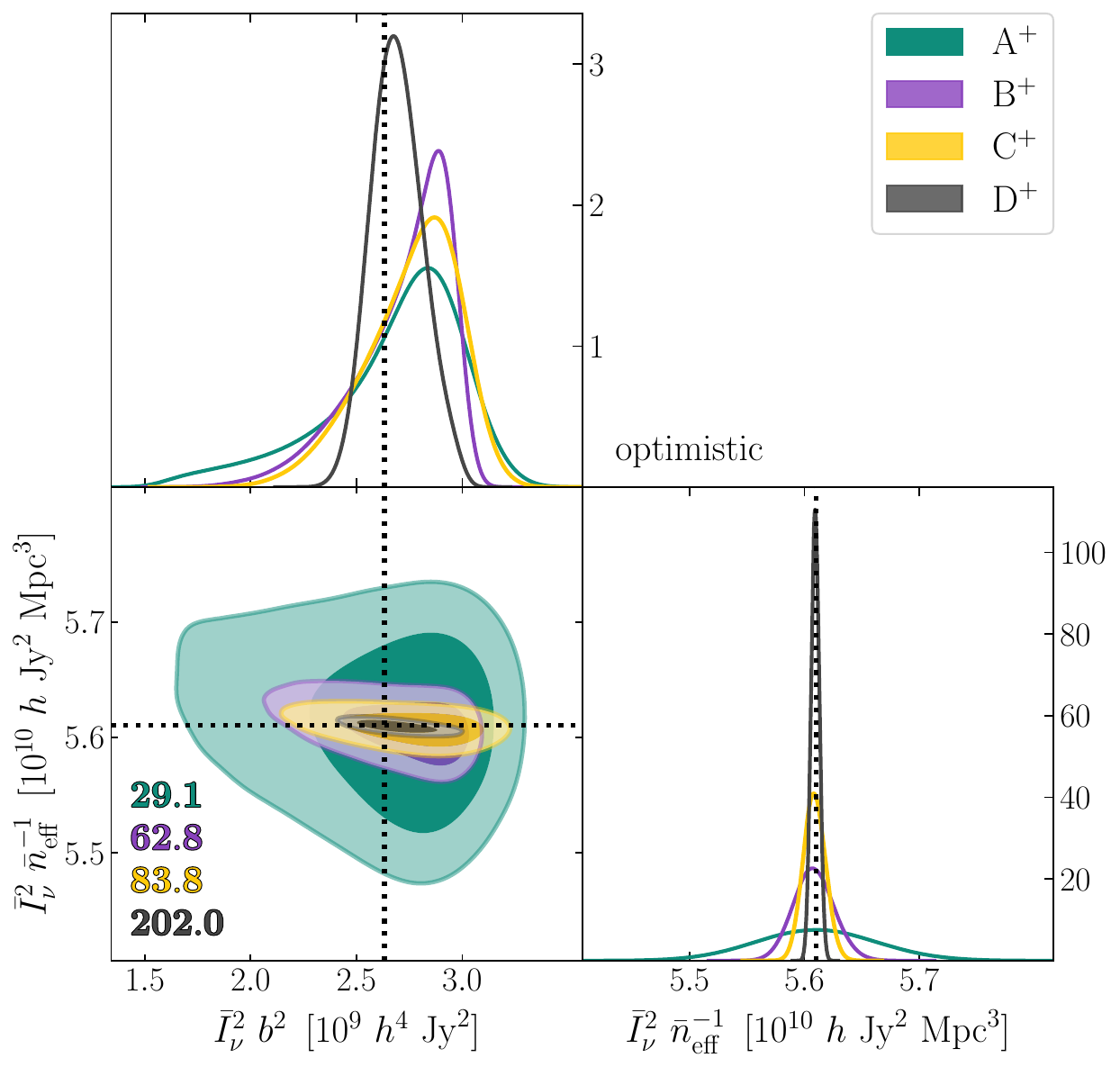} \hspace{0.5cm}
\includegraphics[width=0.4\textwidth]{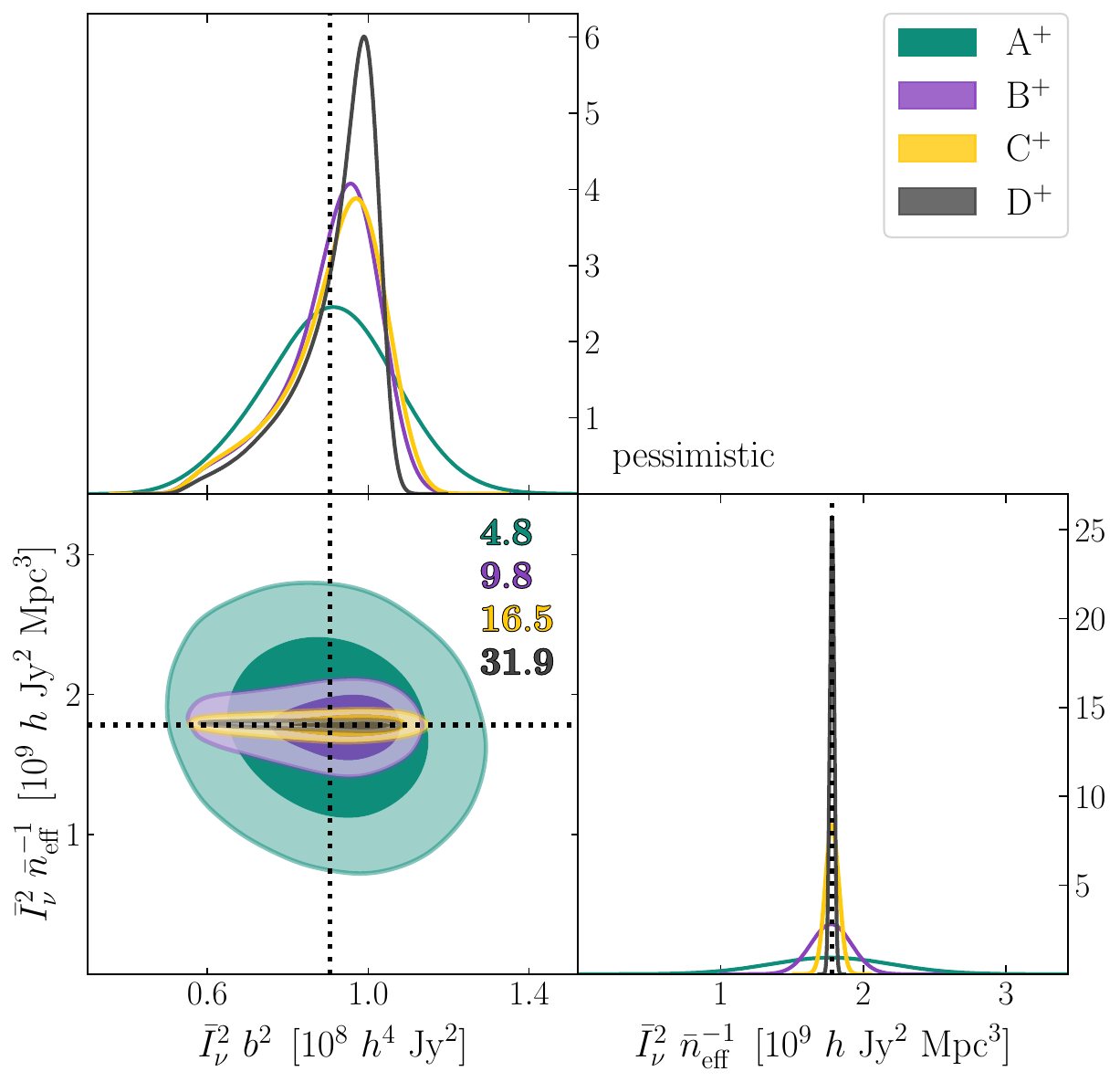} 
\caption{Same as Fig.~\ref{fig:constraints_momentsLF}, but for $R=500$.}
\label{fig:constraints_momentsLF_R500}
\end{figure}

\begin{figure}[h!]
\centering
\includegraphics[width=0.4\textwidth]{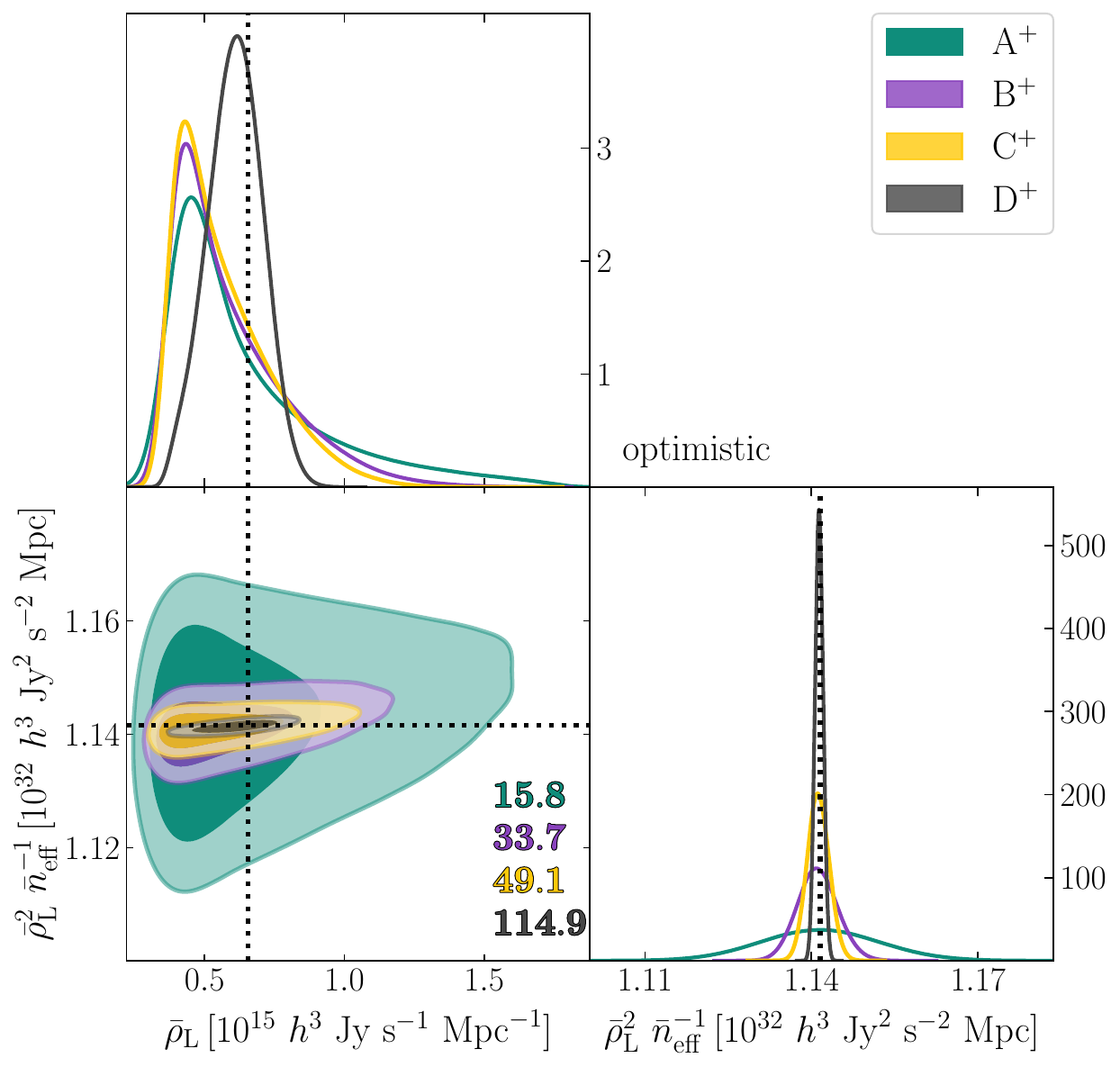} \hspace{0.5cm}
\includegraphics[width=0.4\textwidth]{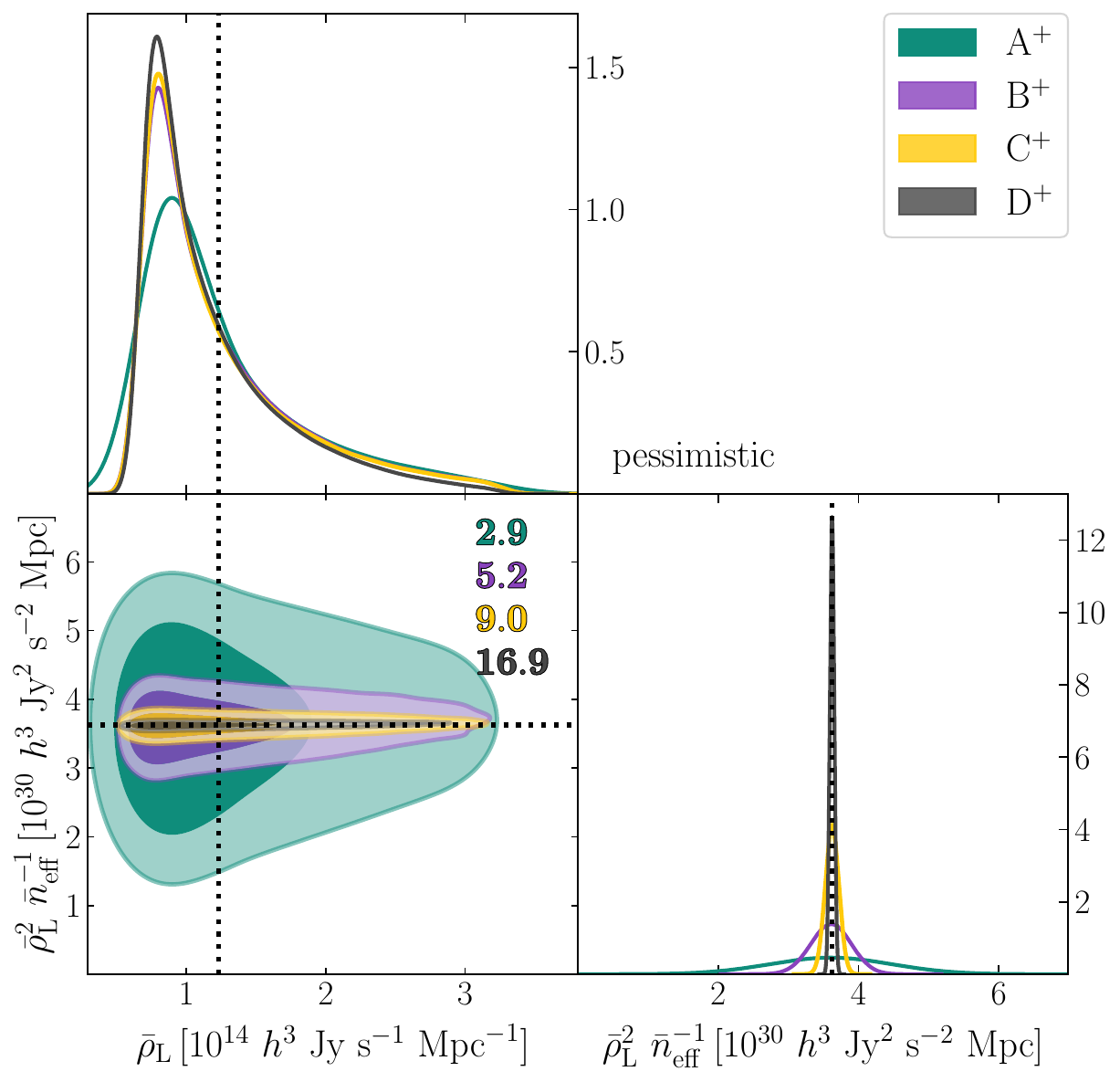} 
\caption{Same as Fig.~\ref{fig:constraints_derived_momentsLF}, but for $R=500$.}
\label{fig:constraints_derived_momentsLF_R500}
\end{figure}

\begin{figure}[h!]
\centering
\includegraphics[width=0.31\textwidth]{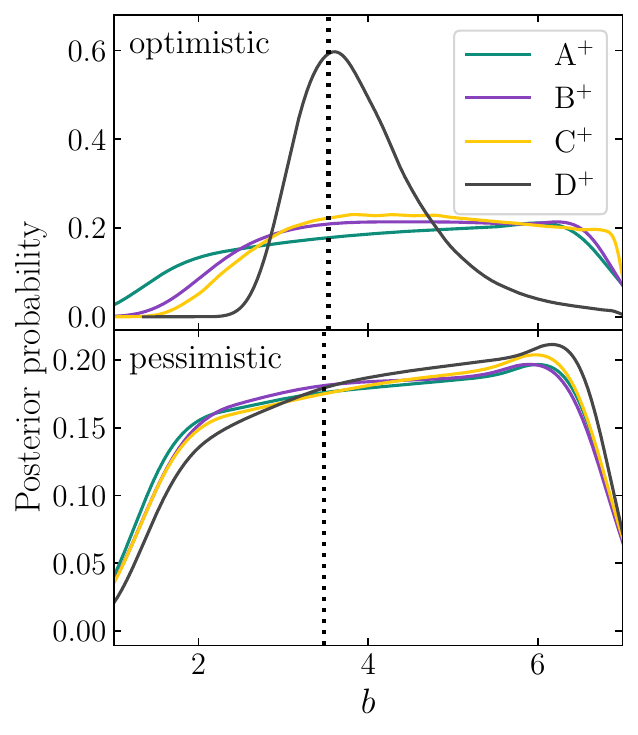}
\caption{Same as Fig.~\ref{fig:bposterior}, but for $R=500$.}
\label{fig:bposterior_R500}
\end{figure}

\begin{figure}[h!]
\centering
\includegraphics[width=0.3\textwidth]{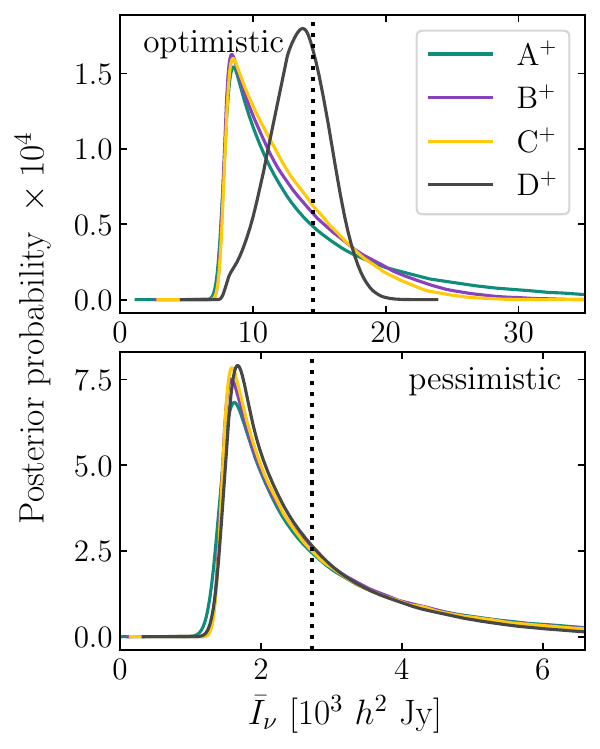} \hspace{1.5cm}
\includegraphics[width=0.3\textwidth]{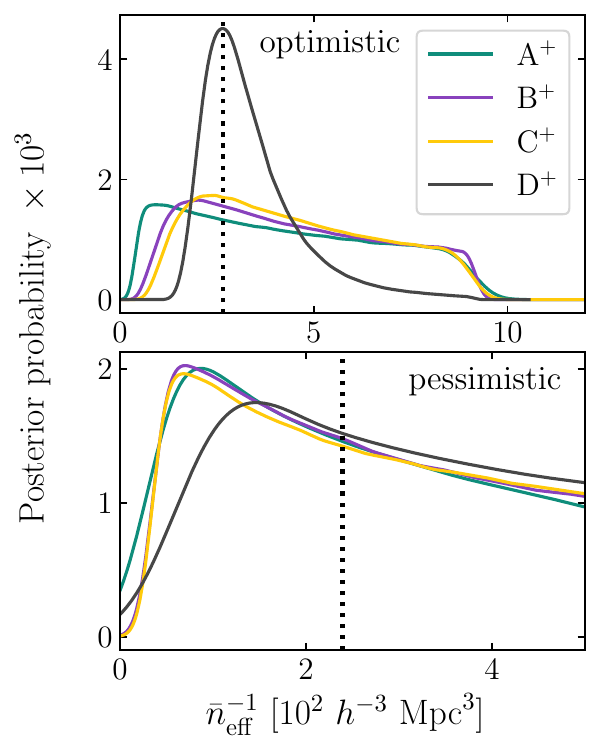} 
\caption{Same as Fig.~\ref{fig:Iposterior}, but for $R=500$.}
\label{fig:Iposterior_R500}
\end{figure}

\vspace*{20cm}

\section{Marginal prior for the mean intensity}
\label{sec:I-prior}
Let us consider two model parameters \(x>0\) and \(y>0\) with  
independent uniform priors on \( x \in [x_\mathrm{min}, x_\mathrm{max}] \) and \( y \in [y_\mathrm{min}, y_\mathrm{max}] \).
The effective prior on the derived parameter
\[
z = \sqrt{\frac{x}{y^2}} = \frac{\sqrt{x}}{y}\;,
\]
must be computed by marginalising over the joint prior distribution
\begin{align}
p(z)&=\iint \delta_\mathrm{D}\left(z-\frac{\sqrt{x}}{y}\right)\,p(x,y) \,\der x\, \der y \nonumber\\
&=\iint 2\sqrt{x}\,y\,\delta_\mathrm{D}\left(x-z^2y^2\right)\,p(x,y) \,\der x\, \der y\nonumber \\
&=2z\,\int y^2\,p(z^2y^2,y)\,\der y
\;,
\end{align}
where $p(x,y)$ assumes the constant value $A=[(x_\mathrm{max}-x_\mathrm{min})\,(y_\mathrm{max}-y_\mathrm{min})]^{-1}$ within the prior ranges and vanishes otherwise.
To compute the marginal prior for \( z \), we must integrate over \( y \), applying the constraint that \( x = z^2 y^2 \) remains within the allowed range:
\[
x_\mathrm{min} \leq z^2 y^2 \leq x_\mathrm{max}
\quad \Rightarrow \quad
\sqrt{\frac{x_\mathrm{min}}{z^2}} \leq y \leq \sqrt{\frac{x_\mathrm{max}}{z^2}}.
\]
However, \( y \) must also satisfy \( y \in [y_\mathrm{min}, y_\mathrm{max}] \), so the integration bounds become
\[
y \in \left[
\max\left(y_\mathrm{min}, \sqrt{\frac{x_\mathrm{min}}{z^2}} \right),
\min\left(y_\mathrm{max}, \sqrt{\frac{x_\mathrm{max}}{z^2}} \right)
\right].
\]

Inserting the appropriate values from the main text for $x=\bar{I}_\nu^2 b^2$, $y=b$, and $z=\bar{I}_\nu$,
this yields
\begin{equation}
   p(\bar{I}_\nu)=
   \begin{cases}
       \displaystyle{\frac{2A}{3}}\,342\,\bar{I}_\nu\;, &\text{if $0<\bar{I}_\nu<\displaystyle{\frac{10^5}{7}}\,h^2$ Jy}\;, \\
       \displaystyle{\frac{2A}{3}}\,\left(\frac{10^{15}}{\bar{I}_\nu^2}-\bar{I}_\nu \right) & \text{if $\displaystyle{\frac{10^5}{7}}<\bar{I}_\nu<10^5\,h^2$ Jy}\;, \\
       0 &\text{otherwise}\;,
   \end{cases}
\end{equation}
with $A={(6\times 10^{10}\,h^2\ \text{Jy})}^{-1}$.
The effective prior thus peaks at $\bar{I}_\nu=10^5/7\simeq 14.3\times 10^3\,h^2$ Jy (which is close to the true value for the optimistic case), grows linearly
with $\bar{I}_\nu$ for smaller values,
and drops off approximately as $\bar{I}_\nu^{-2}$ for larger values until it vanishes at $\bar{I}_\nu=10^5\,h^2$ Jy.

A similar approach can be used to derive the marginalised prior on $\bar{n}_\mathrm{eff}^{-1}$; however, we omit the detailed calculations here, as the resulting expressions are rather lengthy and cumbersome.

\end{appendix}
\end{document}